\documentclass[preprint,prd,tightenlines,floatfix,
nofootinbib,eqsecnum,superscriptaddress]{revtex4-1}

\usepackage[T1]{fontenc}		

\usepackage{amsmath,amsfonts,amssymb,amstext,mathrsfs}
\usepackage{mathpazo}

\usepackage[dvips]{graphicx}
\usepackage{epsf,float}
\usepackage{revsymb}

\usepackage{dcolumn}
\usepackage{braket}
\usepackage{color,xcolor}
\usepackage{graphicx}
\usepackage{subfigure}
\usepackage{multirow}
\usepackage{tabularx}
\usepackage{pstricks}
\usepackage[section]{placeins}
\usepackage{booktabs}
\usepackage{array}

\usepackage{tablefootnote}

\usepackage{commath}

\usepackage{hyperref}

\usepackage{lineno}
\usepackage{hyphenat}

\usepackage{bm}

\newcommand{\be}{\begin{equation}}
\newcommand{\ee}{\end{equation}}
\newcommand{\ba}{\begin{align}}
\newcommand{\eda}{\end{align}}
\newcommand{\nn}{\nonumber}

\newcommand{\Pom}{\mathbb{P}}

\newcommand{\Reg}{\mathbb{R}}

\newcommand{\bk}{\mbox{\boldmath $k$}}
\newcommand{\bpa}{\mbox{\boldmath $p_{a}$}}

\newcommand{\bpaa}{\mbox{\boldmath $p_{1}$}}

\newcommand{\bhpaa}{\mbox{\boldmath $\hat{p}_{1}$}}

\newcommand{\blaperp}{\mbox{\boldmath $l_{1 \perp}$}}
\newcommand{\blbperp}{\mbox{\boldmath $l_{2 \perp}$}}
\newcommand{\bkperp}{\mbox{\boldmath $k_{\perp}$}}

\newcommand{\bpaap}{\mbox{\boldmath $p_{1}'$}}
\newcommand{\bpbbp}{\mbox{\boldmath $p_{2}'$}}

\newcommand{\bhk}{\mbox{\boldmath $\hat{k}$}}

\newcommand{\bhpaap}{\mbox{\boldmath $\hat{p}_{1}'$}}

\newcommand{\bhpa}{\mbox{\boldmath $\hat{p}_{a}$}}

\newcommand{\p}{\partial}

\usepackage[normalem]{ulem}  

\begin{document}


\title{\boldmath 
High-energy $\pi\pi$ scattering without and with photon radiation}

\vspace{0.6cm}

\author{Piotr Lebiedowicz}
\email{Piotr.Lebiedowicz@ifj.edu.pl}
\affiliation{Institute of Nuclear Physics Polish Academy of Sciences, 
Radzikowskiego 152, PL-31342 Krak{\'o}w, Poland}

\author{Otto Nachtmann}
\email{O.Nachtmann@thphys.uni-heidelberg.de}
\affiliation{Institut f\"ur Theoretische Physik, Universit\"at Heidelberg,
Philosophenweg 16, D-69120 Heidelberg, Germany}

\author{Antoni Szczurek
\footnote{Also at \textit{College of Natural Sciences, 
Institute of Physics, University of Rzesz{\'o}w, 
ul. Pigonia 1, PL-35310 Rzesz{\'o}w, Poland}.}}
\email{Antoni.Szczurek@ifj.edu.pl}
\affiliation{Institute of Nuclear Physics Polish Academy of Sciences, 
Radzikowskiego 152, PL-31342 Krak{\'o}w, Poland}

\begin{abstract}

We discuss the processes $\pi \pi \to \pi \pi$ and $\pi \pi \to \pi \pi \gamma$ from a general quantum field theory (QFT) point of view. 
We study the soft-photon limit where the photon energy $\omega \to 0$ and where we have the theorems due to F.E. Low and S. Weinberg.
We consider for the radiative amplitude the Laurent expansion
in $\omega$ and calculate the terms of order $\omega^{-1}$ and $\omega^{0}$.
The pole term $\propto \omega^{-1}$ is given by Weinberg's soft-photon theorem.
Then we calculate the amplitudes for the above reactions 
for high center-of-mass energies and small momentum transfers,
that is, in the soft-diffraction regime using the tensor-pomeron model. 
We identify places where ``anomalous'' soft photons could come from. 
Three soft-photon approximations (SPAs) are introduced. 
The corresponding SPA results are compared to those obtained from the full tensor-pomeron model for center-of-mass energies $\sqrt{s} = 10$ GeV and 100 GeV. The kinematic regions where the SPAs are a good representation of the full amplitude are determined. Finally we make some remarks on the type of fundamental information one could obtain from high-energy exclusive hadronic reactions without and with soft photon radiation.

\end{abstract}


\maketitle

\section{Introduction}

In this paper we shall be concerned with photon emission in
some strong-interaction processes.
In particular, we shall consider soft photon emission, that is,
the emission of photons with energy $\omega$ approaching zero.
For this kinematic region there exists Low's theorem \cite{Low:1958sn}
which is based strictly on Quantum Field Theory (QFT).
The theorem states that for $\omega \to 0$ the photons come exclusively
from the external hadrons in the process considered.
But this poses immediately the question: 
how close do we have to come to $\omega = 0$
in order to see the behaviour of the photon-emission amplitude predicted by Low?

There have been a number of experimental studies trying to verify Low's theorem
\cite{Goshaw:1979kq,Chliapnikov:1984ed,Botterweck:1991wf,Banerjee:1992ut,
Antos:1993wv,Tincknell:1996ks,Belogianni:1997rh,Belogianni:2002ib,Belogianni:2002ic,
Abdallah:2005wn,Abdallah:2007aa}. 
For a review of the experimental situation see \cite{Wong:2014pY}.
The result is, that many experiments see rather large deviations
from theoretical calculations in the soft-photon approximation (SPA)
based on Low's theorem. Clearly, this situation is unsatisfactory.
This has motivated the feasibility study of measuring soft-photon phenomena in
a next-generation experiment in the framework of the heavy-ion physics
programme at the LHC for the 2030's \cite{Adamova:2019vkf}.
Clearly, for preparing such soft-photon experiments 
accompanying theoretical studies are needed.

One class of hadronic reactions one can study at the LHC
are exclusive diffractive proton-proton collisions.
Examples are $pp$ elastic scattering and central exclusive production
(CEP) reactions, for instance $pp \to p \pi^{+} \pi^{-} p$.
In these reactions we can, of course, also have photon emission:
\begin{eqnarray}
&&p + p \to p + p + \gamma\,, \nonumber\\
&&p + p \to p + \pi^{+} + \pi^{-} + p + \gamma\,,
\label{pp_reactions}
\end{eqnarray}
and we can study the soft-photon limit.
The advantage of these exclusive diffractive reactions is that
they are ``clean'' from the experimental side and
that we have reasonable theoretical models for them.
We shall work within the tensor-pomeron model as proposed in \cite{Ewerz:2013kda}.
There, the soft pomeron and the charge conjugation
$C = +1$ reggeons are described as 
effective rank-2 symmetric tensor exchanges,
the odderon and the $C = -1$ reggeons
as effective vector exchanges.
The tensor-pomeron model has been applied to quite a number of CEP reactions
\cite{Lebiedowicz:2013ika,Lebiedowicz:2014bea,Lebiedowicz:2016ioh,
Lebiedowicz:2016zka,Lebiedowicz:2018eui,
Lebiedowicz:2018sdt,Lebiedowicz:2019jru,Lebiedowicz:2019boz,Lebiedowicz:2020yre,
Lebiedowicz:2021pzd}
which can and should all be studied by the present RHIC and LHC experiments
\cite{Adam:2020sap,McNulty:2017ejl,Schicker:2019qcn,Sirunyan:2019nog,Sirunyan:2020cmr,Sikora:2020mae}.
The next generation LHC experiment \cite{Adamova:2019vkf}
should be able to study these reactions in even greater detail, in particular,
in the region of low transverse momenta.
Applications of the model of \cite{Ewerz:2013kda} have furthermore 
been made to photoproduction of $\pi^{+} \pi^{-}$ pairs \cite{Bolz:2014mya},
a reaction which is also of interest for the LHC,
and to deep-inelastic lepton-nucleon scattering at low $x$ \cite{Britzger:2019lvc}.
In \cite{Ewerz:2016onn} it was shown that the experimental results
\cite{Adamczyk:2012kn} on the spin dependence of high-energy proton-proton
elastic scattering exclude a scalar character of the pomeron couplings but
are perfectly compatible with the tensor-pomeron model.
A vector coupling for the pomeron could definitely be ruled out 
in \cite{Britzger:2019lvc}.

With the present paper we want to start the theoretical study 
of soft-photon emission in hadronic exclusive diffractive 
high-energy reactions in the TeV energy region in
the framework of the tensor-pomeron model.
Our first example will be, for simplicity,
pion-pion elastic scattering.
This is, of course, not easy to study in experiments but,
as we shall see, we can in this example compare our
``exact'' model results for photon emission to approximations based 
on the soft-photon theorems of \cite{Low:1958sn, Weinberg:1965nx}.

Before coming to our present investigations we make remarks 
on some hadronic processes where photon emission has been studied,
frequently using the soft-photon approximation.

Direct photons (i.e. photons which originate not from hadronic decays, 
but from inelastic scattering processes between partons)
are an important electromagnetic probe of the quark-gluon plasma
as created in heavy-ion collisions.
Since pions are the dominant meson species produced in the heavy-ion collisions,
the photon production via bremsstrahlung in pion-pion elastic collisions 
was found to be a very important source to interpret 
the data on the direct photon spectra
and elliptic flow simultaneously \cite{Linnyk:2013hta,Linnyk:2013wma}.
In \cite{Linnyk:2013hta,Linnyk:2013wma}
the SPA was used and, therefore, 
the resulting yield of the bremsstrahlung photons
depends on some model assumptions.

The description of the photon bremsstrahlung in meson-meson scattering
beyond the SPA, within the one-boson exchange (OBE) model,
was discussed for the first time in \cite{Eggers:1995jq} and applied
to the dilepton bremsstrahlung in pion-pion collisions.
Later on, in \cite{Liu:2007zzw}, it was applied to the low-energy
photon bremsstrahlung in pion-pion and kaon-kaon collisions.
Within the OBE model the interaction of pions is described
by three resonance exchanges
$\sigma$, $\rho$ and $f_{2}(1270)$ 
in the $t$, $u$ and $s$ channels
(the $u$ channel diagrams are needed only in the case of identical pions).

In \cite{Linnyk:2015tha,Linnyk:2015rco} the authors 
applied the covariant OBE effective (chiral) model 
for the pion-pion scattering.
The ``exact'' OBE model result of the invariant rate of photon
bremsstrahlung was compared with that of the SPA.
It was noted there that the accuracy of the SPA approximation
can be significantly improved and the region of its applicability can be extended 
by evaluating the on-shell elastic cross section 
not at the c.m. energy $\sqrt{s}$ of the $\pi \pi \to \pi \pi \gamma$ process
but at a certain smaller energy.
One can see in Fig.~6 of \cite{Linnyk:2015tha}
(or Fig.~21 of \cite{Linnyk:2015rco})
that the ``improved SPA model''
gives a good approximation to the ``exact'' OBE result up to
photon energies $\approx 2$~GeV.
There the dominant contribution to the rates comes 
from low collision energies $\sqrt{s}$. 
The deviation between the OBE result and that calculated 
within the improved SPA is most pronounced 
at high $\sqrt{s}$ and high photon energies.

Whereas the examples of photon radiation discussed above
concerned low energy reactions, there have, of course,
also been studies of photon radiation for exclusive reactions at the LHC.
Exclusive diffractive photon bremsstrahlung
in proton-proton collisions was discussed 
in \cite{Khoze:2010jv,Lebiedowicz:2013xlb}.
Feasibility studies of the measurement 
of the exclusive diffractive bremsstrahlung cross section
in proton-proton collisions
at the center-of-mass energy $\sqrt{s} = 13$~TeV at the LHC
were performed in \cite{Chwastowski:2016jkl,Chwastowski:2016zzl}.

Very interesting general investigations of photon emission
in hadronic reactions have been presented, for instance,
in \cite{Gribov:1966hs,Lipatov:1988ii,
DelDuca:1990gz,Gervais:2017yxv,Bern:2014vva,Lysov:2014csa}.
We shall comment on results given in these papers below,
as far as they have a bearing on our own investigations.

Now we list the high-energy reactions which we want to study in our present paper.
In Sec.~\ref{sec:2} we discuss the reactions $\pi^{-} \pi^{0} \to \pi^{-} \pi^{0}$
and $\pi^{-} \pi^{0} \to \pi^{-} \pi^{0} \gamma$ from a general QFT point of view.
Section~\ref{sec:3} deals with the limit of photon-energy $\omega \to 0$ and
we discuss the terms in the amplitude of orders $\omega^{-1}$ and $\omega^{0}$.
In Sec.~\ref{sec:4} we introduce our model for $\pi^{\mp} \pi^{0}$ and
charged-pion scattering and for the corresponding reactions with photon emission.
Section~\ref{sec:5} is devoted to a comparison of our ``exact'' model results to various
approximations based Low's theorem. 
In Sec.~\ref{sec:6} we give our conclusions and an outlook on further work.
In Appendix~\ref{sec:appendixA} we compare in detail our results
for the terms of order $\omega^{-1}$ and $\omega^{0}$
in the photon emission amplitude to results presented in the literature.
In Appendix~\ref{sec:appendixB} we discuss 
the cross section $d\sigma / d\omega$ for $\omega \to 0$.

\section{General properties of the reactions $\pi \pi \to \pi \pi$ and $\pi \pi \to \pi \pi \gamma$}
\label{sec:2}

Here we study general QFT relations for pion-pion elastic scattering without
and with photon radiation.
We shall work to leading order in the electromagnetic coupling.
For simplicity we shall consider $\pi^{-} \pi^{0}$ scattering,
that is, the reactions
\begin{eqnarray}
&&\pi^{-} (p_{a}) + \pi^{0} (p_{b}) \to \pi^{-} (p_{1}) + \pi^{0} (p_{2})\,,
\label{2.1} \\
&&\pi^{-} (p_{a}) + \pi^{0} (p_{b}) \to \pi^{-} (p_{1}') + \pi^{0} (p_{2}') 
+ \gamma (k, \epsilon)\,.
\label{2.2}
\end{eqnarray}
Here $p_{a}$, $p_{b}$, $p_{1}$, $p_{2}$, $p_{1}'$, $p_{2}'$ and $k$
are the momenta of the particles and $\epsilon$ 
is the polarisation vector of the photon, respectively.
The energy-momentum conservation in (\ref{2.1}) and (\ref{2.2}) requires
\begin{eqnarray}
&&p_{a} + p_{b} = p_{1} + p_{2}\,,
\label{2.3} \\
&&p_{a} + p_{b} = p_{1}' + p_{2}' + k\,.
\label{2.4}
\end{eqnarray}

\begin{figure}[!h]
\includegraphics[width=5.0cm]{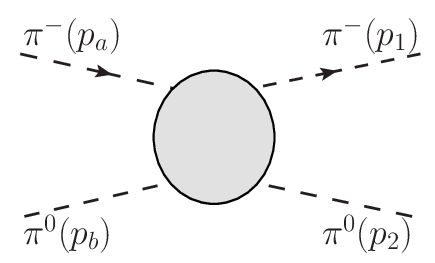}
\caption{Diagram for the off shell $\pi^{-} \pi^{0}$ scattering amplitude.}
\label{fig:pimpi0_pimpi0}
\end{figure}

We denote the amplitude for the reaction (\ref{2.1}) by
\begin{eqnarray}
{\cal T}(p_{a}, p_{b}, p_{1}, p_{2}) = 
\braket{\pi^{-}(p_{1}),\pi^{0}(p_{2})|{\cal T}|\pi^{-}(p_{a}),\pi^{0}(p_{b})} \,.
\label{2.5}
\end{eqnarray}
Since pions have $G$ parity $-1$ all diagrams for (\ref{2.5}) are one-particle irreducible.
In QFT we can extend the amplitude (\ref{2.5}) for off shell pions (Fig.~\ref{fig:pimpi0_pimpi0}).

This off shell scattering amplitude will still satisfy the energy-momentum conservation (\ref{2.3})
and can only depend on the following 6 variables
\begin{eqnarray}
&& s_{L} = p_{a} \cdot p_{b} + p_{1} \cdot p_{2}\,, \nonumber \\
&& t = (p_{a} - p_{1})^{2} = (p_{b} - p_{2})^{2}\,, \nonumber \\
&& m_{a}^{2} = p_{a}^{2}\,, \;\, m_{b}^{2} = p_{b}^{2}\,, \;\,
   m_{1}^{2} = p_{1}^{2}\,, \;\, m_{2}^{2} = p_{2}^{2}\,.
\label{2.6}
\end{eqnarray}
Here we use as squared energy variable $s_{L}$, following \cite{Low:1958sn},
instead of the more usual Mandelstam variable $s$.
We have
\begin{eqnarray}
s = s_{L} + \frac{1}{2} \left(m_{a}^{2} + m_{b}^{2} + m_{1}^{2} + m_{2}^{2} \right)\,.
\label{2.7}
\end{eqnarray}
The on- or off-shell amplitude corresponding to (\ref{2.5}) as a function of the variables (\ref{2.6})
will be denoted by
\begin{eqnarray}
{\cal M}^{(0)}(s_{L},t,m_{a}^{2},m_{b}^{2},m_{1}^{2},m_{2}^{2}) = 
\left.{\cal T}(p_{a}, p_{b}, p_{1}, p_{2})\right|_{\rm on\; shell\; or\; off\; shell}.
\label{2.8}
\end{eqnarray}
The on-shell amplitude (\ref{2.5}) is then
${\cal M}^{(0)}(s_{L},t,m_{\pi}^{2},m_{\pi}^{2},m_{\pi}^{2},m_{\pi}^{2})$.

\begin{figure}[!h]
(a)\includegraphics[width=7.1cm]{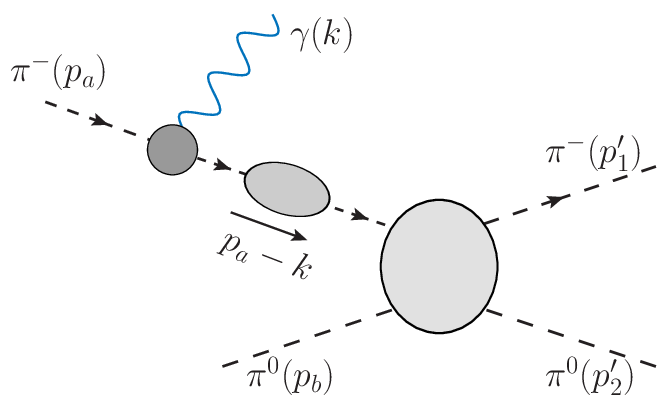}\hspace{0.5cm}
(b)\includegraphics[width=7.2cm]{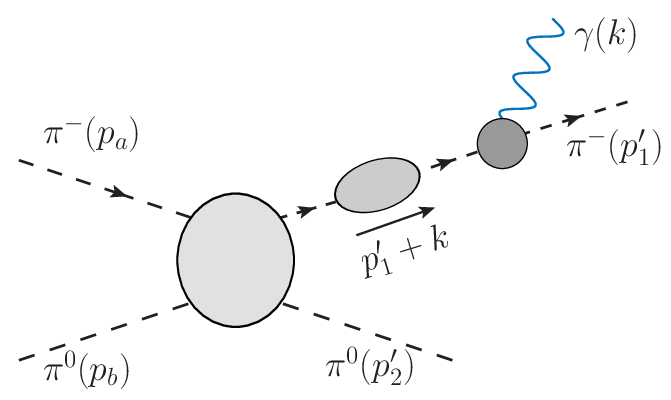}\\
(c)\includegraphics[width=6.2cm]{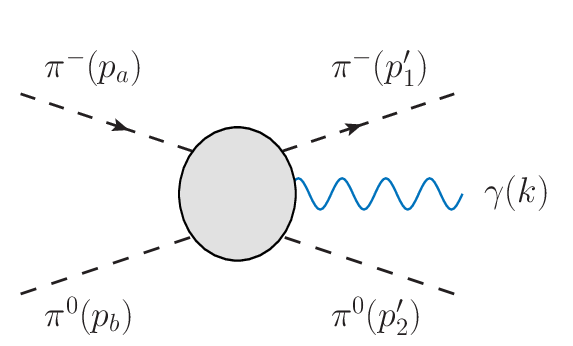}
\caption{One-particle reducible (a, b) and irreducible (c) diagrams
for $\pi^{-} \pi^{0} \to \pi^{-} \pi^{0} \gamma$.}
\label{fig:pimpi0gamma}
\end{figure}

Next we study the reaction (\ref{2.2}) 
where we have two one-particle reducible diagrams 
(Figs.~\ref{fig:pimpi0gamma}~(a), (b)) 
and one irreducible diagram (Fig.~\ref{fig:pimpi0gamma}~(c)).

For the diagrams (a) and (b) we need the off-shell $\pi \pi$ amplitude (\ref{2.8}),
the pion propagator $\Delta(p^{2})$ and the pion-photon vertex function
$\Gamma_{\lambda}(p',p)$:
\begin{eqnarray}
\includegraphics[width=120pt]{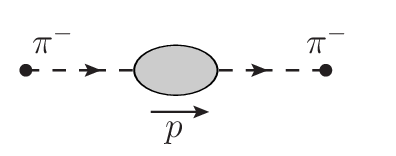} 
&&i\Delta(p^{2})\,,
\label{2.9}\\
\includegraphics[width=120pt]{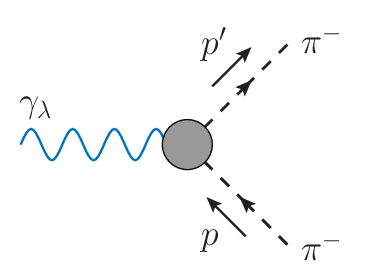}
&&ie\Gamma_{\lambda}(p',p)\,.
\label{2.10}
\end{eqnarray}
We denote by $e = \sqrt{4 \pi \alpha} > 0$ the $\pi^{+}$ charge.

The expressions for the amplitudes of Figs.~\ref{fig:pimpi0gamma}~(a, b)
can be written as follows:
\begin{eqnarray}
&&{\cal M}_{\lambda}^{(a)} = -e \,{\cal M}^{(0,\,a)}\, \Delta[ (p_{a}-k)^{2} ]\,
\Gamma_{\lambda}(p_{a}-k,p_{a})\,, \nonumber \\
&&{\cal M}^{(0,\,a)}= {\cal M}^{(0)}[ (p_{a}-k, p_{b}) + p_{1}' \cdot p_{2}', 
(p_{b}-p_{2}')^{2}, (p_{a}-k)^{2}, m_{\pi}^{2}, m_{\pi}^{2}, m_{\pi}^{2} ],
\label{2.11}
\end{eqnarray}
\begin{eqnarray}
&&{\cal M}_{\lambda}^{(b)} = -e \,\Gamma_{\lambda}(p_{1}',p_{1}'+k) \,\Delta[ (p_{1}'+k)^{2} ]\,
{\cal M}^{(0,\,b)}\,, \nonumber \\
&&{\cal M}^{(0,\,b)}= {\cal M}^{(0)}[ p_{a} \cdot p_{b} + (p_{1}'+k, p_{2}'), 
(p_{b}-p_{2}')^{2}, m_{\pi}^{2}, m_{\pi}^{2}, (p_{1}'+k)^{2},  m_{\pi}^{2} ].
\label{2.12}
\end{eqnarray}
The photon-emission amplitude is
\begin{eqnarray}
\braket{\gamma(k,\epsilon),\pi^{-}(p_{1}'),\pi^{0}(p_{2}')|{\cal T}|\pi^{-}(p_{a}),\pi^{0}(p_{b})} =
( \epsilon^{\lambda} )^{*}\, {\cal M}_{\lambda}
\label{2.13}
\end{eqnarray}
where
\begin{eqnarray}
{\cal M}_{\lambda} =
{\cal M}_{\lambda}^{(a)} + 
{\cal M}_{\lambda}^{(b)} + 
{\cal M}_{\lambda}^{(c)}\,.
\label{2.14}
\end{eqnarray}
${\cal M}_{\lambda}$ also determines the emission of virtual photons of mass
$k^{2} > 0$ which then decay to a lepton pair.
For $k^{2} < 0$ ${\cal M}_{\lambda}$ enters the amplitude for
the 3-body reaction 
$e^{\pm} \pi^{-} \pi^{0} \to e^{\pm} \pi^{-} \pi^{0}$.
The amplitude ${\cal M}_{\lambda}$ must satisfy the gauge-invariance relation,
valid for all $k^{2}$,
\begin{eqnarray}
k^{\lambda}{\cal M}_{\lambda} =
k^{\lambda} \left( {\cal M}_{\lambda}^{(a)} + 
{\cal M}_{\lambda}^{(b)} + 
{\cal M}_{\lambda}^{(c)} \right) = 0 \,,
\label{2.15}
\end{eqnarray}
that is, we have
\begin{eqnarray}
k^{\lambda} {\cal M}_{\lambda}^{(c)} = 
- k^{\lambda}{\cal M}_{\lambda}^{(a)} - k^{\lambda}{\cal M}_{\lambda}^{(b)}\,.
\label{2.16}
\end{eqnarray}

We shall now use (\ref{2.11}), (\ref{2.12}), and (\ref{2.16}),
to get a simple relation
between $k^{\lambda} {\cal M}_{\lambda}^{(c)}$
and ${\cal M}^{(0,\,a)}$, ${\cal M}^{(0,\,b)}$. 
For this we recall the normalisation conditions for the pion propagator
and the vertex function. We have
\begin{eqnarray}
&&\left.\Delta^{-1}(p^{2})\right|_{p^{2} = m_{\pi}^{2}} = 0\,, \quad 
\frac{\p}{\p p^{2}}\left.\Delta^{-1}(p^{2})\right|_{p^{2} = m_{\pi}^{2}} = 1\,, \nonumber\\
&&\left.\Gamma_{\lambda}(p',p)\right|_{p'=p, p^{2} = m_{\pi}^{2}} = 2 p_{\lambda}\,.
\label{2.17}
\end{eqnarray}
Furthermore we have the Ward-Takahashi identity 
\cite{Ward:1950xp,Takahashi:1957xn},
\begin{eqnarray}
(p'-p)^{\lambda} \Gamma_{\lambda}(p',p) =
\Delta^{-1}(p'^{2}) - \Delta^{-1}(p^{2}) \,.
\label{2.18}
\end{eqnarray}
From (\ref{2.17}) and (\ref{2.18}) we obtain for $p_{a}^{2} = m_{\pi}^{2}$
\begin{eqnarray}
&&\Delta[(p_{a}-k)^{2}] \,\Gamma_{\lambda}(p_{a}-k,p_{a})\, k^{\lambda} 
\nonumber \\
&&\qquad \qquad =
-\Delta[(p_{a}-k)^{2}] \,\Gamma_{\lambda}(p_{a}-k,p_{a})\, (p_{a} - k - p_{a})^{\lambda} 
\nonumber \\
&&\qquad \qquad =
-\Delta[(p_{a}-k)^{2}] \,\lbrace \Delta^{-1}[(p_{a}-k)^{2}] - \Delta^{-1}[p_{a}^{2}] \rbrace 
\nonumber \\
&&\qquad \qquad = -1\,.  
\label{2.19}
\end{eqnarray}
Similarly we get for $p_{1}'^{2} = m_{\pi}^{2}$
\begin{eqnarray}
&&k^{\lambda} \,\Gamma_{\lambda}(p_{1}',p_{1}'+k)\, \Delta[(p_{1}'+k)^{2}]
\nonumber \\
&&\qquad \qquad =
-[p_{1}' - (p_{1}' + k)]^{\lambda} \,\Gamma_{\lambda}(p_{1}',p_{1}' + k)\, \Delta[(p_{1}'+ k)^{2}] 
\nonumber \\
&&\qquad \qquad =
-\lbrace \Delta^{-1}[p_{1}'^{2}] - \Delta^{-1}[(p_{1}'+k)^{2}] \rbrace \,\Delta[(p_{1}'+k)^{2}]
\nonumber \\
&&\qquad \qquad = 1\,.  
\label{2.20}
\end{eqnarray}

From (\ref{2.11}), (\ref{2.12}), (\ref{2.16}), (\ref{2.19}), and (\ref{2.20}), 
we obtain
\begin{eqnarray}
&&k^{\lambda} {\cal M}_{\lambda}^{(a)} = e\,{\cal M}^{(0,\,a)}\,, \nonumber\\
&&k^{\lambda} {\cal M}_{\lambda}^{(b)} =-e\,{\cal M}^{(0,\,b)}\,,
\label{2.21}\\
&&k^{\lambda} {\cal M}_{\lambda}^{(c)} = -e\,{\cal M}^{(0,\,a)}
                                         +e\,{\cal M}^{(0,\,b)}\,,
\label{2.22}
\end{eqnarray}
where ${\cal M}^{(0,\,a)}$ and ${\cal M}^{(0,\,b)}$
are given explicitly in (\ref{2.11}) and (\ref{2.12}), respectively.

\section{The expansion of the photon-emission amplitude to the order $\omega^{-1}$ plus $\omega^{0}$}
\label{sec:3}

In this section we discuss the expansion of the amplitude
${\cal M}_{\lambda}$ (\ref{2.14}) to the orders $\omega^{-1}$ and $\omega^{0}$.
Here $\omega = k^{0}$ and, if not stated otherwise, we work in the overall
c.m. system of the reaction (\ref{2.2}).
We shall in the following assume that all components of the photon momentum
are proportional to $\omega$,
$k^{\mu} \propto \omega$, with $\omega \to 0$.
This is perfectly alright theoretically, but can this also be realised in nature?
For real photon emission, $k^{2} = 0$,
this clearly can be realised.
It is also possible for $k^{2} < 0$ in the 3-body collision
\begin{eqnarray}
e^{\pm} + \pi^{-} + \pi^{0} \to e^{\pm} + \pi^{-} + \pi^{0}\,.
\label{3.0}
\end{eqnarray}
For $k^{2} > 0$ we can have $e^{+}e^{-}$ production
\begin{eqnarray}
\pi^{-} + \pi^{0} \to e^{+} + e^{-} + \pi^{-} + \pi^{0}\,.
\label{3.0a}
\end{eqnarray}
But here $\omega \geqslant 2 m_{e}$ and $k^{2} \geqslant 4 m_{e}^{2}$,
with $m_{e}$ the electron mass.
Thus, in (\ref{3.0a}) we cannot reach $\omega = 0$.
But the electron mass is very small on a hadronic scale, $m_{e} \simeq 0.5$~MeV,
and, therefore, the limit $\omega \to 0$ should also be of relevance
for the reaction (\ref{3.0a}).
In the following we shall consider only
$k^{2} \geqslant 0$, $k^{0} \geqslant 0$.

We start our investigation of the small $\omega$ limit
with the pion propagator (\ref{2.9}).
We are working to lowest order in the electromagnetic coupling.
Thus, $\Delta^{-1}(p^{2})$ is for us a purely hadronic object.
Its nearest singularity to $p^{2} = 0$ is at $p^{2} = (3 m_{\pi})^{2}$
as we see from the Landau conditions
(cf. for instance \cite{Bjorken:1965}).
Therefore, we can expand $\Delta^{-1}(p^{2})$ around $p^{2} = m_{\pi}^{2}$
as follows with $c$ a constant:
\begin{eqnarray}
\Delta^{-1}(p^{2}) = p^{2} - m_{\pi}^{2} + c (p^{2} - m_{\pi}^{2})^{2} + \ldots \,.
\label{3.1}
\end{eqnarray}
This gives for $p_{a}^{2} = m_{\pi}^{2}$ and $p_{1}'^{2} = m_{\pi}^{2}$
the following
\begin{eqnarray}
&&\Delta^{-1}[(p_{a}-k)^{2}] =  (-2 p_{a} \cdot k + k^{2})
[1 + c (-2 p_{a} \cdot k + k^{2}) + {\cal O}(\omega^{2})]
\,, \nonumber\\
&&\Delta[(p_{a}-k)^{2}] =  \frac{1}{-2 p_{a} \cdot k + k^{2}}
[1 - c (-2 p_{a} \cdot k + k^{2}) + {\cal O}(\omega^{2})]
\,,
\label{3.2}\\
&&\Delta^{-1}[(p_{1}'+k)^{2}] =  (2 p_{1}' \cdot k + k^{2})
[1 + c (2 p_{1}' \cdot k + k^{2}) + {\cal O}(\omega^{2})]
\,, \nonumber\\
&&\Delta[(p_{1}'+k)^{2}] =  \frac{1}{2 p_{1}' \cdot k + k^{2}}
[1 - c (2 p_{1}' \cdot k + k^{2}) + {\cal O}(\omega^{2})]
\,.
\label{3.3}
\end{eqnarray}
From (\ref{2.11}), (\ref{2.12}) and (\ref{2.14}) 
we see that we must now expand
$\Gamma_{\lambda}$, ${\cal M}^{(0,\,a)}$ and ${\cal M}^{(0,\,b)}$
up to order $\omega$ and ${\cal M}_{\lambda}^{(c)}$ up to order
$\omega^{0}$ for getting the total amplitude ${\cal M}_{\lambda}$
expanded up to order $\omega^{0}$.

We start with $\Gamma_{\lambda}(p',p)$ which has the general expansion
\begin{eqnarray}
\Gamma_{\lambda}(p',p) &=& (p'+p)_{\lambda}\, 
A[p'^{2}-m_{\pi}^{2},p^{2}-m_{\pi}^{2},(p'-p)^{2}] \nonumber\\
&&+ (p'-p)_{\lambda}\, 
B[p'^{2}-m_{\pi}^{2},p^{2}-m_{\pi}^{2},(p'-p)^{2}]
\,.
\label{3.4}
\end{eqnarray}
The functions $A$ and $B$ are analytic in their variables
in the region of interest to us as we see again
from the Landau conditions.
The Ward-Takahashi identity gives
\begin{eqnarray}
(p'-p)^{\lambda}\, \Gamma_{\lambda}(p',p) &=&
(p'^{2}-p^{2}) \,A[p'^{2}-m_{\pi}^{2},p^{2}-m_{\pi}^{2},(p'-p)^{2}] \nonumber\\
&&+ (p'-p)^{2}\,
B[p'^{2}-m_{\pi}^{2},p^{2}-m_{\pi}^{2},(p'-p)^{2}] \nonumber\\
&=& \Delta^{-1}(p'^{2}) - \Delta^{-1}(p^{2}) \,.
\label{3.5}
\end{eqnarray}
Now we set in (\ref{3.5}) $p = p_{a}$, $p' = p_{a} - k$,
$p_{a}^{2} = m_{\pi}^{2}$ and get
\begin{eqnarray}
&&(p'^{2}-m_{\pi}^{2})\,A(p'^{2}-m_{\pi}^{2},0,k^{2}) + 
k^{2}\,B(p'^{2}-m_{\pi}^{2},0,k^{2}) \nonumber \\
&& \qquad = \Delta^{-1}(p'^{2}) = 
p'^{2} - m_{\pi}^{2} + c (p'^{2} - m_{\pi}^{2})^{2} + \ldots \,.
\label{3.6}
\end{eqnarray}
Therefore, we must have
\begin{eqnarray}
B(p'^{2}-m_{\pi}^{2},0,k^{2}) = 
(p'^{2} - m_{\pi}^{2}) \, \tilde{B}(p'^{2}-m_{\pi}^{2},k^{2})
\label{3.7}
\end{eqnarray}
and we get with $p'^{2} - m_{\pi}^{2} = -2 p_{a} \cdot k + k^{2}$
\begin{eqnarray}
&&A(-2 p_{a} \cdot k + k^{2},0,k^{2}) = 
1 + c (-2 p_{a} \cdot k + k^{2}) 
+ {\cal O}(\omega^{2}) \,, \nonumber\\
&&B(-2 p_{a} \cdot k + k^{2},0,k^{2}) = {\cal O}(\omega) \,.
\label{3.8}
\end{eqnarray}
Inserting (\ref{3.8}) in (\ref{3.4}) we find
\begin{eqnarray}
\Gamma_{\lambda}(p_{a}-k,p_{a}) = (2 p_{a} - k)_{\lambda}
[1 + c (-2 p_{a} \cdot k + k^{2})] + {\cal O}(\omega^{2})\,.
\label{3.9}
\end{eqnarray}
In a completely analogous way we get for $p_{1}'^{2} = m_{\pi}^{2}$
\begin{eqnarray}
\Gamma_{\lambda}(p_{1}',p_{1}'+k) = (2 p_{1}' + k)_{\lambda}
[1 + c (2 p_{1}' \cdot k + k^{2})] + {\cal O}(\omega^{2})\,.
\label{3.10}
\end{eqnarray}
From (\ref{3.2}), (\ref{3.3}), (\ref{3.9}) and (\ref{3.10}) we get
\begin{eqnarray}
&&\Delta[(p_{a}-k)^{2}] \, \Gamma_{\lambda}(p_{a}-k,p_{a}) = 
\frac{(2 p_{a} - k)_{\lambda}}{-2p_{a} \cdot k + k^{2}}
+ {\cal O}(\omega)\,,
\label{3.11} \\
&&\Gamma_{\lambda}(p_{1}',p_{1}'+k) \, \Delta[(p_{1}'+k)^{2}] = 
\frac{(2 p_{1}' + k)_{\lambda}}{2p_{1}' \cdot k + k^{2}}
+ {\cal O}(\omega)\,.
\label{3.12}
\end{eqnarray}

Next we investigate the energy-momentum conservation conditions
(\ref{2.3}) and (\ref{2.4}) 
for the reactions (\ref{2.1}) and (\ref{2.2}), respectively.
It is clear that for $k \neq 0$ we cannot have $p_{1} = p_{1}'$
and $p_{2} = p_{2}'$ since
\begin{eqnarray}
p_{a} + p_{b} \neq p_{1} + p_{2} + k\,.
\label{3.13}
\end{eqnarray}
This means that when going from (\ref{2.1}) to (\ref{2.2}) we must
have a change of momenta $p_{1} \to p_{1}' \neq p_{1}$ and 
$p_{2} \to p_{2}' \neq p_{2}$. In fact, choosing for the reaction (\ref{2.2})
some $k \neq 0$, even a small momentum $k$,
this does not fix $p_{1}'$ and $p_{2}'$.
This is best seen in the rest system of the four-vector $p_{a} + p_{b} - k$.
There we have $\bpaap + \bpbbp = 0$,
$|\bpaap|$ is fixed and thus $\bpaap$ can still vary on
a sphere of radius $|\bpaap|$.
For the following we work, however, 
in the overall c.m. system of reaction (\ref{2.2}).

\begin{figure}[!ht]
\includegraphics[width=8.0cm]{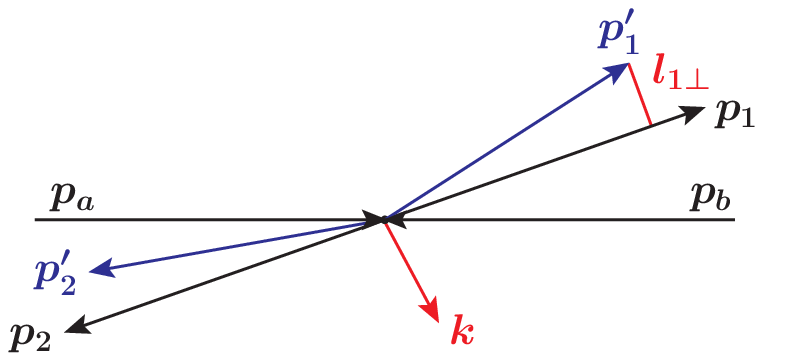}
\caption{\label{fig:100}
\small
Illustration of the momentum configurations 
for $\pi^{-} \pi^{0} \to \pi^{-} \pi^{0}$ (\ref{2.1})
and $\pi^{-} \pi^{0} \to \pi^{-} \pi^{0} \gamma$ (\ref{2.2})
in the c.m. system.}
\end{figure}

We write
\begin{eqnarray}
p_{1}' = p_{1} - l_{1}\,, \quad p_{2}' = p_{2} - l_{2}\,,
\label{3.14}
\end{eqnarray}
and get from (\ref{2.3}) and (\ref{2.4}) the conditions
\begin{eqnarray}
l_{1} + l_{2} = k\,, \quad (p_{1} - l_{1})^{2} = m_{\pi}^{2}\,,
\quad (p_{2} - l_{2})^{2} = m_{\pi}^{2}\,.
\label{3.15}
\end{eqnarray}
For given $k$ these are 6 conditions for the 8 unknowns $l_{1}$, $l_{2}$
giving a 2-parameter solution as it should be. 
Working in the common c.m. system of the reactions 
(\ref{2.1}) and (\ref{2.2}) we set with 
$\bhpaa = \bpaa / |\bpaa|$
\begin{eqnarray}
&&(l_{1}^{\mu}) =
\left( \begin{array}{l}
 l_{1}^{0} \\
 l_{1 \parallel} \bhpaa + \blaperp \\
\end{array} \right)\,, \quad \blaperp \cdot \bhpaa = 0\,, \nonumber\\
&&(l_{2}^{\mu}) =
\left( \begin{array}{l}
 l_{2}^{0} \\
 l_{2 \parallel} \bhpaa + \blbperp \\
\end{array} \right)\,, \quad \blbperp \cdot \bhpaa = 0\,,\nonumber\\
&&(k^{\mu}) =
\left( \begin{array}{l}
 \omega \\
 k_{\parallel} \bhpaa + \bkperp \\
\end{array} \right)\,, \quad \bkperp \cdot \bhpaa = 0\,.
\label{3.16}
\end{eqnarray} 
Inserting this in (\ref{3.15}) we get the system of equations
\begin{eqnarray}
&&l_{2} = k - l_{1}\,, \nonumber\\
&&p_{1}^{0}l_{1}^{0} - |\bpaa| l_{1 \parallel} = \frac{1}{2}l_{1}^{2}\,,\nonumber\\
&&p_{1}^{0}l_{1}^{0} + |\bpaa| l_{1 \parallel} = 
  p_{1}^{0}k^{0} + |\bpaa| k_{\parallel} - \frac{1}{2}(k-l_{1})^{2}\,.
\label{3.17}
\end{eqnarray}
Now we make an \underline{important choice} for the following.
We assume that together with the soft photon emitted 
with energy $\omega \to 0$ we consider only slight changes of the momenta $p_{1} \to p_{1}'$ and $p_{2} \to p_{2}'$.
That is, we assume
\begin{eqnarray}
l_{1}^{\mu} = {\cal O}(\omega)\,, \quad l_{2}^{\mu} = {\cal O}(\omega)\,.
\label{3.18}
\end{eqnarray}
With this we can neglect the quadratic terms in $l_{1}$, $l_{2}$, $k$
in (\ref{3.17}). The solution of the resulting equations is
\begin{eqnarray}
&&(l_{1}^{\mu}) =
\left( \begin{array}{l}
 \dfrac{1}{2 p_{1}^{0}} (p_{2} \cdot k) \\
 \dfrac{1}{2 |\bpaa|} \bhpaa (p_{2} \cdot k) + \blaperp \\
\end{array} \right), \nonumber\\
&&(l_{2}^{\mu}) =
\left( \begin{array}{l}
 \dfrac{1}{2 p_{1}^{0}} (p_{1} \cdot k) \\
 \bk - \dfrac{1}{2 |\bpaa|} \bhpaa (p_{2} \cdot k) - \blaperp \\
\end{array} \right).
\label{3.19}
\end{eqnarray} 
Here $\blaperp$ stays undetermined, corresponding to 
the 2-parameter freedom of the momenta $p_{1}'$, $p_{2}'$ for given $k$.
This is illustrated in Fig.~\ref{fig:100}.
In the order of $\omega$ considered we get
\begin{eqnarray}
p_{1} \cdot l_{1} = 0\,, \quad p_{2} \cdot l_{2} = 0\,.
\label{3.20}
\end{eqnarray}
%

Now we can expand ${\cal M}^{(0,\,a)}$ (\ref{2.11}) and 
${\cal M}^{(0,\,b)}$ (\ref{2.12}) up to order $\omega$.
We get with $s_{L}$ and~$t$ from (\ref{2.6}),
\begin{eqnarray}
{\cal M}^{(0,\,a)}&=& {\cal M}^{(0)}[ (p_{a}-k, p_{b}) + p_{1}' \cdot p_{2}', 
(p_{b}-p_{2}')^{2}, (p_{a}-k)^{2}, m_{\pi}^{2}, m_{\pi}^{2}, m_{\pi}^{2}] \nonumber\\
&=&{\cal M}^{(0)}[s_{L} - (p_{b}+p_{1},k) - (p_{2} \cdot l_{1}), 
t-2 (p_{a}-p_{1},k-l_{1}), m_{\pi}^{2} - 2 (p_{a} \cdot k), m_{\pi}^{2}, m_{\pi}^{2}, m_{\pi}^{2}] 
 \nonumber\\
&& + {\cal O}(\omega^{2})\nonumber\\
&=& 
\big\lbrace 1 - \left[ (p_{b}+p_{1},k) +  (p_{2} \cdot l_{1}) \right] \frac{\partial}{\partial s_{L}}
         -\left[2(p_{a}-p_{1},k) - 2(p_{a} \cdot l_{1}) \right] \frac{\partial}{\partial t} 
        - 2 (p_{a} \cdot k) \frac{\partial}{\partial m_{a}^{2}} \big\rbrace \nonumber\\
&& \times
        \left.{\cal M}^{(0)}(s_{L},t,m_{a}^{2},m_{\pi}^{2}, m_{\pi}^{2}, m_{\pi}^{2})\right|_{m_{a}^{2} = m_{\pi}^{2}} + {\cal O}(\omega^{2})\,,
\label{3.21}\\
{\cal M}^{(0,\,b)}&=& {\cal M}^{(0)}[ p_{a} \cdot p_{b} + (p_{1}'+k, p_{2}'), 
(p_{b}-p_{2}')^{2}, m_{\pi}^{2}, m_{\pi}^{2}, (p_{1}'+k)^{2}, m_{\pi}^{2}] \nonumber\\
&=&{\cal M}^{(0)}[s_{L} - (p_{1} \cdot k), 
t-2 (p_{a}-p_{1},k) + 2 (p_{a} \cdot l_{1}), m_{\pi}^{2}, m_{\pi}^{2}, m_{\pi}^{2} + 2(p_{1} \cdot k), m_{\pi}^{2}]
 \nonumber\\
&& + {\cal O}(\omega^{2})\nonumber\\
&=& 
\big\lbrace 1 - (p_{1} \cdot k) \frac{\partial}{\partial s_{L}} -
          \left[2(p_{a}-p_{1},k) - 2(p_{a} \cdot l_{1}) \right] \frac{\partial}{\partial t} 
        + 2 (p_{1} \cdot k) \frac{\partial}{\partial m_{1}^{2}} \big\rbrace \nonumber\\
&& \times
        \left.{\cal M}^{(0)}(s_{L},t,m_{\pi}^{2}, m_{\pi}^{2},m_{1}^{2},m_{\pi}^{2})\right|_{m_{1}^{2} = m_{\pi}^{2}} + {\cal O}(\omega^{2})\,.
\label{3.22}
\end{eqnarray}

To determine ${\cal M}^{(c)}_{\lambda}$ to order $\omega^{0}$
we use (\ref{2.22}).
To order $\omega$ we get, inserting (\ref{3.21}) and (\ref{3.22})
in (\ref{2.22}),
\begin{eqnarray}
k^{\lambda} {\cal M}^{(c)}_{\lambda} &=& 
e \big\lbrace (p_{b} + p_{2},k) \frac{\partial}{\partial s_{L}}
+ 2(p_{a} \cdot k) \frac{\partial}{\partial m_{a}^{2}} 
+ 2(p_{1} \cdot k) \frac{\partial}{\partial m_{1}^{2}} \big\rbrace 
\nonumber\\
&& \times
\left.{\cal M}^{(0)}(s_{L},t,m_{a}^{2}, m_{\pi}^{2},m_{1}^{2},m_{\pi}^{2})\right|_{m_{a}^{2} = m_{1}^{2} = m_{\pi}^{2}} + {\cal O}(\omega^{2})\,.
\label{3.23}
\end{eqnarray}
From (\ref{3.23}) we can read off the term of order $\omega^{0}$
for ${\cal M}^{(c)}_{\lambda}$:
\begin{eqnarray}
{\cal M}^{(c)}_{\lambda} &=& 
e \big\lbrace (p_{b} + p_{2})_{\lambda} \frac{\partial}{\partial s_{L}}
+ 2 p_{a \lambda} \frac{\partial}{\partial m_{a}^{2}} 
+ 2 p_{1 \lambda} \frac{\partial}{\partial m_{1}^{2}} \big\rbrace 
\nonumber\\
&& \times
\left.{\cal M}^{(0)}(s_{L},t,m_{a}^{2}, m_{\pi}^{2},m_{1}^{2},m_{\pi}^{2})\right|_{m_{a}^{2} = m_{1}^{2} = m_{\pi}^{2}} + {\cal O}(\omega)\,.
\label{3.24}
\end{eqnarray}

Now we collect everything together and we obtain from
(\ref{2.14}), (\ref{3.11}), (\ref{3.12}), (\ref{3.21}), (\ref{3.22}),
and (\ref{3.24}) the following expansion for the amplitude
$\pi^{-} \pi^{0} \to \pi^{-} \pi^{0} \gamma$:
\begin{eqnarray}
{\cal M}_{\lambda} &=&
{\cal M}_{\lambda}^{(a)} + 
{\cal M}_{\lambda}^{(b)} + 
{\cal M}_{\lambda}^{(c)} \nonumber \\
&=& e {\cal M}^{(0)}(s_{L},t,m_{\pi}^{2},m_{\pi}^{2},m_{\pi}^{2},m_{\pi}^{2})
\big[ \frac{(2 p_{a} -k)_{\lambda}}{2 (p_{a} \cdot k)-k^{2}} 
-      \frac{(2 p_{1}'+k)_{\lambda}}{2 (p_{1}' \cdot k)+k^{2}} \big] \nonumber\\
&&+ 2 e \frac{\partial}{\partial s_{L}} {\cal M}^{(0)}(s_{L},t,m_{\pi}^{2},m_{\pi}^{2},m_{\pi}^{2},m_{\pi}^{2})
\big[ - (p_{b} \cdot k) \frac{p_{a\lambda}}{(p_{a} \cdot k)} 
+      p_{b\lambda} \big] \nonumber\\
&&- 2 e \frac{\partial}{\partial t} {\cal M}^{(0)}(s_{L},t,m_{\pi}^{2},m_{\pi}^{2},m_{\pi}^{2},m_{\pi}^{2})
\big[ (p_{a} - p_{1},k) - (p_{a} \cdot l_{1}) 
\big] \big[\frac{p_{a\lambda}}{(p_{a} \cdot k)} 
- \frac{p_{1\lambda}}{(p_{1} \cdot k)} \big] \nonumber\\
&&+ {\cal O}(\omega)\,.
\label{3.25}
\end{eqnarray}
In the first term on the right-hand side (r.h.s.) of (\ref{3.25}) we should,
for consistency of the expansion in $\omega$ up to $\omega^{0}$,
make the following replacements:
\begin{eqnarray}
&&\frac{(2 p_{a} -k)_{\lambda}}{2 (p_{a} \cdot k)-k^{2}}
\to \frac{p_{a\lambda}}{(p_{a} \cdot k)} + \frac{1}{2(p_{a} \cdot k)^{2}}
[p_{a \lambda}k^{2} - k_{\lambda}(p_{a} \cdot k)]\,, \nonumber\\
&&\frac{(2 p_{1}'+k)_{\lambda}}{2 (p_{1}' \cdot k)+k^{2}}
\to \frac{p_{1\lambda}}{(p_{1} \cdot k)} + \frac{1}{2(p_{1} \cdot k)^{2}}
[p_{1 \lambda}(2(l_{1} \cdot k)-k^{2})-(2 l_{1 \lambda}-k_{\lambda})(p_{1} \cdot k)] \,. \qquad
\label{3.26}
\end{eqnarray}

Now we set
\begin{align}
(k^{\mu})=
\omega \left( \begin{array}{l}
1
\vspace{0.1cm}\\
\bm{\tilde{k}}
\end{array}\right), \quad 
\omega \geqslant 0, \quad
\bm{\tilde{k}}^{2} \leqslant 1\,, \quad
\bm{l_{1 \perp}} = \omega \;\bm{\tilde{l}_{1 \perp}}\,,
\quad |\bm{\tilde{l}_{1 \perp}}| = {\mathcal O}(1)\,.
\label{3.29}
\end{align}
We consider the limit $\omega \to 0$ keeping
$\bm{\tilde{k}}$ and $\bm{\tilde{l}_{1 \perp}}$ fixed.
Note that this implies from (\ref{3.19}) that also
$l_{1}$ and $l_{2}$ are proportional to $\omega$.
Inserting (\ref{3.26}) in (\ref{3.25}) we then find, 
\begin{align}
&{\cal M}_{\lambda}(p_{1}',p_{2}',k,p_{a},p_{b}) = 
e {\cal M}^{(0)}(s_{L},t,m_{\pi}^{2},m_{\pi}^{2},m_{\pi}^{2},m_{\pi}^{2})
\bigg[ \frac{p_{a \lambda}}{p_{a} \cdot k} 
-      \frac{p_{1 \lambda}}{p_{1} \cdot k} \bigg] \nn \\
&\qquad + e \bigg\lbrace \frac{1}{2(p_{a} \cdot k)^{2}}
[p_{a \lambda}k^{2} - k_{\lambda}(p_{a} \cdot k)] 
-
\frac{1}{2(p_{1} \cdot k)^{2}}
[p_{1 \lambda}(2l_{1} \cdot k-k^{2})
-(2 l_{1 \lambda}-k_{\lambda})p_{1} \cdot k]
\nn \\
&\qquad
-2 \bigg[ (p_{b} \cdot k)\frac{p_{a\lambda}}{p_{a} \cdot k} - p_{b\lambda} \bigg]
\frac{\partial}{\partial s_{L}} 
- 2 [(p_{a} - p_{1},k) - p_{a} \cdot l_{1}] 
\bigg[ \frac{p_{a \lambda}}{p_{a} \cdot k} 
-      \frac{p_{1 \lambda}}{p_{1} \cdot k} \bigg] 
\frac{\partial}{\partial t}
\bigg\rbrace
\nn \\
&\qquad \times
{\cal M}^{(0)}(s_{L},t,m_{\pi}^{2},m_{\pi}^{2},m_{\pi}^{2},m_{\pi}^{2}) + {\cal O}(\omega)\,,
\label{3.30}
\end{align}
$s_{L} = p_{a} \cdot p_{b} + p_{1} \cdot p_{2}$, 
$t = (p_{a} - p_{1})^{2} = (p_{b} - p_{2})^{2}$.

With (\ref{3.30}) we have given the first two terms of the Laurent
expansion in $\omega$
of ${\cal M}_{\lambda}$ around $\omega = 0$.
The first term on the r.h.s. of (\ref{3.30}) is the pole term
$\propto \omega^{-1}$ which, for $k^{2} = 0$,
is exactly the soft-photon term as given by S.~Weinberg;
see Sec.~II~1 of \cite{Weinberg:1965nx}.
The~term on the r.h.s. of (\ref{3.30})
with curly brackets is the term of order~$\omega^{0}$.
Note that in (\ref{3.30}) we give the expansion of the radiative amplitude
${\cal M}_{\lambda} \equiv {\cal M}_{\lambda}(p_{1}',p_{2}',k,p_{a},p_{b})$ 
[see~(\ref{2.13})] around
the phase-space point of zero radiation
($p_{1}, p_{2}, k = 0$).

Now we discuss the relation of (\ref{3.30}) to Low's theorem,
which gives an approximate expression for
${\cal M}_{\lambda}(p_{1}',p_{2}',k,p_{a},p_{b})$
for this phase-space point ($p_{1}'$, $p_{2}'$, $k$).
Low's formula, (1.7) of \cite{Low:1958sn},
which is for real photon emission, reads,
using our metric convention and notation,
as follows:
\begin{align}
{\cal M}_{\lambda}(p_{1}',p_{2}',k,p_{a},p_{b}) =& \; 
e {\cal M}^{(0)}(s_{L}',t_{2},m_{\pi}^{2},m_{\pi}^{2},m_{\pi}^{2},m_{\pi}^{2})
\bigg[ \frac{p_{a \lambda}}{p_{a} \cdot k} 
-      \frac{p_{1 \lambda}'}{p_{1}' \cdot k} \bigg] \nn \\
&
-e \bigg[  (p_{b} \cdot k)\frac{p_{a\lambda}}{p_{a} \cdot k}
         + (p_{2}' \cdot k)\frac{p_{1\lambda}'}{p_{1}' \cdot k}
         - p_{b\lambda} - p_{2\lambda}' \bigg]
\frac{\partial}{\partial s_{L}'}
\nn \\
&\times
{\cal M}^{(0)}(s_{L}',t_{2},m_{\pi}^{2},m_{\pi}^{2},m_{\pi}^{2},m_{\pi}^{2})
+ {\cal O}(\omega)\,,
\label{3.31}
\end{align}
$s_{L}' = p_{a} \cdot p_{b} + p_{1}' \cdot p_{2}'$, 
$t_{2} = (p_{b} - p_{2}')^{2}$.

This looks quite different from our Eqs. (\ref{3.25}) and (\ref{3.30}),
setting there $k^{2} = 0$,
but this is alright since the meaning of (\ref{3.25}) and (\ref{3.30}) versus
(\ref{3.31}) is \underline{different}.
As emphasized above, (\ref{3.30}) gives the first two terms
of the Laurent expansion of the radiative amplitude around the phase-space point
of no radiation ($p_{1}$, $p_{2}$, $k = 0$).
Low's formula (\ref{3.31}) is valid only at the given phase-space point
($p_{1}'$, $p_{2}'$, $k$).
That is, we can use (\ref{3.31}) \underline{only} 
for the value of $k$
dictated by energy-momentum conservation:
$k = p_{a} + p_{b} - p_{1}' - p_{2}'$.
If (\ref{3.31}) is used for a different $k$ we go outside
of the physical region of 
the radiative amplitude~${\cal M}_{\lambda}$.
Therefore, we prefer to call (\ref{3.31}) Low's approximate expression for
${\cal M}_{\lambda}$ in order to distinguish it 
from the Laurent expansion (\ref{3.25}) and (\ref{3.30}).
We can then construct the Laurent expansion of Low's expression (\ref{3.31})
around a phase-space point of no radiation 
($p_{1}, p_{2}, k = 0$),
where, of course, ($p_{1}$, $p_{2}$) must be near to ($p_{1}'$, $p_{2}'$).
In this way we get the relation between Low's formula (\ref{3.31})
and our Laurent expansion (\ref{3.25}) and (\ref{3.30}).
The details of this are given 
in Refs.~\cite{Nachtmann_talk, Lebiedowicz:2023ell}.


\section{The reactions $\pi \pi \to \pi \pi$ and $\pi \pi \to \pi \pi \gamma$ in the tensor-pomeron model}
\label{sec:4}

In this section we shall discuss elastic $\pi \pi$ scattering,
without and with photon emission,
in the tensor-pomeron model \cite{Ewerz:2013kda}.
Let us make some remarks on the tensor-pomeron model
which is a special Regge-type model. 
Of course, Regge theory for high-energy reactions 
has a long history, starting from papers around the 1960s.
Some early papers are \cite{Regge:1959mz,Chew:1962eu,Gribov:1963gx}, 
for reviews see \cite{Collins:1977,Collins:1984,Caneschi,
Donnachie:2002en,Gribov:2009}.
In the tensor-pomeron model of \cite{Ewerz:2013kda} the assumption is made
that the pomeron and the charge-conjugation $C = +1$ reggeons
$f_{2 \Reg}$, $a_{2 \Reg}$ couple to hadrons like 
symmetric tensors of rank~2,
the odderon and the $C = -1$ reggeons 
$\omega_{\Reg}$, $\rho_{\Reg}$ as vectors.
The idea, that the pomeron couplings could be related to
a tensor coupling was, to our knowledge, 
first proposed in \cite{Freund:1962}.
There the pomeron couplings were related to the couplings
of the energy-momentum tensor.
For more historical remarks on the tensor pomeron ideas
we refer to \cite{Ewerz:2016onn}.
The tensor-pomeron model \cite{Ewerz:2013kda}, 
which we shall use in the following,
is for \underline{soft} hadronic high-energy reactions
and has its origin in general investigations of the soft,
nonperturbative, pomeron in QCD using 
functional-integral techniques \cite{Nachtmann:1991ua}.
We note that also in holographic QCD a tensor character
of the pomeron couplings is preferred
\cite{Domokos:2009hm,Iatrakis:2016rvj}.
In \cite{Ewerz:2013kda} the constants in the vertex functions
describing the pomeron-hadron couplings were, 
as far as possible, determined from comparisons of theory and experiment.

Now we shall first, for simplicity, discuss the reactions
$\pi^{-} \pi^{0} \to \pi^{-} \pi^{0}$ and
$\pi^{-} \pi^{0} \to \pi^{-} \pi^{0} \gamma$
[see (\ref{2.1}), (\ref{2.2})] 
and then turn to charged-pion scattering.

\subsection{The reactions $\pi^{-} \pi^{0} \to \pi^{-} \pi^{0}$ and $\pi^{-} \pi^{0} \to \pi^{-} \pi^{0} \gamma$}
\label{sec:4.1}

We consider the elastic $\pi \pi$ scattering at high c.m. energy $\sqrt{s}$
where pomeron ($\Pom$) exchange dominates.
The amplitude for the subleading reggeon
($f_{2 \Reg}$, $\rho_{\Reg}$) exchanges will be treated in
Sec.~\ref{sec:4.2}.
The propagator and the pion couplings of the tensor pomeron are given in
(3.10), (3.11) and (3.34), (3.45), (3.46)
of \cite{Ewerz:2013kda}, respectively,
\newline
\includegraphics[width=140pt]{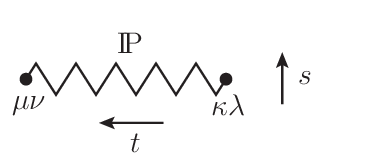} 
\vspace*{-0.4cm}
\begin{eqnarray}
&&i\Delta^{(\Pom)}_{\mu\nu,\kappa\lambda} (s,t) 
= \frac{1}{4s} \left(g_{\mu\kappa} g_{\nu\lambda} + g_{\mu\lambda} g_{\nu\kappa} 
- \frac{1}{2} g_{\mu\nu} g_{\kappa\lambda} \right) 
\, (-i s \alpha'_{\Pom})^{\alpha_{\Pom}(t)-1}\,,
\label{4.1}
\end{eqnarray}
\begin{eqnarray}
&&\alpha_{\Pom}(t) = \alpha_{\Pom}(0) + \alpha'_{\Pom} t \,,\quad
\alpha_{\Pom}(0) = 1 + \epsilon_{\Pom} \,,\nonumber\\
&&\epsilon_{\Pom} = 0.0808 \,,\quad \alpha'_{\Pom} = 0.25 \;\mbox{GeV}^{-2} \,;
\label{4.2}
\end{eqnarray}

\includegraphics[width=130pt]{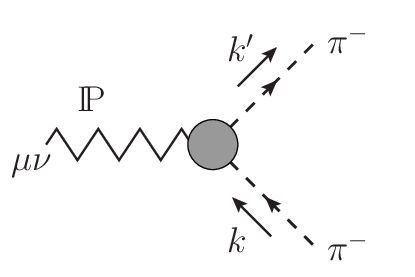} 
\includegraphics[width=130pt]{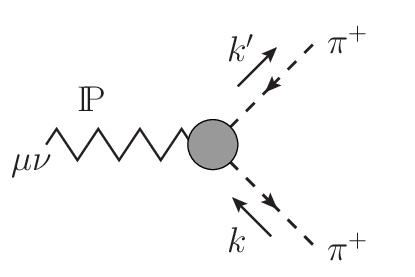} 
\includegraphics[width=130pt]{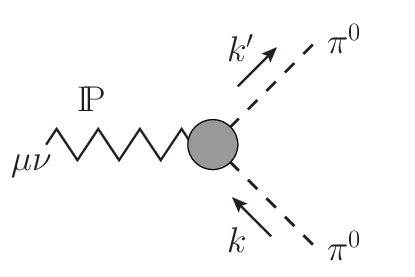}
\vspace*{-0.2cm}
\begin{eqnarray}
i \Gamma_{\mu\nu}^{(\Pom \pi\pi)} (k',k) 
= -i \, 2 \beta_{\Pom \pi\pi} F_M[(k'-k)^2] 
\left[ (k'+k)_\mu (k'+k)_\nu - \frac{1}{4} \, g_{\mu\nu} (k'+k)^2 \right]\,, \qquad 
\label{4.3}
\end{eqnarray}
\begin{eqnarray}
\beta_{\Pom \pi\pi} = 1.76 \;\mbox{GeV}^{-1}\,, \quad
F_{M}(t) = \frac{m_0^2}{m_0^2 -t}\,, \quad
m_{0}^{2} = 0.50 \;{\rm GeV}^2\,.
\label{4.4}
\end{eqnarray}
%

\begin{figure}[!h]
\includegraphics[width=5.0cm]{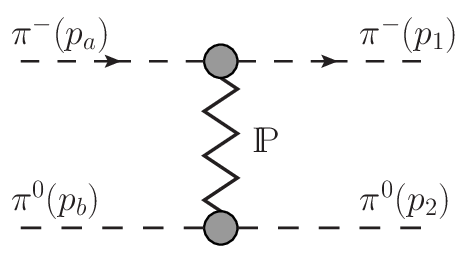}
\caption{Diagram with pomeron exchange
for $\pi^{-} \pi^{0} \to \pi^{-} \pi^{0}$ in the tensor-pomeron model.}
\label{fig:pimpi0_pimpi0_pom}
\end{figure}
The pomeron-exchange diagram for the reaction (\ref{2.1}) 
$\pi^{-} \pi^{0} \to \pi^{-} \pi^{0}$,
allowing the pions to be off-shell, 
is shown in Fig.~\ref{fig:pimpi0_pimpi0_pom}, 
and easily evaluated.
We get with the kinematic variables 
of (\ref{2.6}) and (\ref{2.7}) for (\ref{2.8}):
\begin{eqnarray}
&&{\cal M}^{(0)}_{\Pom}(s_{L},t, m_{a}^{2}, m_{b}^{2}, m_{1}^{2}, m_{2}^{2})
= i {\cal F}_{\Pom}(s,t)\left[ 2 (p_{a}+p_{1},p_{b}+p_{2})^{2}-\frac{1}{2}(p_{a}+p_{1})^{2}(p_{b}+p_{2})^{2} \right] \nonumber \\
&&\qquad \qquad \qquad = i {\cal F}_{\Pom}(s,t)\left[ 2(2 s_{L} + t)^{2} - \frac{1}{2}(-t+2 m_{a}^{2}+2m_{1}^{2})(-t+2 m_{b}^{2}+2m_{2}^{2})\right]\,.
\label{4.5}
\end{eqnarray}
Here we set
\begin{eqnarray}
{\cal F}_{\Pom}(s,t) &=& 
{\cal F}_{\Pom}\left[s_{L} + 
\frac{1}{2}(m_{a}^{2}+m_{b}^{2}+m_{1}^{2}+m_{2}^{2}),t\right] \nonumber\\
&=& \left[ 2 \beta_{\Pom \pi \pi} F_{M}(t) \right]^{2}
\frac{1}{4s}(-is \alpha'_{\Pom})^{\alpha_{\Pom}(t)-1} \,.
\label{4.6}
\end{eqnarray}

For the scattering of $\pi^{-} \pi^{0} \to \pi^{-} \pi^{0}$
with on-shell pions this gives
\begin{eqnarray}
&&\left.\braket{\pi^{-}(p_{1}),\pi^{0}(p_{2})|{\cal T}|\pi^{-}(p_{a}),\pi^{0}(p_{b})}\right|_{\rm on \; shell} =
{\cal M}^{(0)}_{\Pom}(s_{L},t, m_{\pi}^{2}, m_{\pi}^{2}, m_{\pi}^{2}, m_{\pi}^{2}) \nonumber \\
&& \qquad \qquad = i {\cal F}_{\Pom}(s,t)\left[ 2 (p_{a}+p_{1},p_{b}+p_{2})^{2}-\frac{1}{2}(p_{a}+p_{1})^{2}(p_{b}+p_{2})^{2} \right]
\nonumber \\
&& \qquad \qquad = 8is^{2} {\cal F}_{\Pom}(s,t)
\Big[ 1-\frac{4m_{\pi}^{2}-t}{s}+\frac{3}{16s^{2}} (4m_{\pi}^{2}-t)^{2} \Big]\,,
\label{4.7}
\end{eqnarray}
\begin{eqnarray}
&&\sigma_{\rm tot}(\pi^{-}\pi^{0}) =
\frac{1}{\sqrt{s(s-4 m_{\pi}^{2})}}\,
{\rm Im} \braket{\pi^{-}(p_{a}),\pi^{0}(p_{b})|{\cal T}|\pi^{-}(p_{a}),\pi^{0}(p_{b})}
\nonumber \\
&& \qquad = 2 \left( 2 \beta_{\Pom \pi \pi} \right)^{2} 
(s \alpha'_{\Pom})^{\epsilon_{\Pom}}
\cos\left( \frac{\pi}{2} \epsilon_{\Pom} \right)
\Big( 1-\frac{4m_{\pi}^{2}}{s} \Big)^{-1/2}
\Big[ 1-\frac{4m_{\pi}^{2}}{s} 
+\frac{3}{16} \Big(\frac{4m_{\pi}^{2}}{s} \Big)^{2} \Big]\,.\qquad
\label{4.8}
\end{eqnarray}

Now we come to the photon-emission process (\ref{2.2})
\begin{eqnarray}
\pi^{-} (p_{a}) + \pi^{0} (p_{b}) \to \pi^{-} (p_{1}') + \pi^{0} (p_{2}') 
+ \gamma (k, \epsilon)\,.
\label{4.9}
\end{eqnarray}
The relevant kinematic variables are here
\begin{eqnarray}
&& s = (p_{a} + p_{b})^{2} = (p_{1}' + p_{2}' + k)^{2}\,, \nonumber \\
&& t_{1} = (p_{a} - p_{1}')^{2} = (p_{b} - p_{2}' - k)^{2}\,, \nonumber \\
&& t_{2} = (p_{b} - p_{2}')^{2} = (p_{a} - p_{1}' - k)^{2}\,.
\label{4.10}
\end{eqnarray}
%
\begin{figure}[!h]
(a)\includegraphics[width=4.6cm]{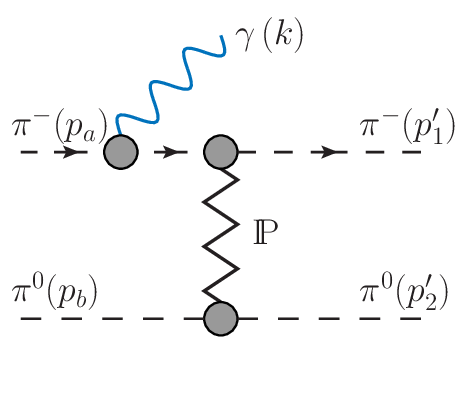}\quad
(b)\includegraphics[width=5.1cm]{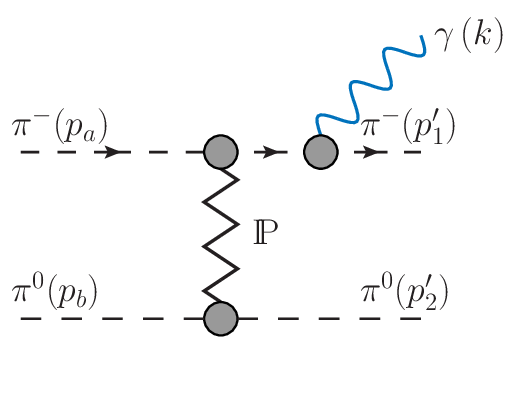}
(c)\includegraphics[width=4.6cm]{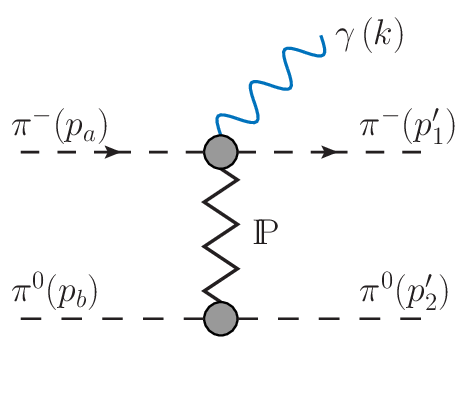}
\caption{Pomeron-exchange diagrams for $\pi^{-} \pi^{0} \to \pi^{-} \pi^{0} \gamma$
in the tensor-pomeron model.}
\label{fig:pimpi0_pimpi0gamma_diagrams}
\end{figure}
We have to calculate ${\cal M}_{\lambda \Pom}$ (\ref{2.13}), (\ref{2.14})
from the diagrams of Fig.~\ref{fig:pimpi0_pimpi0gamma_diagrams}.
First we calculate ${\cal M}^{(a)}_{\lambda \Pom}$ and ${\cal M}^{(b)}_{\lambda \Pom}$
from (\ref{2.11}) and (\ref{2.12}), respectively,
inserting for ${\cal M}^{(0)}$ the tensor-pomeron expression (\ref{4.5}).
Furthermore, we use the standard pion propagator
and the standard $\gamma \pi \pi$ vertex function
(see e.g. \cite{Ewerz:2013kda,Bolz:2014mya}). 
This gives
\begin{eqnarray}
&&\Delta[(p_{a}-k)^{2}] \, \Gamma_{\lambda}(p_{a}-k,p_{a}) = 
\frac{(2 p_{a} - k)_{\lambda}}{-2(p_{a} \cdot k) + k^{2}}\,, \nonumber \\
&&\Gamma_{\lambda}(p_{1}',p_{1}'+k) \, \Delta[(p_{1}'+k)^{2}] = 
\frac{(2 p_{1}' + k)_{\lambda}}{2(p_{1}' \cdot k) + k^{2}}\,.
\label{4.11}
\end{eqnarray}
From (\ref{3.11}) and (\ref{3.12}) we see that in QFT these relations
are exact for $\omega \to 0$ up to corrections of order $\omega$.
For us (\ref{4.11}) is part of our model assumptions.

With (\ref{4.5}) and (\ref{4.11}) we get from (\ref{2.11}) the following
amplitude ${\cal M}^{(a)}_{\lambda \Pom}$ corresponding to the diagram
of Fig.~\ref{fig:pimpi0_pimpi0gamma_diagrams}~(a):
\begin{eqnarray}
&&{\cal M}_{\lambda \Pom}^{(a)} = -e \,{\cal M}_{\Pom}^{(0,\,a)}\, 
\frac{(2 p_{a} - k)_{\lambda}}{-2(p_{a} \cdot k) + k^{2}}\,, \nonumber \\
&&{\cal M}_{\Pom}^{(0,\,a)}= i {\cal F}_{\Pom}[(p_{a}+p_{b}-k)^{2},t_{2}]
\left[ 2 (p_{a}+p_{1}'-k,p_{b}+p_{2}')^{2}
-\frac{1}{2}(p_{a}+p_{1}'-k)^{2}(p_{b}+p_{2}')^{2} \right].\nonumber\\
\label{4.12}
\end{eqnarray}
From (\ref{2.12}) we get for ${\cal M}^{(b)}_{\lambda \Pom}$ 
corresponding to the diagram
of Fig.~\ref{fig:pimpi0_pimpi0gamma_diagrams}~(b):
\begin{eqnarray}
&&{\cal M}_{\lambda \Pom}^{(b)} = -e \, 
\frac{(2 p_{1}' + k)_{\lambda}}{2(p_{1}' \cdot k) + k^{2}}\,
{\cal M}_{\Pom}^{(0,\,b)}\,, \nonumber \\
&&{\cal M}_{\Pom}^{(0,\,b)}= i {\cal F}_{\Pom}(s,t_{2})
\left[ 2 (p_{a}+p_{1}'+k,p_{b}+p_{2}')^{2}
-\frac{1}{2}(p_{a}+p_{1}'+k)^{2}(p_{b}+p_{2}')^{2} \right].
\label{4.13}
\end{eqnarray}
For ${\cal M}^{(c)}_{\lambda \Pom}$ we get from (\ref{2.22})
\begin{eqnarray}
k^{\lambda} {\cal M}_{\lambda \Pom}^{(c)} &=& 
-e {\cal M}_{\Pom}^{(0,\,a)} + e {\cal M}_{\Pom}^{(0,\,b)} \nonumber \\
&&-ie \big\lbrace {\cal F}_{\Pom}(s,t_{2})\big[ -8 (k,p_{b}+p_{2}')(p_{a}+p_{1}',p_{b}+p_{2}')
+2(k,p_{a}+p_{1}')(p_{b}+p_{2}')^{2} \big]
 \nonumber\\
&&+\big[ {\cal F}_{\Pom}[(p_{a}+p_{b}-k)^{2},t_{2})] - {\cal F}_{\Pom}(s,t_{2}) \big]
\nonumber\\
&&
\times \big[ 2 (p_{a}+p_{1}'-k,p_{b}+p_{2}')^{2}
-\frac{1}{2}(p_{a}+p_{1}'-k)^{2}(p_{b}+p_{2}')^{2} \big] \big\rbrace\,.
\label{4.14}
\end{eqnarray}
Using the explicit expression for ${\cal F}_{\Pom}(s,t_{2})$ (\ref{4.6}) we get
\begin{eqnarray}
{\cal F}_{\Pom}[(p_{a}+p_{b}-k)^{2},t_{2})] 
- {\cal F}_{\Pom}(s,t_{2}) =
{\cal F}_{\Pom}(s,t_{2})\, \left(2 - \alpha_{\Pom}(t_2)\right)\,
\frac{2(p_{a} + p_{b},k)-k^{2}}{s} \, g_{\Pom}(\varkappa, t_{2})
\,, \nonumber\\
\label{4.15}
\end{eqnarray}
where we define
\begin{eqnarray}
\varkappa &=& \frac{2(p_{a} + p_{b},k)-k^{2}}{s}\,, 
\label{4.16}\\
g_{\Pom}(\varkappa, t_{2}) &=& \frac{1}{\left(2 - \alpha_{\Pom}(t_2)\right)\varkappa}
\big[ (1-\varkappa)^{\alpha_{\Pom}(t_2)-2}-1
\big] \nonumber\\
&=& 
1 + \frac{\varkappa}{2!}\left(3-\alpha_{\Pom}(t_2)\right) + 
\frac{\varkappa^{2}}{3!}\left(3-\alpha_{\Pom}(t_2)\right)\left(4-\alpha_{\Pom}(t_2)\right) + \ldots \,.
\label{4.17}
\end{eqnarray}
The series expansion in (\ref{4.17}) is absolutely convergent 
for $|\varkappa| < 1$ which is the only region of interest for us.

Inserting (\ref{4.15}) in (\ref{4.14}) we get
\begin{eqnarray}
k^{\lambda} {\cal M}^{(c)}_{\lambda \Pom} &=& 
-ie {\cal F}_{\Pom}(s,t_{2}) \big\lbrace
-8 (k,p_{b}+p_{2}')(p_{a}+p_{1}',p_{b}+p_{2}') + 2(k,p_{a}+p_{1}')(p_{b}+p_{2}')^{2}
\nonumber\\
&&+ 
\frac{2(p_{a} + p_{b},k)-k^{2}}{s} \, \left(2 - \alpha_{\Pom}(t_2)\right)\, g_{\Pom}(\varkappa, t_{2}) \nonumber\\
&&
\times \big[ 2 (p_{a}+p_{1}'-k,p_{b}+p_{2}')^{2}
-\frac{1}{2}(p_{a}+p_{1}'-k)^{2}(p_{b}+p_{2}')^{2} \big] \big\rbrace\,.
\label{4.18}
\end{eqnarray}
From this we see that a \underline{simple} solution of (\ref{4.18}) for
${\cal M}^{(c)}_{\lambda \Pom}$ is
\begin{eqnarray}
{\cal M}^{(c)}_{\lambda \Pom} &=& 
-ie {\cal F}_{\Pom}(s,t_{2}) \big\lbrace
-8 (p_{b}+p_{2}')_{\lambda}(p_{a}+p_{1}',p_{b}+p_{2}') 
+2 (p_{a}+p_{1}')_{\lambda}(p_{b}+p_{2}')^{2}
\nonumber\\
&&+ 
(2p_{a} + 2p_{b} -k)_{\lambda} \, \left(2 - \alpha_{\Pom}(t_2)\right)\, g_{\Pom}(\varkappa, t_{2}) \nonumber\\
&&
\times \frac{1}{s}
\big[ 2 (p_{a}+p_{1}'-k,p_{b}+p_{2}')^{2}
-\frac{1}{2}(p_{a}+p_{1}'-k)^{2}(p_{b}+p_{2}')^{2} \big] \big\rbrace\,.
\label{4.19}
\end{eqnarray}
However, we could add to ${\cal M}^{(c)}_{\lambda \Pom}$ from (\ref{4.19}), for instance,
terms proportional to
\begin{eqnarray}
p_{a \lambda} (p_{1}' \cdot k) - p'_{1 \lambda} (p_{a} \cdot k)\,,
\label{4.19a}
\end{eqnarray}
or
\begin{eqnarray}
\varepsilon_{\lambda \mu \nu \rho} p_{a}^{\mu} p_{b}^{\nu} k^{\rho}
\big( \varepsilon_{\alpha \beta \gamma \delta} p_{a}^{\alpha} p_{b}^{\beta} 
p_{1}'^{\gamma} p_{2}'^{\delta} \big)\,,
\label{4.20}
\end{eqnarray}
and still have a solution of (\ref{4.18}).
Thus, the solution (\ref{4.19}) for ${\cal M}^{(c)}_{\lambda \Pom}$
is in general not unique
as it is to order $\omega^{0}$; see (\ref{3.24}).
This fact is well known in the literature;
see for instance \cite{Bern:2014vva}.

Collecting now everything together we get for the amplitude
of reaction (\ref{4.9}) in our model
\begin{eqnarray}
&&{\cal M}_{\lambda \Pom}^{(\pi^{-} \pi^{0} \to \pi^{-} \pi^{0} \gamma)} = {\cal M}_{\lambda \Pom}^{(a)} +
{\cal M}_{\lambda \Pom}^{(b)} + {\cal M}_{\lambda \Pom}^{(c)}
\label{4.21}
\end{eqnarray}
with ${\cal M}_{\lambda \Pom}^{(c)}$ given in (\ref{4.19})
and ${\cal M}_{\lambda \Pom}^{(a)}$ and ${\cal M}_{\lambda \Pom}^{(b)}$
obtained from (\ref{4.12}), (\ref{4.13}) and (\ref{4.15}),
as follows:
\begin{eqnarray}
&&{\cal M}_{\lambda \Pom}^{(a)} = ie {\cal F}_{\Pom}(s,t_{2})
\Big[ 1+ \left(2 - \alpha_{\Pom}(t_2)\right)\, 
\frac{2(p_{a} + p_{b},k) - k^{2}}{s}\,
g_{\Pom}(\varkappa, t_{2}) \Big] \nonumber\\
&&\qquad \quad \;\;\times \left[ 2 (p_{a}+p_{1}'-k,p_{b}+p_{2}')^{2}
-\frac{1}{2}(p_{a}+p_{1}'-k)^{2}(p_{b}+p_{2}')^{2} \right]
\frac{(2p_{a} -k)_{\lambda}}{2(p_{a} \cdot k)-k^{2}}\,, \qquad
\label{4.22}\\
&&{\cal M}_{\lambda \Pom}^{(b)} = -ie {\cal F}_{\Pom}(s,t_{2})
\left[ 2 (p_{a}+p_{1}'+k,p_{b}+p_{2}')^{2}
-\frac{1}{2}(p_{a}+p_{1}'+k)^{2}(p_{b}+p_{2}')^{2} \right]
\frac{(2p_{1}'+k)_{\lambda}}{2(p_{1}' \cdot k)+k^{2}}.
\nonumber \\
\label{4.23}
\end{eqnarray}
%
In the soft-photon limit $\omega \to 0$
according to (\ref{3.29}) the amplitude
${\cal M}_{\lambda \Pom}^{(\pi^{-} \pi^{0} \to \pi^{-} \pi^{0} \gamma)}$
from (\ref{4.19}), (\ref{4.21})--(\ref{4.23}),
must satisfy our general QFT formulas (\ref{3.25}), (\ref{3.30}).
We have checked this by explicit calculation,
setting in (\ref{4.19}), (\ref{4.22}), (\ref{4.23}),
$p_{1}' = p_{1} - l_{1}$, $p_{2}' = p_{2} - l_{2}$
from (\ref{3.14}), expanding in $\omega$,
and using ${\cal M}_{\Pom}^{(0)}$ from (\ref{4.7}).

The results for ${\cal M}_{\lambda \Pom}^{(\pi^{-} \pi^{0} \to \pi^{-} \pi^{0} \gamma)}$, (\ref{4.19}), (\ref{4.21})--(\ref{4.23}), 
hold for $k^{0} \geqslant 0$, 
$k^{2} \geqslant 0$, that is, for real- and virtual-photon emission.

Below in Sec.~\ref{sec:5}
we shall consider only real 
photon emission where we have $k^{2} = 0$.

Some comments on these results are in order.
We are interested in soft photon emission 
where $\omega \ll \sqrt{s}$.
We have then from (\ref{4.16}) and (\ref{4.17})
$|\varkappa| = {\cal O}(\omega/\sqrt{s})$ 
and $g_{\Pom}(\varkappa, t_{2}) \approx 1$.
Looking at ${\cal M}_{\lambda \Pom}^{(a)}$ we see that there the term
proportional to $g_{\Pom}(\varkappa, t_{2})$
is a correction of order $\omega/\sqrt{s}$ 
relative to the leading term.
On the other hand, in ${\cal M}_{\lambda \Pom}^{(c)}$ the term
proportional to $g_{\Pom}(\varkappa, t_{2})$ is not suppressed
relative to the first term in the wavy brackets of (\ref{4.19}).
But in the soft photon region ${\cal M}_{\lambda \Pom}^{(c)}$
is, anyway, only of order $\omega/\sqrt{s}$ relative to
${\cal M}_{\lambda \Pom}^{(a)}$ and ${\cal M}_{\lambda \Pom}^{(b)}$.
Thus, in the soft-photon region our model 
should give reliable results. 
But the question arises how high we can go in $\omega$ and still
trust the model.
We have, as basis of the model, used the high-energy approximation,
given by the pomeron-exchange term,
for the $\pi \pi$ scattering amplitude.
Therefore, in ${\cal M}^{(0)}$ (\ref{4.5}), (\ref{4.6})
the c.m. energy squared $s$ should be large enough,
above the resonance region, say
\begin{equation}
s \geqslant s_{0} = (5\; {\rm GeV})^{2}\,.
\label{4.23a}
\end{equation}
But in the reaction $\pi \pi \to \pi \pi \gamma$ we need
the off-shell amplitudes ${\cal M}^{(0,a)}$ (\ref{4.12})
and ${\cal M}^{(0,b)}$ (\ref{4.13}) where
the squared c.m. energies are, respectively,
\begin{eqnarray}
&&s_{a} = (p_{a} + p_{b} -k)^{2} = (p_{1}' + p_{2}')^{2}\,,
\label{4.23b_sa}\\
&&s_{b} = s\,.
\label{4.23b_sb}
\end{eqnarray}
Surely, in order to apply our Regge model also for 
${\cal M}^{(0,a)}$ we should require
\begin{eqnarray}
s_{a} = (p_{a} + p_{b} -k)^{2} 
= s - 2 (p_{a} + p_{b},k) + k^{2} \geqslant s_{0}\,.
\label{4.23c}
\end{eqnarray}
In the overall c.m. system this means
\begin{eqnarray}
\omega \leqslant \frac{1}{2 \sqrt{s}}\,
\left(s - s_{0} + k^{2} \right)\,.
\label{4.23d}
\end{eqnarray}
Below, in Sec.~\ref{sec:5}, we shall take this constraint into account.

In \cite{Bolz:2014mya} vertices for the coupling of 
$\gamma \pi \pi$ and $\Pom \gamma \pi \pi$ were derived
from a Lagrangian; see (B.66)--(B.71) there.
Using these vertices for evaluating the diagrams of 
Fig.~\ref{fig:pimpi0_pimpi0gamma_diagrams} and using
in all three diagrams the pomeron propagator
$\Delta^{(\Pom)}_{\mu \nu, \kappa \lambda}(s,t_{2})$
with the \underline{common} value $s = (p_{a} + p_{b})^{2}$
gives ${\cal M}_{\lambda \Pom}^{(a)}$, ${\cal M}_{\lambda \Pom}^{(b)}$
and ${\cal M}_{\lambda \Pom}^{(c)}$ as in (\ref{4.22}),
(\ref{4.23}), and (\ref{4.19}), respectively, but setting
$g_{\Pom}(\varkappa, t_{2}) = 0$.
Thus, our full results for ${\cal M}_{\lambda \Pom}^{(a)}$, 
${\cal M}_{\lambda \Pom}^{(b)}$, ${\cal M}_{\lambda \Pom}^{(c)}$ above
are an improvement of the simple results,
as we respect now the general QFT structure of the amplitudes
shown in Fig.~\ref{fig:pimpi0gamma}. 
As discussed above, for soft photons the improvement amounts to
suitable additions of non-leading terms of relative order
$\omega/\sqrt{s}$.

What about anomalous soft photons in this framework?
Given the amplitude for $\pi^{-} \pi^{0} \to \pi^{-} \pi^{0}$
we have constructed ${\cal M}_{\lambda \Pom}^{(a)}$ and 
${\cal M}_{\lambda \Pom}^{(b)}$ in a straightforward way.
Of course, we had to extrapolate to off-shell pions and to assume
(\ref{4.11}) to hold not only for $\omega \to 0$.
But by and large we think that ${\cal M}_{\lambda \Pom}^{(a)}$ and 
${\cal M}_{\lambda \Pom}^{(b)}$ leave little room for anomalous soft photons.
This is quite different for ${\cal M}_{\lambda \Pom}^{(c)}$ 
which we determined here as the simplest solution 
of the gauge-invariance condition (\ref{4.18}).
Clearly, other solutions of (\ref{4.18}) for ${\cal M}_{\lambda \Pom}^{(c)}$
are possible which could describe ``anomalous'' production
of soft photons. One of the present authors has been involved in
a suggestion for the origin of such anomalous soft photons:
``synchrotron radiation from the vacuum'' 
\cite{Nachtmann:1983uz,Botz:1994bg,
Nachtmann:ELFE,
Nachtmann:Lectures,Nachtmann:2014qta}.
For a list of suggestions by other authors we refer to 
\cite{Wong:2014pY}.

\subsection{Charged-pion scattering without and with photon radiation}\label{sec:4.2}

In this section we consider the following reactions at high energies
in the tensor-pomeron model:
\begin{eqnarray}
&&\pi^{-}(p_{a}) + \pi^{+}(p_{b}) \to \pi^{-}(p_{1}) + \pi^{+}(p_{2})\,,
\label{4.24}\\
&&\pi^{-}(p_{a}) + \pi^{+}(p_{b}) \to \pi^{-}(p_{1}') + \pi^{+}(p_{2}')
+ \gamma (k, \epsilon)\,,
\label{4.25}
\end{eqnarray}
and
\begin{eqnarray}
&&\pi^{\pm}(p_{a}) + \pi^{\pm}(p_{b}) \to \pi^{\pm}(p_{1}) + \pi^{\pm}(p_{2})\,,
\label{4.26}\\
&&\pi^{\pm}(p_{a}) + \pi^{\pm}(p_{b}) \to \pi^{\pm}(p_{1}') + \pi^{\pm}(p_{2}')
+ \gamma (k, \epsilon)\,.
\label{4.27}
\end{eqnarray}
Again we leave $k$ arbitrary and do not require $k^{2} = 0$.

\begin{figure}[!h]
\includegraphics[width=5.0cm]{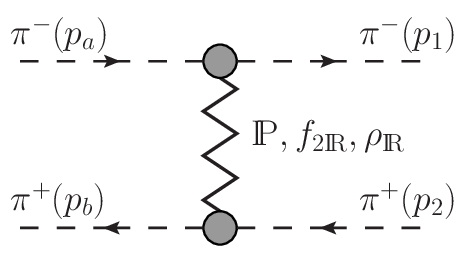}
\caption{The diagram for $\pi^{-} \pi^{+} \to \pi^{-} \pi^{+}$ 
elastic scattering with exchange of the pomeron,
the $f_{2 \Reg}$, and the $\rho_{\Reg}$ reggeons.}
\label{fig:pimpip_pimpip_PFR}
\end{figure}
The diagrams for the elastic scattering processes 
(\ref{4.24}) and (\ref{4.26}) are analogous to the one 
in Fig.~\ref{fig:pimpi0_pimpi0_pom} 
but now we include the subleading $f_{2 \Reg}$ and $\rho_{\Reg}$ reggeon exchanges;
see Fig.~\ref{fig:pimpip_pimpip_PFR}.
To evaluate these diagrams we need the effective
$f_{2 \Reg}$ and $\rho_{\Reg}$ propagators and their couplings to pions.
In our model these are given in (3.12)--(3.15) and
(3.53), (3.54), (3.63), (3.64) of \cite{Ewerz:2013kda}, respectively.
The $f_{2 \Reg}$ propagator and the $f_{2 \Reg} \pi \pi$
couplings are as in (\ref{4.1})--(\ref{4.3})
with the replacements
\begin{eqnarray}
&&\alpha_{\Pom}(t) \to 
\alpha_{f_{2 \Reg}}(t) = \alpha_{f_{2 \Reg}}(0) + \alpha'_{f_{2 \Reg}} t \,,\nonumber\\
&&\alpha_{f_{2 \Reg}}(0) = 0.5475 \,,
\quad \alpha'_{f_{2 \Reg}} = 0.9 \;\mbox{GeV}^{-2} \,,\nonumber\\
&&2 \beta_{\Pom \pi \pi} \to \frac{g_{f_{2 \Reg}\pi \pi}}{2 M_{0}}\,,\nonumber\\
&&g_{f_{2 \Reg}\pi \pi} = 9.30 \,,
\quad M_{0} = 1 \;\mbox{GeV}\,.
\label{4.27a}
\end{eqnarray}
For the effective $\rho_{\Reg}$ propagator and the $\rho_{\Reg} \pi \pi$ coupling we have
\newline
\includegraphics[width=140pt]{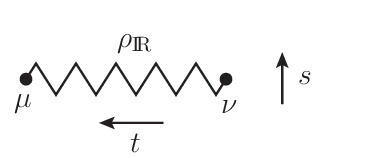} 
\vspace*{-0.4cm}
\begin{eqnarray}
&&i\Delta^{(\rho_{\Reg})}_{\mu\nu} (s,t) 
= i g_{\mu\nu} \frac{1}{M_{-}^{2}} 
\, (-i s \alpha'_{\rho_{\Reg}})^{\alpha_{\rho_{\Reg}}(t)-1}\,,
\label{4.27b}\\
&&\alpha_{\rho_{\Reg}}(t) = \alpha_{\rho_{\Reg}}(0) + \alpha'_{\rho_{\Reg}} t \,,\nonumber\\
&&\alpha_{\rho_{\Reg}}(0) = 0.5475 \,,
\quad \alpha'_{\rho_{\Reg}} = 0.9 \;\mbox{GeV}^{-2} \,,\nonumber\\
&&M_{-} = 1.41 \;\mbox{GeV}\,.
\label{4.27c}
\end{eqnarray}
\includegraphics[width=130pt]{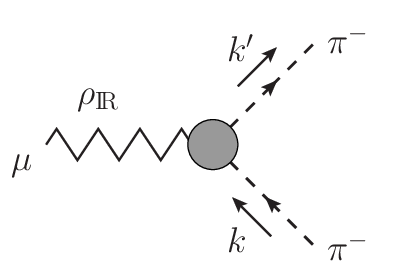} 
\includegraphics[width=130pt]{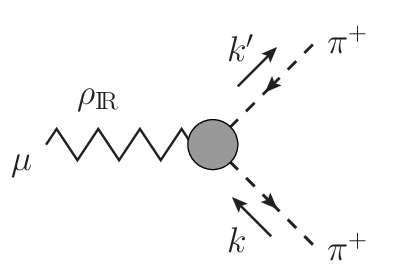} 
\vspace*{-0.2cm}
\begin{eqnarray} 
&&i \Gamma_{\mu}^{(\rho_{\Reg} \pi^{-}\pi^{-})} (k',k) =
-i \Gamma_{\mu}^{(\rho_{\Reg} \pi^{+}\pi^{+})} (k',k)
= \frac{i}{2} \, g_{\rho_{\Reg} \pi\pi} F_{M}[(k'-k)^2]\,
(k'+k)_\mu\,, \nonumber \\
&&g_{\rho_{\Reg} \pi\pi} = 15.63\,,
\label{4.27d}
\end{eqnarray}
where $F_{M}(t)$ is defined in (\ref{4.4}).

Now everything is prepared to evaluate the diagram of 
Fig.~\ref{fig:pimpip_pimpip_PFR} for the general off-shell
$\pi^{-} \pi^{+}$ scattering amplitude. 
We get [cf. (\ref{2.8}) and (\ref{4.5})]
\begin{eqnarray}
\left.\braket{\pi^{-}(p_{1}),\pi^{+}(p_{2})|{\cal T}|\pi^{-}(p_{a}),\pi^{+}(p_{b})}\right|_{\rm off \; shell} &=&
{\cal M}^{(0)\pi^{-}\pi^{+}}(s_{L},t,m_{a}^{2},m_{b}^{2},m_{1}^{2},m_{2}^{2})\nonumber \\
&=&{\cal M}^{(0)}_{\Pom} +
{\cal M}^{(0)}_{f_{2 \Reg}} +
{\cal M}^{(0)}_{\rho_{\Reg}}\,,
\label{4.27e}
\end{eqnarray}
where
\begin{eqnarray}
{\cal M}^{(0)}_{\Pom}
&=& i {\cal F}_{\Pom}(s,t)\left[ 2 (p_{a}+p_{1},p_{b}+p_{2})^{2}-\frac{1}{2}(p_{a}+p_{1})^{2}(p_{b}+p_{2})^{2} \right] \nonumber \\
&=& i {\cal F}_{\Pom}(s,t)\left[ 2(2 s_{L} + t)^{2} - \frac{1}{2}(-t+2 m_{a}^{2}+2m_{1}^{2})(-t+2 m_{b}^{2}+2m_{2}^{2})\right]\,,
\label{4.27f} \\
{\cal M}^{(0)}_{f_{2 \Reg}}
&=& i {\cal F}_{f_{2 \Reg}}(s,t)\left[ 2 (p_{a}+p_{1},p_{b}+p_{2})^{2}-\frac{1}{2}(p_{a}+p_{1})^{2}(p_{b}+p_{2})^{2} \right] \nonumber \\
&=& i {\cal F}_{f_{2 \Reg}}(s,t)\left[ 2(2 s_{L} + t)^{2} - \frac{1}{2}(-t+2 m_{a}^{2}+2m_{1}^{2})(-t+2 m_{b}^{2}+2m_{2}^{2})\right]\,,
\label{4.27g} \\
{\cal M}^{(0)}_{\rho_{\Reg}}
&=& {\cal F}_{\rho_{\Reg}}(s,t)(p_{a}+p_{1},p_{b}+p_{2})\nonumber \\
&=& {\cal F}_{\rho_{\Reg}}(s,t) (2 s_{L} + t)\,.
\label{4.27h} 
\end{eqnarray}
Here ${\cal F}_{\Pom}(s,t)$ is defined in (\ref{4.6}) and we have set
\begin{eqnarray}
&&{\cal F}_{f_{2 \Reg}}(s,t) =
\left[ \frac{g_{f_{2 \Reg}\pi \pi}}{2 M_{0}} F_{M}(t) \right]^{2}
\frac{1}{4s}(-is \alpha'_{f_{2 \Reg}})^{\alpha_{f_{2 \Reg}}(t)-1} \,,
\label{4.27i} \\
&&{\cal F}_{\rho_{\Reg}}(s,t) =
\left[ \frac{g_{\rho_{\Reg}\pi \pi}}{2 M_{-}} F_{M}(t) \right]^{2}
(-is \alpha'_{\rho_{\Reg}})^{\alpha_{\rho_{\Reg}}(t)-1} \,.
\label{4.27j}
\end{eqnarray}

For the on-shell elastic $\pi^{-} \pi^{+}$ scattering we get,
setting $m_{a}^{2} = m_{b}^{2} = m_{1}^{2} = m_{2}^{2} = m_{\pi}^{2}$
in (\ref{2.6}), (\ref{2.7}) and (\ref{4.27f})--(\ref{4.27h})
\begin{eqnarray}
&&\braket{\pi^{-}(p_{1}),\pi^{+}(p_{2})|{\cal T}|\pi^{-}(p_{a}),\pi^{+}(p_{b})} 
={\cal M}^{(0)\pi^{-}\pi^{+}}(s_{L},t, m_{\pi}^{2}, m_{\pi}^{2}, m_{\pi}^{2}, m_{\pi}^{2})\nonumber \\
&& \qquad 
\equiv 
{\cal M}^{(0)\pi^{-}\pi^{+}}(s,t) \nonumber \\
&& \qquad = i \Big[{\cal F}_{\Pom}(s,t) + {\cal F}_{f_{2 \Reg}}(s,t) \Big]
\Big[ 2 (p_{a}+p_{1},p_{b}+p_{2})^{2}-\frac{1}{2}(p_{a}+p_{1})^{2}(p_{b}+p_{2})^{2} \Big] \nonumber \\
&& \qquad \quad + {\cal F}_{\rho_{\Reg}}(s,t) (p_{a}+p_{1},p_{b}+p_{2})
\nonumber \\
&& \qquad = 8is^{2} \Big[{\cal F}_{\Pom}(s,t) + {\cal F}_{f_{2 \Reg}}(s,t) \Big]
\Big[ 1-\frac{4m_{\pi}^{2}-t}{s}+\frac{3}{16s^{2}} (4m_{\pi}^{2}-t)^{2} \Big] \nonumber \\
&& \qquad \quad + 2s {\cal F}_{\rho_{\Reg}}(s,t) 
\Big[ 1 - \frac{4 m_{\pi}^{2}-t}{2s}\Big]\,.
\label{4.28}
\end{eqnarray}
For brevity of notation we use in the following the notation
${\cal M}^{(0)\pi^{-}\pi^{+}}(s,t)$ 
for the on-shell pion-pion elastic scattering amplitude.

Turning now to the reactions (\ref{4.26}) of like sign
$\pi \pi$ scattering we get from the diagrams analogous to
Fig.~\ref{fig:pimpip_pimpip_PFR} the following for on-shell pions
\begin{eqnarray}
&&\braket{\pi^{+}(p_{1}),\pi^{+}(p_{2})|{\cal T}|\pi^{+}(p_{a}),\pi^{+}(p_{b})} =
\braket{\pi^{-}(p_{1}),\pi^{-}(p_{2})|{\cal T}|\pi^{-}(p_{a}),\pi^{-}(p_{b})} 
\nonumber \\
&& \qquad = 8is^{2} \Big[{\cal F}_{\Pom}(s,t) + {\cal F}_{f_{2 \Reg}}(s,t) \Big]
\Big[ 1-\frac{4m_{\pi}^{2}-t}{s}+\frac{3}{16s^{2}} (4m_{\pi}^{2}-t)^{2} \Big] \nonumber \\
&& \qquad \quad - 2s {\cal F}_{\rho_{\Reg}}(s,t) 
\Big[ 1 - \frac{4 m_{\pi}^{2}-t}{2s}\Big] 
+ (p_{1} \leftrightarrow p_{2})\,.
\label{4.28a}
\end{eqnarray}
The exchange $p_{1} \leftrightarrow p_{2}$ implies 
$t \leftrightarrow u$ where $u = -s -t + 4m_{\pi}^{2}$.

The total cross sections for $\pi \pi$ scattering are obtained
from the forward-scattering amplitudes using the optical theorem.
In this way we get from (\ref{4.28})
for $\pi^{-} \pi^{+}$ scattering
\begin{eqnarray}
&&\sigma_{{\rm tot} \,,\pi^{-}\pi^{+}}(s) =
\frac{1}{\sqrt{s(s-4 m_{\pi}^{2})}}\,
{\rm Im} \braket{\pi^{-}(p_{a}),\pi^{+}(p_{b})|{\cal T}|\pi^{-}(p_{a}),\pi^{+}(p_{b})}
\nonumber \\
&& \qquad = 2 \Big( 1-\frac{4m_{\pi}^{2}}{s} \Big)^{-1/2}
\Big\lbrace 
\Big[
\left( 2 \beta_{\Pom \pi \pi} \right)^{2} 
(s \alpha'_{\Pom})^{\alpha_{\Pom}(0)-1}
\cos\left( \frac{\pi}{2} (\alpha_{\Pom}(0)-1) \right)
\nonumber \\
&& \qquad 
+
\left( \frac{g_{f_{2 \Reg} \pi \pi}}{2 M_{0}} \right)^{2} 
(s \alpha'_{f_{2 \Reg}})^{\alpha_{f_{2 \Reg}}(0)-1}
\cos\left( \frac{\pi}{2} (\alpha_{f_{2 \Reg}}(0)-1) \right) 
\Big]
\Big( 1-\frac{4m_{\pi}^{2}}{s} +\frac{3m_{\pi}^{4}}{s^{2}} \Big)
\nonumber \\
&& \qquad +
\left( \frac{g_{\rho_{\Reg} \pi \pi}}{2 M_{-}} \right)^{2} 
(s \alpha'_{\rho_{\Reg}})^{\alpha_{\rho_{\Reg}}(0)-1}
\sin\left( \frac{\pi}{2} (1-\alpha_{\rho_{\Reg}}(0)) \right)
\Big( 1-\frac{2m_{\pi}^{2}}{s}\Big)
\Big\rbrace
\,.
\label{4.28b}
\end{eqnarray}
The total cross sections for $\pi^{+}\pi^{+}$ and $\pi^{-}\pi^{-}$
scattering are obtained from (\ref{4.28a}) for $t = 0$.
Here for $s \gg 4 m_{\pi}^{2}$ and $t = 0$ the term
$(p_{1} \leftrightarrow p_{2})$ is highly suppressed and,
thus, very small.
Neglecting the term $(p_{1} \leftrightarrow p_{2})$ for $t = 0$
we get the total cross sections for $\pi^{+}\pi^{+}$ 
and $\pi^{-}\pi^{-}$ scattering as in (\ref{4.28b})
but with a sign change in the $\rho_{\Reg}$ term.

\begin{figure}[!h]
(a)\includegraphics[width=4.6cm]{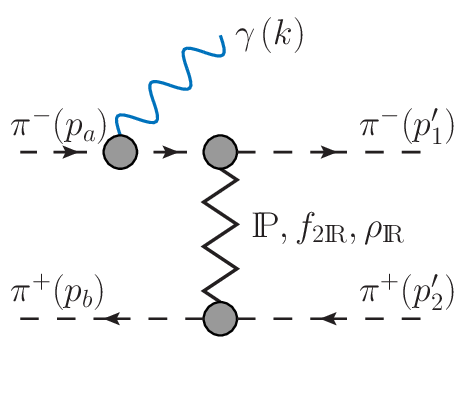}\quad
(b)\includegraphics[width=5.2cm]{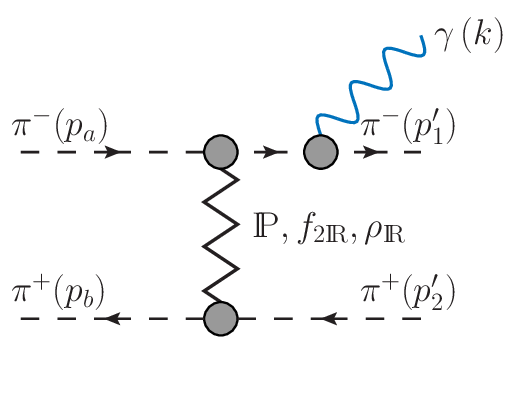}
(c)\includegraphics[width=4.6cm]{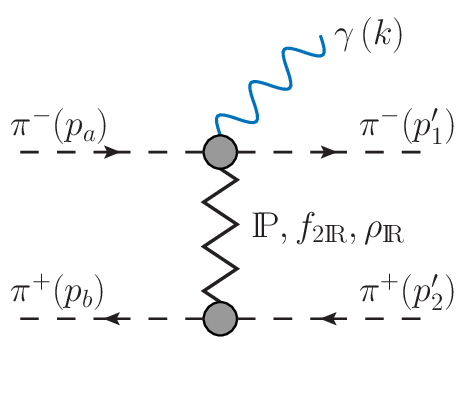}
(d)\includegraphics[width=4.6cm]{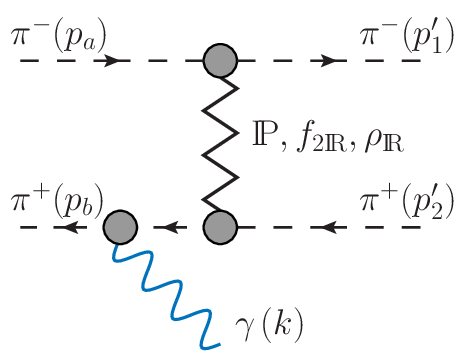}\quad
(e)\includegraphics[width=5.2cm]{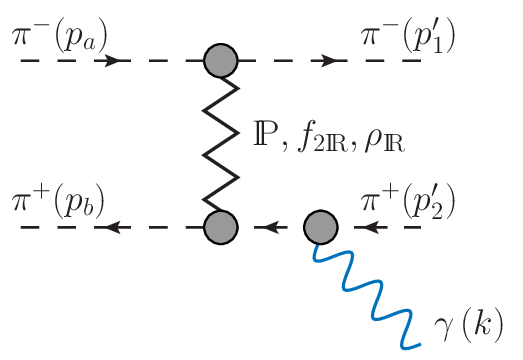}
(f)\includegraphics[width=4.6cm]{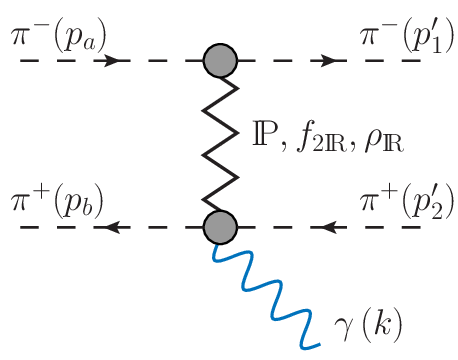}
\caption{Diagrams for the reaction 
$\pi^{-} \pi^{+} \to \pi^{-} \pi^{+} \gamma$
with exchange of the pomeron,
the $f_{2 \Reg}$, and the $\rho_{\Reg}$ reggeons.}
\label{fig:pipigam}
\end{figure}

For the photon emission process (\ref{4.25})
we have 6 diagrams shown in Fig.~\ref{fig:pipigam}.
The diagrams for (\ref{4.27}) are analogous
but in addition we have the diagrams with
$p_{1}'$ and $p_{2}'$ interchanged.
The kinematic variables for these reactions are 
as in (\ref{4.10}).
We have
\begin{eqnarray}
&&\braket{\pi^{-}(p_{1}'),\pi^{+}(p_{2}'),\gamma(k,\epsilon)|{\cal T}|
\pi^{-}(p_{a}),\pi^{+}(p_{b})} 
= (\epsilon^{\lambda})^{*}
{\cal M}_{\lambda}^{(\pi^{-} \pi^{+} \to \pi^{-} \pi^{+} \gamma)}\,,
\label{4.29} \\
&&\braket{\pi^{+}(p_{1}'),\pi^{+}(p_{2}'),\gamma(k,\epsilon)|{\cal T}|
\pi^{+}(p_{a}),\pi^{+}(p_{b})} 
= (\epsilon^{\lambda})^{*}
{\cal M}_{\lambda}^{(\pi^{+} \pi^{+} \to \pi^{+} \pi^{+} \gamma)}\,.
\label{4.30}
\end{eqnarray}
Our building blocks for these ${\cal M}_{\lambda}$ amplitudes
are ${\cal M}_{\lambda}^{(a)}, \ldots, {\cal M}_{\lambda}^{(f)}$ 
corresponding to the diagrams (a)--(f) from Fig.~\ref{fig:pipigam}.
We have here
\begin{eqnarray}
{\cal M}_{\lambda}^{(a)} = 
{\cal M}_{\lambda \Pom}^{(a)} +
{\cal M}_{\lambda f_{2 \Reg}}^{(a)} + 
{\cal M}_{\lambda \rho_{\Reg}}^{(a)}
\label{4.30a}
\end{eqnarray}
and similarly for
${\cal M}_{\lambda}^{(b)}, \ldots, {\cal M}_{\lambda}^{(f)}$.
The amplitudes
${\cal M}_{\lambda \Pom}^{(a)}$, 
${\cal M}_{\lambda \Pom}^{(b)}$, and 
${\cal M}_{\lambda \Pom}^{(c)}$ are as in 
(\ref{4.22}), (\ref{4.23}), and (\ref{4.19}), respectively.
From these we obtain the amplitudes
${\cal M}_{\lambda f_{2 \Reg}}^{(a)}$,
${\cal M}_{\lambda f_{2 \Reg}}^{(b)}$, and
${\cal M}_{\lambda f_{2 \Reg}}^{(c)}$
with the replacements (\ref{4.27a}).
For $\rho_{\Reg}$ exchange we get
\begin{eqnarray}
{\cal M}_{\lambda \rho_{\Reg}}^{(a)} &=&
e {\cal M}_{\rho_{\Reg}}^{(0,a)} 
\frac{(2p_{a} - k)_{\lambda}}{2 (p_{a} \cdot k) - k^{2}}\,,
\nonumber\\
{\cal M}_{\rho_{\Reg}}^{(0,a)} &=& 
{\cal F}_{\rho_{\Reg}}(s,t_{2})
\Big[ 1+ \left(1 - \alpha_{\rho_{\Reg}}(t_2)\right)\, 
\frac{2(p_{a} + p_{b},k) - k^{2}}{s}\,
g_{\rho_{\Reg}}(\varkappa, t_{2}) \Big]
\nonumber\\
&&
\times (p_{a}+p_{1}'-k,p_{b}+p_{2}')\,,
\label{4.30b}\\
{\cal M}_{\lambda \rho_{\Reg}}^{(b)} &=&
-e  
\frac{(2p_{1}' + k)_{\lambda}}{2 (p_{1}' \cdot k) + k^{2}}
{\cal M}_{\rho_{\Reg}}^{(0,b)}\,,
\nonumber\\
{\cal M}_{\rho_{\Reg}}^{(0,b)} &=& 
{\cal F}_{\rho_{\Reg}}(s,t_{2}) \,(p_{a}+p_{1}'+k,p_{b}+p_{2}')\,,
\label{4.30c}\\
{\cal M}_{\lambda \rho_{\Reg}}^{(c)} &=&
e
{\cal F}_{\rho_{\Reg}}(s,t_{2}) 
\Big\lbrace 2(p_{b} + p_{2}')_{\lambda} -
\frac{(2p_{a} + 2p_{b}-k)_{\lambda} }{s}
\left(1 - \alpha_{\rho_{\Reg}}(t_2)\right)
g_{\rho_{\Reg}}(\varkappa, t_{2})
\nonumber\\
&&
\times (p_{a}+p_{1}'-k,p_{b}+p_{2}') \Big\rbrace\,.
\label{4.30d}
\end{eqnarray}
Here $\varkappa$ and $g_{\Pom}(\varkappa, t)$
are defined in (\ref{4.16}) and (\ref{4.17}), respectively,
$g_{f_{2 \Reg}}(\varkappa, t)$ is defined analogously
\begin{eqnarray}
g_{f_{2 \Reg}}(\varkappa, t) = 
\frac{1}{\left(2 - \alpha_{f_{2 \Reg}}(t)\right)\varkappa}
\big[ (1-\varkappa)^{\alpha_{f_{2 \Reg}}(t)-2}-1
\big]\,,
\label{4.32e}
\end{eqnarray}
and $g_{\rho_{\Reg}}(\varkappa, t)$ is defined as
\begin{eqnarray}
g_{\rho_{\Reg}}(\varkappa, t) &=&
\frac{1}{\left(1 - \alpha_{\rho_{\Reg}}(t)\right)\varkappa}
\big[ (1-\varkappa)^{\alpha_{\rho_{\Reg}}(t)-1}-1
\big]\, \nonumber \\
&=& 1 + \frac{\varkappa}{2!}\left(2-\alpha_{\rho_{\Reg}}(t)\right) + 
\frac{\varkappa^{2}}{3!}\left(2-\alpha_{\rho_{\Reg}}(t)\right)
\left(3-\alpha_{\rho_{\Reg}}(t)\right) + \ldots \,.
\label{4.32f}
\end{eqnarray}

We emphasize that ${\cal M}_{\lambda \Pom}^{(c)}$,
${\cal M}_{\lambda f_{2 \Reg}}^{(c)}$ and
${\cal M}_{\lambda \rho_{\Reg}}^{(c)}$ are obtained as
the simplest solution of the gauge-invariance relation
\begin{eqnarray}
&&k^{\lambda}
\left({\cal M}_{\lambda}^{(a)}+
{\cal M}_{\lambda}^{(b)}+
{\cal M}_{\lambda}^{(c)}\right) = 0\,.
\label{4.32g}
\end{eqnarray}

For the diagrams of Fig.~\ref{fig:pipigam}~(d)--(f) we find
\begin{eqnarray}
&&{\cal M}_{\lambda}^{(d)} = 
\left.-{\cal M}_{\lambda}^{(a)}\right|_{p_{a}, p_{1}' 
\leftrightarrow p_{b}, p_{2}'} \,,
\label{4.31}\\
&&{\cal M}_{\lambda}^{(e)}=
\left.-{\cal M}_{\lambda}^{(b)}\right|_{p_{a}, p_{1}' 
\leftrightarrow p_{b}, p_{2}'} \,,
\label{4.32}\\
&&{\cal M}_{\lambda}^{(f)}=
\left.-{\cal M}_{\lambda}^{(c)}\right|_{p_{a}, p_{1}' 
\leftrightarrow p_{b}, p_{2}'}\,.
\label{4.33}
\end{eqnarray}
Note that $(p_{a}, p_{1}') \leftrightarrow (p_{b}, p_{2}')$
implies $t_{1} \leftrightarrow t_{2}$; see (\ref{4.10}).
We have also here
\begin{eqnarray}
k^{\lambda}
\left({\cal M}_{\lambda}^{(d)}+
{\cal M}_{\lambda}^{(e)}+
{\cal M}_{\lambda}^{(f)}\right) = 0\,.
\label{4.34}
\end{eqnarray}
For the amplitudes (\ref{4.29}) and (\ref{4.30}) we get finally
\begin{eqnarray}
{\cal M}_{\lambda}^{(\pi^{-} \pi^{+} \to \pi^{-} \pi^{+} \gamma)}
&=&{\cal M}_{\lambda}^{(a)}
+{\cal M}_{\lambda}^{(b)}
+{\cal M}_{\lambda}^{(c)}
+{\cal M}_{\lambda}^{(d)}
+{\cal M}_{\lambda}^{(e)}
+{\cal M}_{\lambda}^{(f)}\,,
\label{4.35} \\
{\cal M}_{\lambda}^{(\pi^{+} \pi^{+} \to \pi^{+} \pi^{+} \gamma)}
&=& -\left(
\hat{\cal M}_{\lambda}^{(a)}
+\hat{\cal M}_{\lambda}^{(b)}
+\hat{\cal M}_{\lambda}^{(c)} \right)
+\hat{\cal M}_{\lambda}^{(d)}
+\hat{\cal M}_{\lambda}^{(e)}
+\hat{\cal M}_{\lambda}^{(f)} \nonumber\\
&&+ (p_{1}' \leftrightarrow p_{2}')\,.
\label{4.36}
\end{eqnarray}
Here we define
\begin{eqnarray}
\hat{\cal M}_{\lambda}^{(a)} =
{\cal M}_{\lambda \Pom}^{(a)}+
{\cal M}_{\lambda f_{2 \Reg}}^{(a)}-
{\cal M}_{\lambda \rho_{\Reg}}^{(a)}
\label{4.346a}
\end{eqnarray}
and similarly for 
$\hat{\cal M}_{\lambda}^{(b)}, \ldots, 
\hat{\cal M}_{\lambda}^{(f)}$.

The inclusive cross section for the real-photon yield of
the reaction (\ref{4.25}) is as follows
\begin{eqnarray}
&&d\sigma({\pi^{-}\pi^{+} \to \pi^{-}\pi^{+} \gamma(k)}) =
\frac{1}{2\sqrt{s(s-4 m_{\pi}^{2})}}\,
\frac{d^{3}k}{(2 \pi)^{3} \,2 k^{0}}
\int \frac{d^{3}p_{1}'}{(2 \pi)^{3} \,2 p_{1}'^{0}}
\frac{d^{3}p_{2}'}{(2 \pi)^{3} \,2 p_{2}'^{0}} 
\nonumber \\
&&\qquad \times (2 \pi)^{4} \delta^{(4)}(p_{1}'+p_{2}'+k-p_{a}-p_{b})
{\cal M}_{\lambda}^{(\pi^{-}\pi^{+} \to \pi^{-}\pi^{+} \gamma)}
\big( {\cal M}_{\rho}^{(\pi^{-}\pi^{+} \to \pi^{-}\pi^{+} \gamma)} \big)^{*} (-g^{\lambda \rho}) \nonumber\\
\label{4.37}
\end{eqnarray}
and similarly for $\pi^{+} \pi^{+} \to \pi^{+} \pi^{+} \gamma$,
including a statistic factor 1/2.

In the following we shall compare our ``exact'' model results,
which we shall call ``standard'' results,
for (\ref{4.35}) and (\ref{4.36}), 
using (\ref{4.22}), (\ref{4.23}), (\ref{4.19}),
(\ref{4.30a})--(\ref{4.30d}),
and (\ref{4.31})--(\ref{4.33}), 
to various soft-photon approximations (SPAs).
Below we list the explicit expressions for photon
emission in $\pi^{-} \pi^{+}$ scattering.

\begin{enumerate}
\item[\underline{SPA1}:] 

Here we keep only the pole terms $\propto \omega^{-1}$ for
${\cal M}_{\lambda}^{(a)} \cdots {\cal M}_{\lambda}^{(f)}$
in (\ref{4.35}).
From (\ref{4.19}), (\ref{4.22}), (\ref{4.23}), (\ref{4.28}),
(\ref{4.30a})--(\ref{4.30d}),
and (\ref{4.31})--(\ref{4.33}) we see that this amounts 
to the following replacements, using $k^{2} = 0$,
and $p_{1}' \to p_{1}$, $p_{2}' \to p_{2}$:
\begin{eqnarray}
&&{\cal M}_{\lambda}^{(a)}
\to e {\cal M}^{(0)\pi^{-}\pi^{+}}(s,t) \frac{p_{a \lambda}}{(p_{a} \cdot k)}\,,
\nonumber \\
&&{\cal M}_{\lambda}^{(b)}
\to -e {\cal M}^{(0)\pi^{-}\pi^{+}}(s,t) \frac{p_{1 \lambda}}{(p_{1} \cdot k)}\,,
\nonumber \\
&&{\cal M}_{\lambda}^{(c)}
\to 0\,,
\nonumber \\
&&{\cal M}_{\lambda}^{(d)}
\to -e {\cal M}^{(0)\pi^{-}\pi^{+}}(s,t) \frac{p_{b \lambda}}{(p_{b} \cdot k)}\,,
\nonumber \\
&&{\cal M}_{\lambda}^{(e)}
\to e {\cal M}^{(0)\pi^{-}\pi^{+}}(s,t) \frac{p_{2 \lambda}}{(p_{2} \cdot k)}\,,
\nonumber \\
&&{\cal M}_{\lambda}^{(f)}
\to 0\,.
\label{4.38}
\end{eqnarray}
From (\ref{4.35}) and (\ref{4.38}) we get then
\begin{eqnarray}
&&{\cal M}_{\lambda}^{(\pi^{-}\pi^{+} \to \pi^{-}\pi^{+} \gamma)}
\to {\cal M}_{\lambda, \;{\rm SPA1}}^{(\pi^{-}\pi^{+} \to \pi^{-}\pi^{+} \gamma)}
\nonumber \\
&& \qquad \qquad = e{\cal M}^{(0)\pi^{-}\pi^{+}}(s,t)
\Big[ 
\frac{p_{a \lambda}}{(p_{a} \cdot k)}
-\frac{p_{1 \lambda}}{(p_{1} \cdot k)}
-\frac{p_{b \lambda}}{(p_{b} \cdot k)}
+\frac{p_{2 \lambda}}{(p_{2} \cdot k)} \Big].
\label{4.39}
\end{eqnarray}
Inserting this in (\ref{4.37}) we get the following SPA1 result 
for the inclusive photon cross section where,
for consistency, we neglect the photon momentum $k$
in the energy-momentum conserving $\delta^{(4)}(.)$ function:
\begin{eqnarray}
&&d\sigma({\pi^{-}\pi^{+} \to \pi^{-}\pi^{+} \gamma(k)})_{\,\rm SPA1} =
\frac{d^{3}k}{(2 \pi)^{3} \,2 k^{0}}
\int d^{3}p_{1} \,d^{3}p_{2}\,e^{2}\nonumber \\
&&\qquad \times 
\Big[ 
\frac{p_{a \lambda}}{(p_{a} \cdot k)}
-\frac{p_{1 \lambda}}{(p_{1} \cdot k)}
-\frac{p_{b \lambda}}{(p_{b} \cdot k)}
+\frac{p_{2 \lambda}}{(p_{2} \cdot k)} \Big]\nonumber \\
&&\qquad \times 
\Big[ 
\frac{p_{a \rho}}{(p_{a} \cdot k)}
-\frac{p_{1 \rho}}{(p_{1} \cdot k)}
-\frac{p_{b \rho}}{(p_{b} \cdot k)}
+\frac{p_{2 \rho}}{(p_{2} \cdot k)} \Big] (-g^{\lambda \rho})
\nonumber \\
&&\qquad \times \frac{d\sigma(\pi^{-}\pi^{+} \to \pi^{-}\pi^{+})}{d^{3}p_{1}d^{3}p_{2}}\,,
\label{4.40}
\end{eqnarray}
where
\begin{eqnarray}
\frac{d\sigma(\pi^{-}\pi^{+} \to \pi^{-}\pi^{+})}{d^{3}p_{1}d^{3}p_{2}} &=&
\frac{1}{2\sqrt{s(s-4 m_{\pi}^{2})}}\,
\frac{1}{(2 \pi)^{3} \,2 p_{1}^{0}\,(2 \pi)^{3} \,2 p_{2}^{0}} \nonumber\\
&&\times (2 \pi)^{4} \delta^{(4)}(p_{1}+p_{2}-p_{a}-p_{b})\,
|{\cal M}^{(0)\pi^{-}\pi^{+}}(s,t)|^{2}\,.\qquad
\label{4.41}
\end{eqnarray}
In (\ref{4.40}), (\ref{4.41}) we have a frequently used SPA.
One takes the distribution of the particles without radiation
[see (\ref{4.41})] and multiplies with the square of the emission
factor in the square brackets in (\ref{4.39}).

\item[\underline{SPA2}:] 
Here we take into account that the squared momentum
transfer is $t_{2}$ for the diagrams of 
Fig.~\ref{fig:pipigam}~(a)--(c)
and $t_{1}$ for those of Fig.~\ref{fig:pipigam}~(d)--(f), where
$t_{1,2}$ are defined in (\ref{4.10}).
We make in (\ref{4.37}) the replacement:
\begin{eqnarray}
&&{\cal M}_{\lambda}^{(\pi^{-}\pi^{+} \to \pi^{-}\pi^{+} \gamma)}
\to {\cal M}_{\lambda, \;{\rm SPA2}}^{(\pi^{-}\pi^{+} \to \pi^{-}\pi^{+} \gamma)}
\nonumber \\
&& \qquad \qquad= e{\cal M}^{(0)\pi^{-}\pi^{+}}(s,t_{2})
\Big[ 
\frac{p_{a \lambda}}{(p_{a} \cdot k)}
-\frac{p_{1 \lambda}'}{(p_{1}' \cdot k)} \Big]
\nonumber \\
&& \qquad \qquad \quad
+ e{\cal M}^{(0)\pi^{-}\pi^{+}}(s,t_{1})
\Big[
-\frac{p_{b \lambda}}{(p_{b} \cdot k)}
+\frac{p_{2 \lambda}'}{(p_{2}' \cdot k)} \Big].\nonumber\\
\label{4.42}
\end{eqnarray}
In the calculation of the photon distribution we keep
the correct energy-momentum conserving $\delta^{(4)}(.)$ function
in (\ref{4.37}).

\item[\underline{SPA3}:] 
In our third example we make in (\ref{4.37}) the replacement
\begin{eqnarray}
&&{\cal M}_{\lambda}^{(\pi^{-}\pi^{+} \to \pi^{-}\pi^{+} \gamma)}
\to {\cal M}_{\lambda, \;{\rm SPA3}}^{(\pi^{-}\pi^{+} \to \pi^{-}\pi^{+} \gamma)}
\nonumber \\
&& \qquad \qquad= e{\cal M}^{(0)\pi^{-}\pi^{+}}(s,t')
\Big[ 
\frac{p_{a \lambda}}{(p_{a} \cdot k)}
-\frac{p_{1 \lambda}'}{(p_{1}' \cdot k)}
-\frac{p_{b \lambda}}{(p_{b} \cdot k)}
+\frac{p_{2 \lambda}'}{(p_{2}' \cdot k)} \Big],
\label{4.43}
\end{eqnarray}
where we choose
\begin{eqnarray}
t' = {\rm min}(t_{1},t_{2})\,.
\label{4.44}
\end{eqnarray}
Also here we keep the correct energy-momentum conserving $\delta^{(4)}(.)$ function in the evaluation of (\ref{4.37}).
\end{enumerate}
 
We shall also consider approximations which we shall call
``improved SPA1'' and ``improved SPA2'', respectively.
For this we consider Fig.~\ref{fig:pipigam}.
In the diagrams (a) and (d) the squared c.m. energy
of the off-shell $\pi \pi \to \pi \pi$ amplitude is
$s_{a} = (p_{a} + p_{b} - k)^{2}$,
in the diagrams (b) and (e) it is
$s = (p_{a} + p_{b})^{2}$; see (\ref{4.23b_sa}), (\ref{4.23b_sb}).
For real photons, $k^{2} = 0$, and working in the overall
c.m. system we have
\begin{eqnarray}
s_{a} = s - 2 \omega \sqrt{s} \,.
\label{4.72}
\end{eqnarray}
Now we take, as a compromise, the average value $\tilde{s}$
of $s_{a}$ and $s$,
\begin{eqnarray}
\tilde{s} = s - \omega \sqrt{s} \,,
\label{4.73}
\end{eqnarray}
as squared c.m. energy of the $\pi \pi \to \pi \pi$ amplitudes.
With the amplitudes ${\cal M}^{(0)\pi^{-}\pi^{+}}(\tilde{s},t)$
and ${\cal M}^{(0)\pi^{-}\pi^{+}}(\tilde{s},t_{1,2})$
in (\ref{4.39}) and (\ref{4.42}), respectively, 
we get what we call the improved SPA1 and SPA2 results.
The prescription to replace $s$ by $\tilde{s}$ (\ref{4.73})
has been advocated by Linnyk \textit{et al.} \cite{Linnyk:2015tha,Linnyk:2015rco}.

\section{Results}
\label{sec:5}

Below we show our results 
for elastic $\pi \pi \to \pi\pi$ scattering 
(subsection~\ref{sec:5A}) 
and results for the $\pi \pi \to \pi \pi \gamma$ reaction (subsection~\ref{sec:5B}).

\subsection{Comparison with the total and elastic $\pi \pi$ cross sections}
\label{sec:5A}

Here we compare our model results with 
the $\pi^{-} \pi^{+}$
and $\pi^{\pm}\pi^{\pm}$ total 
and total elastic cross section data.

\begin{figure}[!ht]
\includegraphics[width=0.45\textwidth]{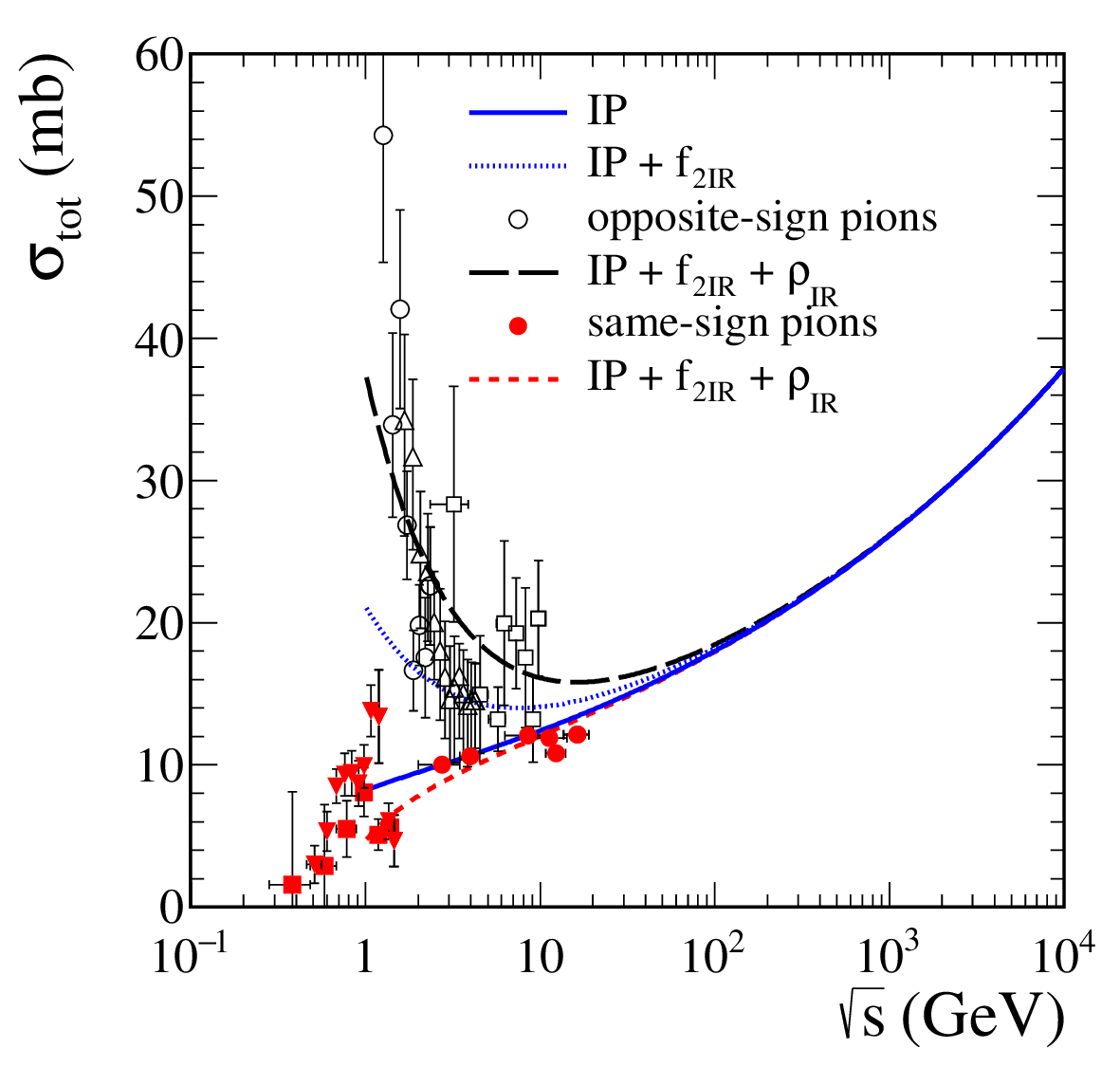}
\includegraphics[width=0.45\textwidth]{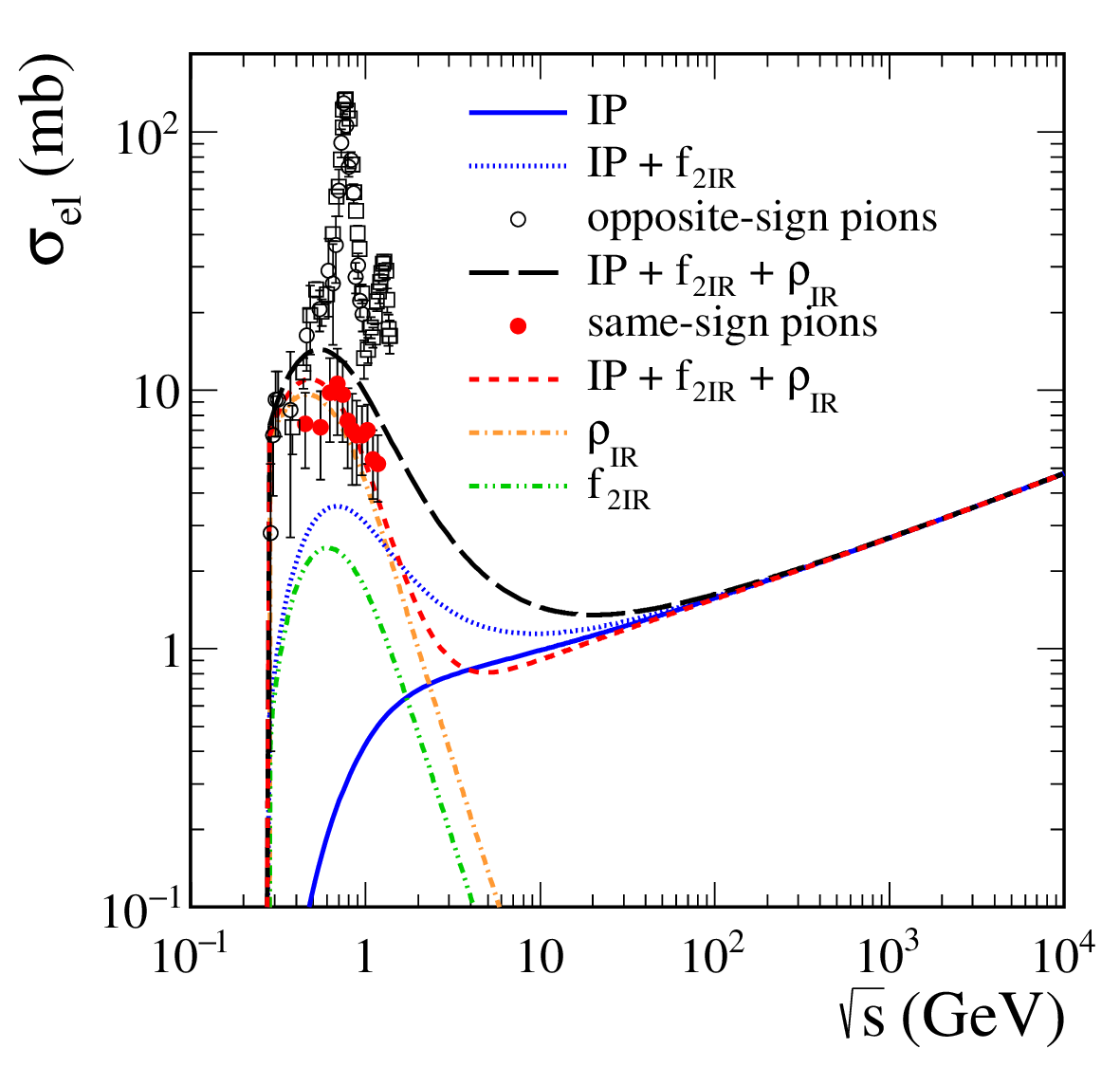}
\caption{\label{fig:XS}
\small
Results
for $\pi^{-} \pi^{+}$
and $\pi^{+}\pi^{+}$ or $\pi^{-}\pi^{-}$
total (left panel) and total elastic (right panel)
cross sections as a function of $\sqrt{s}$
together with the experimental data.
The single pomeron exchange is given by the blue solid line, 
the pomeron plus $f_{2 \Reg}$ reggeon exchanges by the blue dotted line, 
the complete result ($\Pom + f_{2 \Reg} + \rho_{\Reg}$)
for the opposite-sign pions and the same-sign pions
is given
by the black long-dashed line and the red short-dashed line,
respectively.
In the right panel we show also the $f_{2 \Reg}$ reggeon
and the $\rho_{\Reg}$ reggeon terms separately.}
\end{figure}

First we briefly review the experimental results 
for the $\pi \pi$ total and elastic cross sections.
There are no direct measurements of total and elastic 
$\pi \pi$ cross sections at present.
However, indirect data at low and intermediate $\sqrt{s}$,
the pion-pion center-of-mass energy, have been extracted
from reactions like $\pi^- p \to \pi^+ \pi^- n$, $\pi^- \pi^- \Delta^{++}$
\cite{Biswas:1967mpl,Cohen:1973yx,Losty:1973et,Robertson:1973tk} 
and $\pi^{\pm} p \to \Delta^{++} X$ and $\pi^{\pm} n \to p X$ 
\cite{Hanlon:1976ct,Abramowicz:1979ca}.
They are compared with our predictions in Fig.~\ref{fig:XS}.
In the left panel the experimental data are from
\cite{Biswas:1967mpl,Cohen:1973yx,Losty:1973et,Robertson:1973tk,Hanlon:1976ct,Hoogland:1977kt,Abramowicz:1979ca}\footnote{There are also
the data of the total $\pi^{-} \pi^{-}$ cross section
from \cite{Biswas:1967mpl,Robertson:1973tk} 
(see, e.g., Fig.~3 of \cite{Pelaez:2003ky}
or Fig.~2 of \cite{Caprini:2011ky}).
It was stated in \cite{Pelaez:2003ky} that these results
are not consistent with other data at lower energies
probably due to incorrect treatment of final state interactions.
The uncertainties of these data are therefore very large and hence
we do not show them in Fig.~\ref{fig:XS}.}
while in the right panel 
from
\cite{Srinivasan:1975tj,Alekseeva:1982uy}.

We present for the scattering of
$\pi^{-} \pi^{+}$ (opposite-sign pions)
and $\pi^{\pm}\pi^{\pm}$ (same-sign pions)
the total (left panel) and total elastic (right panel)
cross sections versus $\sqrt{s}$.
The results for the single pomeron exchange ($\Pom$),
for the pomeron and $f_{2 \Reg}$ reggeon exchanges ($\Pom + f_{2 \Reg}$),
and the complete results ($\Pom + f_{2 \Reg} + \rho_{\Reg}$) are shown.
The corresponding theoretical 
expressions are given in (\ref{4.27a})--(\ref{4.28b}).
According to our model we treat the $\rho_{\Reg}$ reggeon 
as effective vector exchange
and the pomeron and $f_{2 \Reg}$ reggeon as effective tensor exchanges.
Thus, in the Regge parametrization of the $\pi^{\pm}\pi^{\pm}$ cross section, 
the $\rho_{\Reg}$ contributes with a sign opposite to $\Pom$ and $f_{2 \Reg}$.

We find good agreement with the experimental data
taking into account the default values 
from \cite{Ewerz:2013kda} for the parameters
of the propagators and vertices.
One has to keep in mind that for the subleading exchanges
the errors of the coupling constants are quite large,
in particular for the coupling $g_{\rho_{\Reg} \pi \pi}$,
as was discussed in Sec.~7.1 of \cite{Ewerz:2013kda}.
In addition one also has to keep in mind 
that there should be a smooth transition from reggeon
to particle exchanges when going to very low energies.
Note that the same-sign-pions channels 
do not contain $s$ channel resonances
in contrast to the opposite-sign-pions channel.
Thus, our theoretical results, which include only
$t$-channel exchanges, are in better agreement with
the experimental data for 
$\sigma^{\pi^{\pm} \pi^{\pm}}$ than for
$\sigma^{\pi^{-} \pi^{+}}$.
Moreover, such effects as absorption corrections 
and multiple soft and hard exchanges,
discussed in \cite{Szczurek:2001py}, 
were not included in our calculation.
Clearly, all these topics deserve careful analyses,
but this goes beyond the scope of the present paper.

There are also the data of $\pi^{\pm} \pi^{-}$ total cross sections
from the analysis performed in~\cite{Zakharov_data}.
In that work, a triple reggeon model with absorption
was used to extract 
$\sigma_{\rm tot}^{\pi^{\pm} \pi^{-}}$
from the $\pi^{\pm} p \to \Delta^{++} X$ and $\pi^{\pm} n \to p X$ processes.
The authors of \cite{Zakharov_data} found that the inclusion
of absorptive corrections in these two reactions 
decreases the results by about 10$\,\%$ to 15$\,\%$.
The uncertainty of these results is large
and therefore we do not show these data 
in Fig.~\ref{fig:XS} and instead we refer to
\cite{Szczurek:2001py,Pelaez:2004ab}.
In \cite{Szczurek:2001py} the effect of absorption corrections
(double-scattering effect) on the total cross section 
for $\pi \pi$ scattering as a function of $\sqrt{s}$ was discussed.
The $t$-dependence of the elastic $\pi \pi$ cross sections
was also discussed there.
The authors of \cite{Szczurek:2001py} found that 
the absorption is much weaker for the same-sign pions 
than for the opposite-sign pions; see, e.g.,
Figs.~5, 9 and Table~2 of \cite{Szczurek:2001py}.

The total $\pi^+\pi^-$ and $\pi^\pm \pi^\pm$ cross sections 
including subleading reggeon exchanges
were also discussed
in~\cite{Pelaez:2003ky,Halzen:2011xc,Caprini:2011ky}.
There is the question of the reliability of the Regge model
down to low energies and 
whether in the region of low $\sqrt{s}$ but not low $|t|$ 
the Regge parametrization can be properly applied.
On general grounds, one expects Regge theory 
to work when $s \gg |t|$, $s_{0}$ [see (\ref{4.23a})] 
and $|t| \lesssim 1$~GeV$^{2}$
and, in fact, the Regge parametrization
for $\pi \pi$ becomes unreliable at large $|t|$.
The interested reader may consult Refs.~\cite{Pelaez:2003ky,Caprini:2011ky}
for the detailed discussion of this and other related issues.

In the next subsection we shall discuss soft-photon emission in
$\pi \pi$ scattering for c.m. energies $\sqrt{s} = 10$~GeV and 100~GeV.
We see from Fig.~\ref{fig:XS} that at $\sqrt{s} \gtrsim 100$~GeV
the $\pi \pi$ cross sections are completely dominated 
by the pomeron-exchange contribution.
At least, this is the result of our model.
Therefore, in Sec.~\ref{sec:5B} we shall take into account
only the pomeron-exchange term for the reactions
$\pi \pi \to \pi \pi \gamma$ at $\sqrt{s} \geqslant 100$~GeV.
At $\sqrt{s} \simeq 10$~GeV we will show results
including the pomeron exchange alone and 
in addition
the $\rho_{\Reg}$ and $f_{2 \Reg}$ reggeon exchanges.
As we will show below in Fig.~\ref{fig:regge}, 
the secondary reggeon exchanges play a significant role there.

\subsection{Comparison of our ``exact'' model or ``standard'' results for the $\pi \pi \to \pi \pi \gamma$ reactions with various soft-photon approximations}
\label{sec:5B}

\begin{figure}[!ht]
(a)\includegraphics[width=0.45\textwidth]{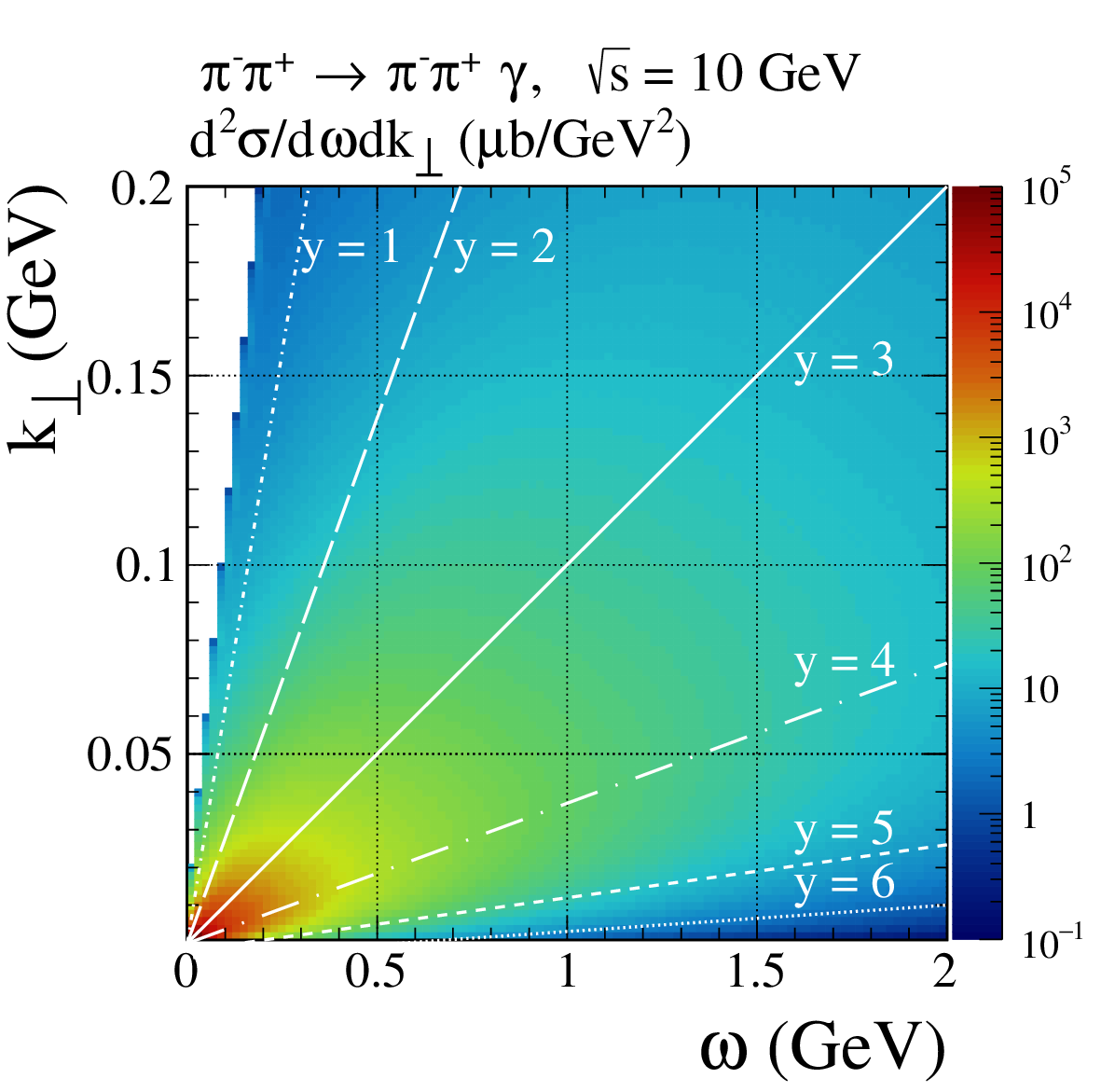}\\
(b)\includegraphics[width=0.45\textwidth]{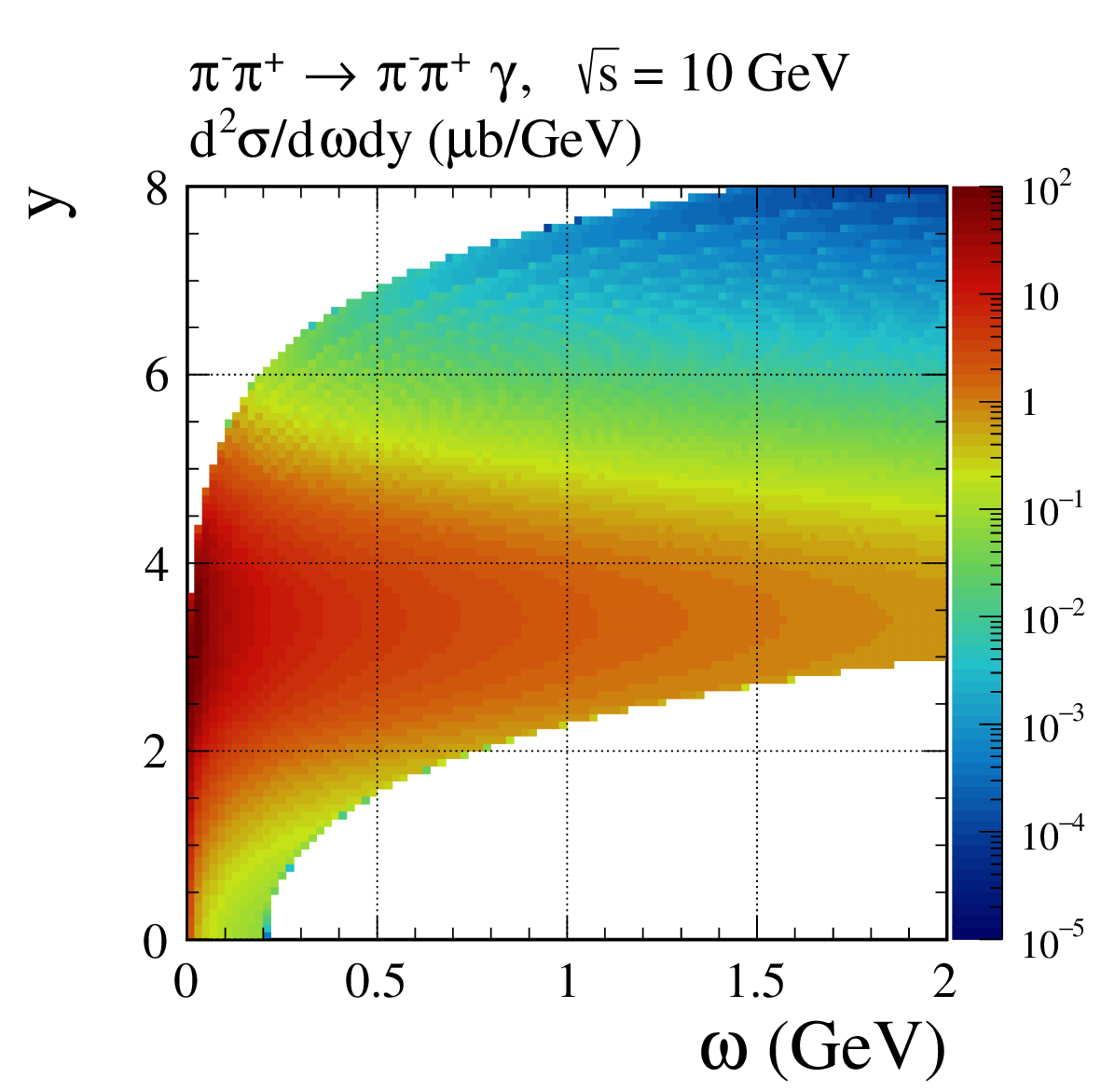}
(c)\includegraphics[width=0.45\textwidth]{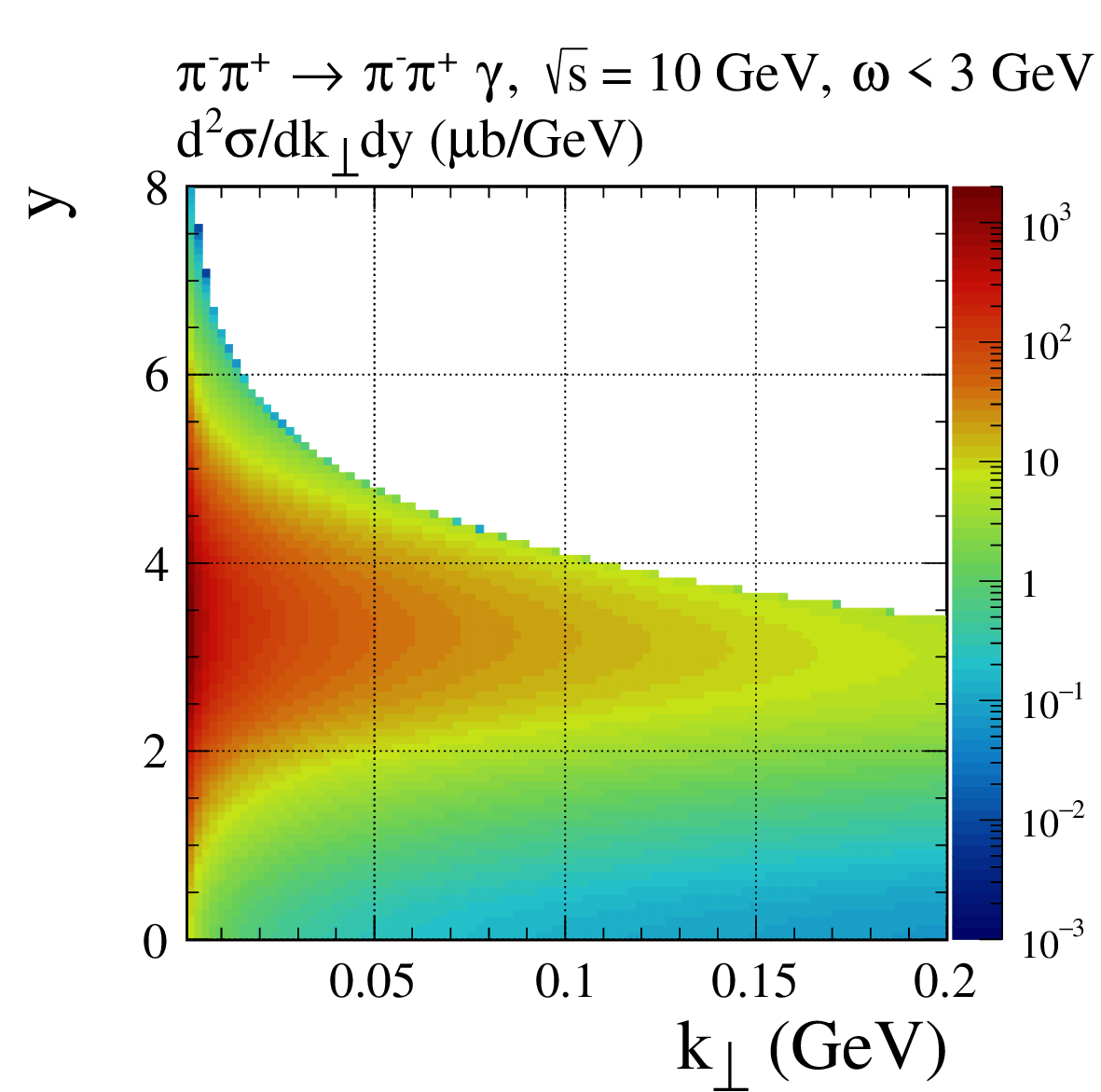}
\caption{\label{fig:2dim_exact}
\small
The two-dimensional distributions
in ($\omega$, $k_{\perp}$), ($\omega$, $\rm{y}$), and
($k_{\perp}$, $\rm{y}$),
for the $\pi^{-} \pi^{+} \to \pi^{-} \pi^{+} \gamma$ reaction
including only the pomeron exchange.
Calculations were done for $\sqrt{s} = 10$~GeV,
$0.001\; {\rm GeV} < k_{\perp} < 0.2\; {\rm GeV}$,
$\omega < 3\; {\rm GeV}$, and $|\rm{y}| < 8$.
The lines plotted in the panel~(a) correspond to the photon rapidities 
${\rm y} = 1, 2, \ldots, 6$.
In panels (b) and (c) we show the results only 
for $0 < \rm{y} < 8$ since these distributions are symmetric under $\rm{y} \to -\rm{y}$.}
\end{figure}

First, in Fig.~\ref{fig:2dim_exact},
we present the two-dimensional distributions
in \mbox{($\omega$, $k_{\perp}$)}, 
\mbox{($\omega$, $\rm{y}$)}, and
\mbox{($k_{\perp}$, $\rm{y}$)},
for the $\pi^{-} \pi^{+} \to \pi^{-} \pi^{+} \gamma$ reaction
for our ``standard'' result (\ref{4.35}), (\ref{4.37}),
including only the pomeron exchange.
Calculations were done for 
the pion-pion collision energy $\sqrt{s} = 10$~GeV.
Here, $\omega = k^{0}$ is the center-of-mass photon energy,
$k_{\perp}$ is the absolute value of the photon 
transverse momentum, and $\rm{y}$ is the rapidity of the photon.
We must remember here, that in order to stay with
all amplitudes in the Regge regime we certainly
have to require (\ref{4.23d}) which reads here,
with $k^{2} = 0$ and $s_{0} = 25\;{\rm GeV}^{2}$,
\begin{eqnarray}
\omega \leqslant \frac{1}{2 \sqrt{s}}\,
\left(s - s_{0}\right) = 3.75\;{\rm GeV}\,.
\label{5.0}
\end{eqnarray}
To be on the safe side, we shall in the following 
only show results for $\omega < 3$~GeV.
In the panel (a) we show the lines corresponding to 
the absolute value of the rapidity of the photon 
${\rm y} = 1, 2, \ldots, 6$.
Large ${\rm y}$ is near the $\omega$ axis 
and ${\rm y} = 0$ on the $k_{\perp}$ axis.
There are in all three plots also regions
that are not accessible kinematically.
From the panel (b) we see that an upper 
cut on $\omega$ is effecting the upper limit
of the allowed ${\rm y}$ range.

Now we compare our ``exact'' model or ``standard'' result
for the $\pi^{-} \pi^{+} \to \pi^{-} \pi^{+} \gamma$ reaction
to various soft-photon approximations (SPAs) 
discussed in Sec.~\ref{sec:4.2}.
We consider $\sqrt{s} = 10$~GeV
and include only the pomeron exchange.

\begin{figure}[!ht]
(a)\includegraphics[width=0.44\textwidth]{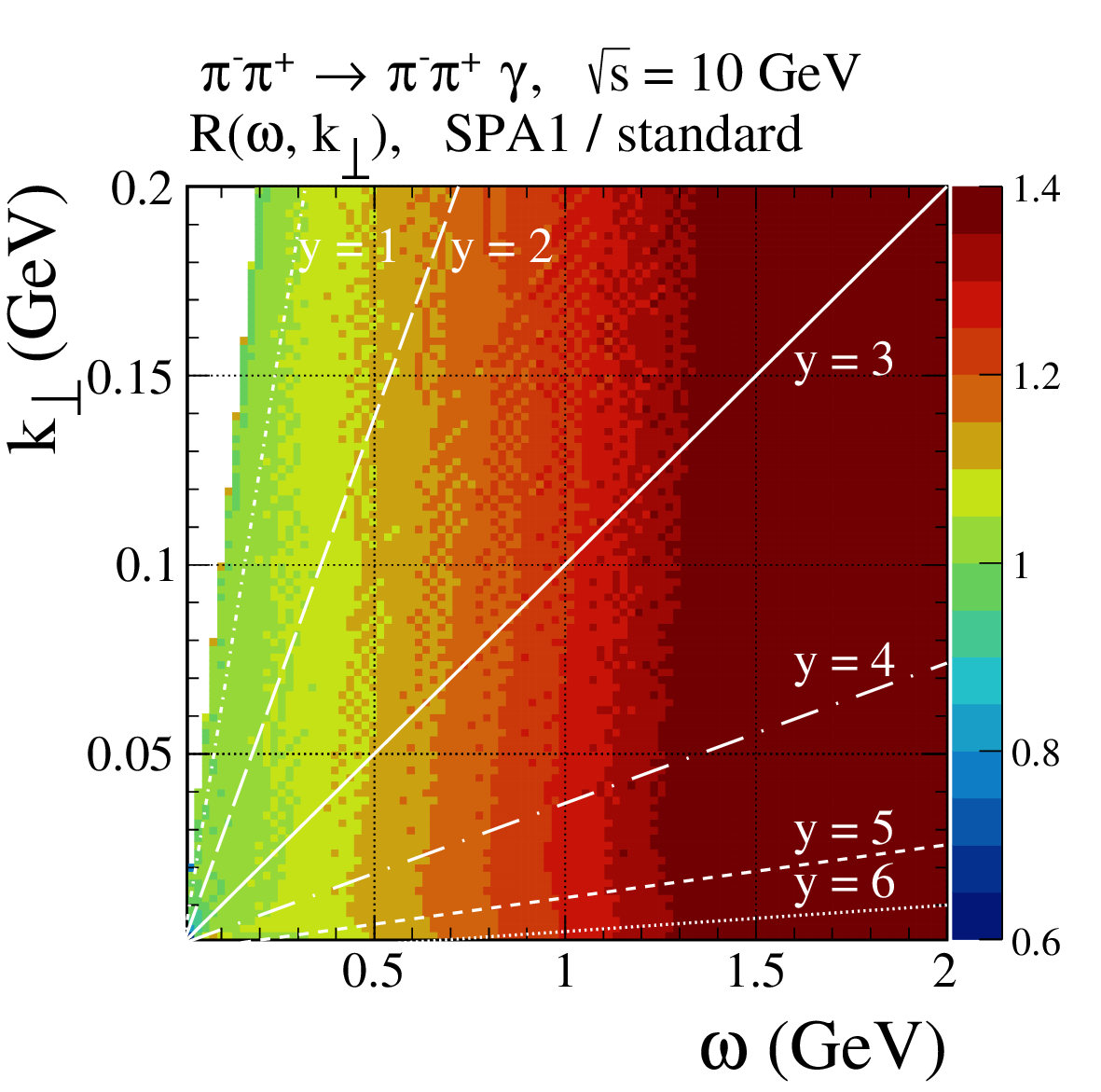}
   \includegraphics[width=0.44\textwidth]{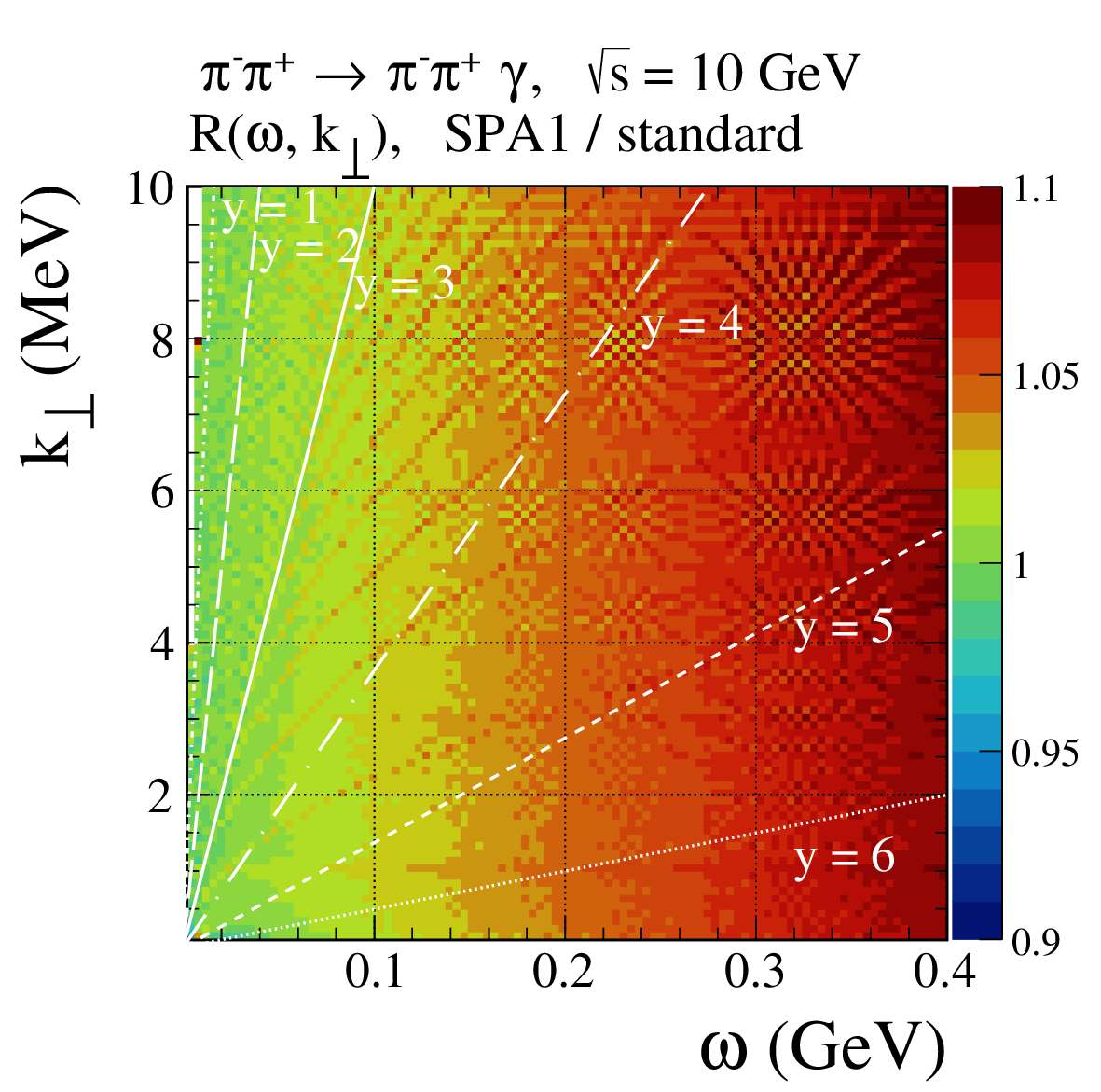}
(b)\includegraphics[width=0.44\textwidth]{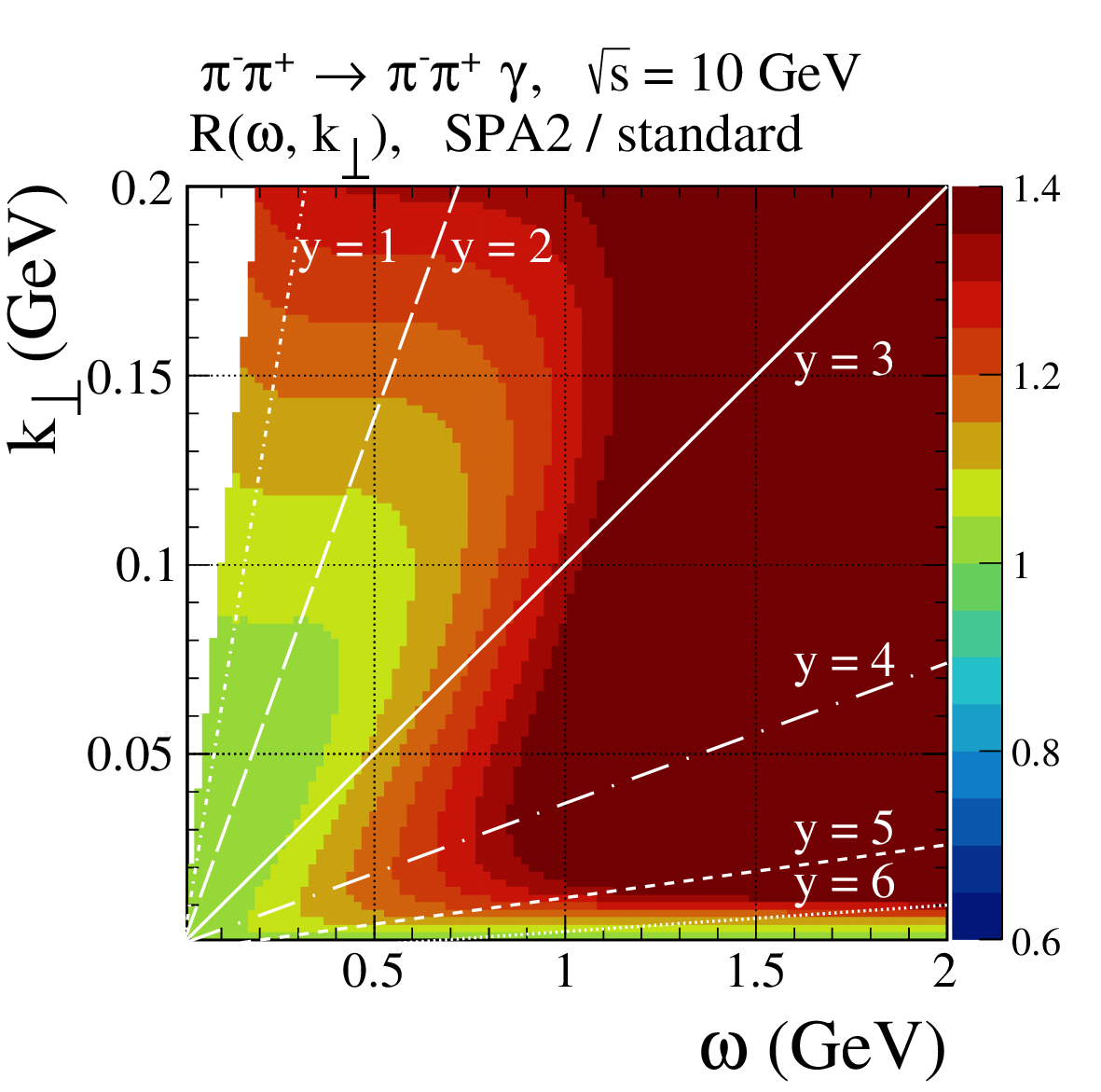}
   \includegraphics[width=0.44\textwidth]{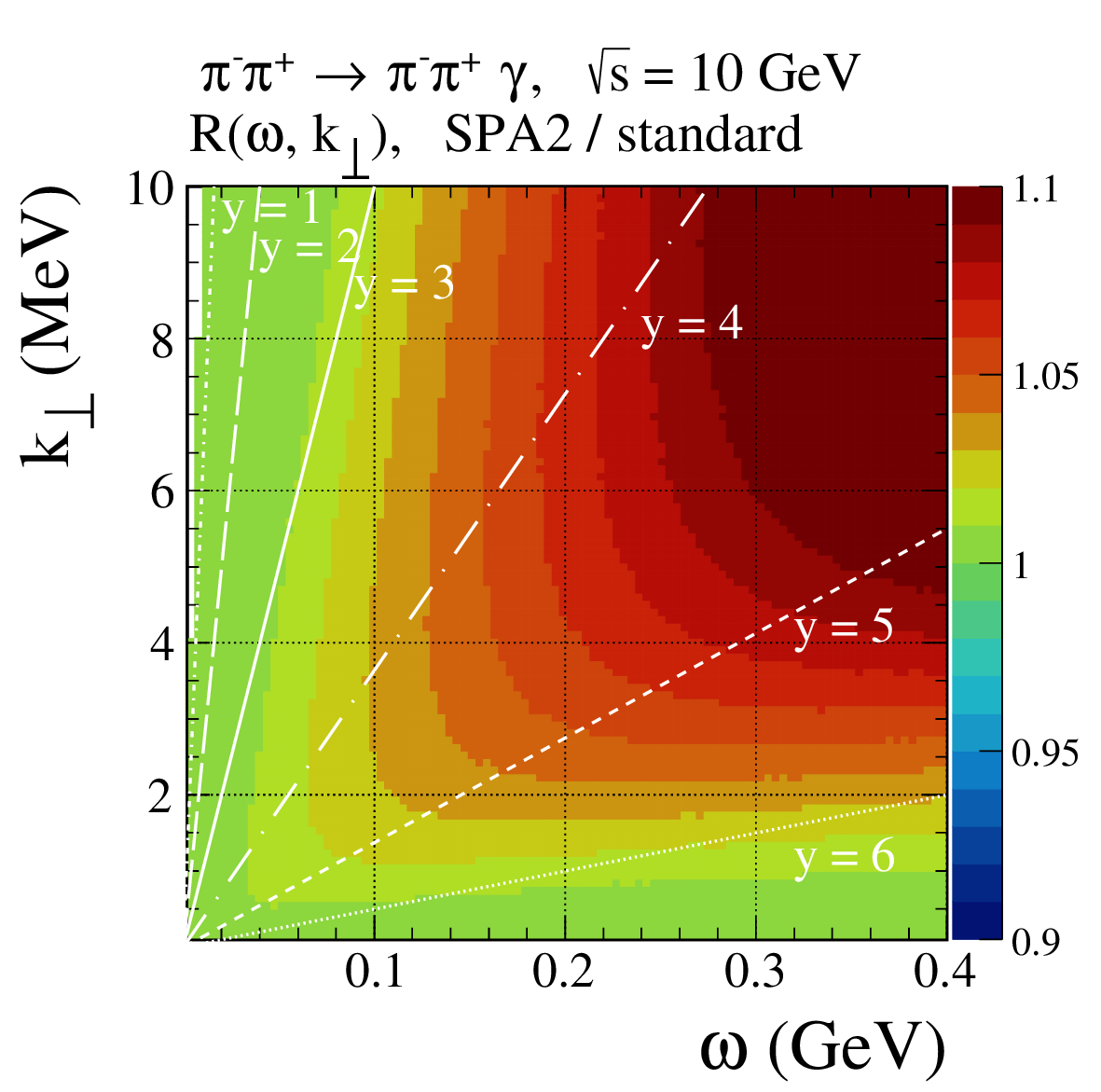}
(c)\includegraphics[width=0.44\textwidth]{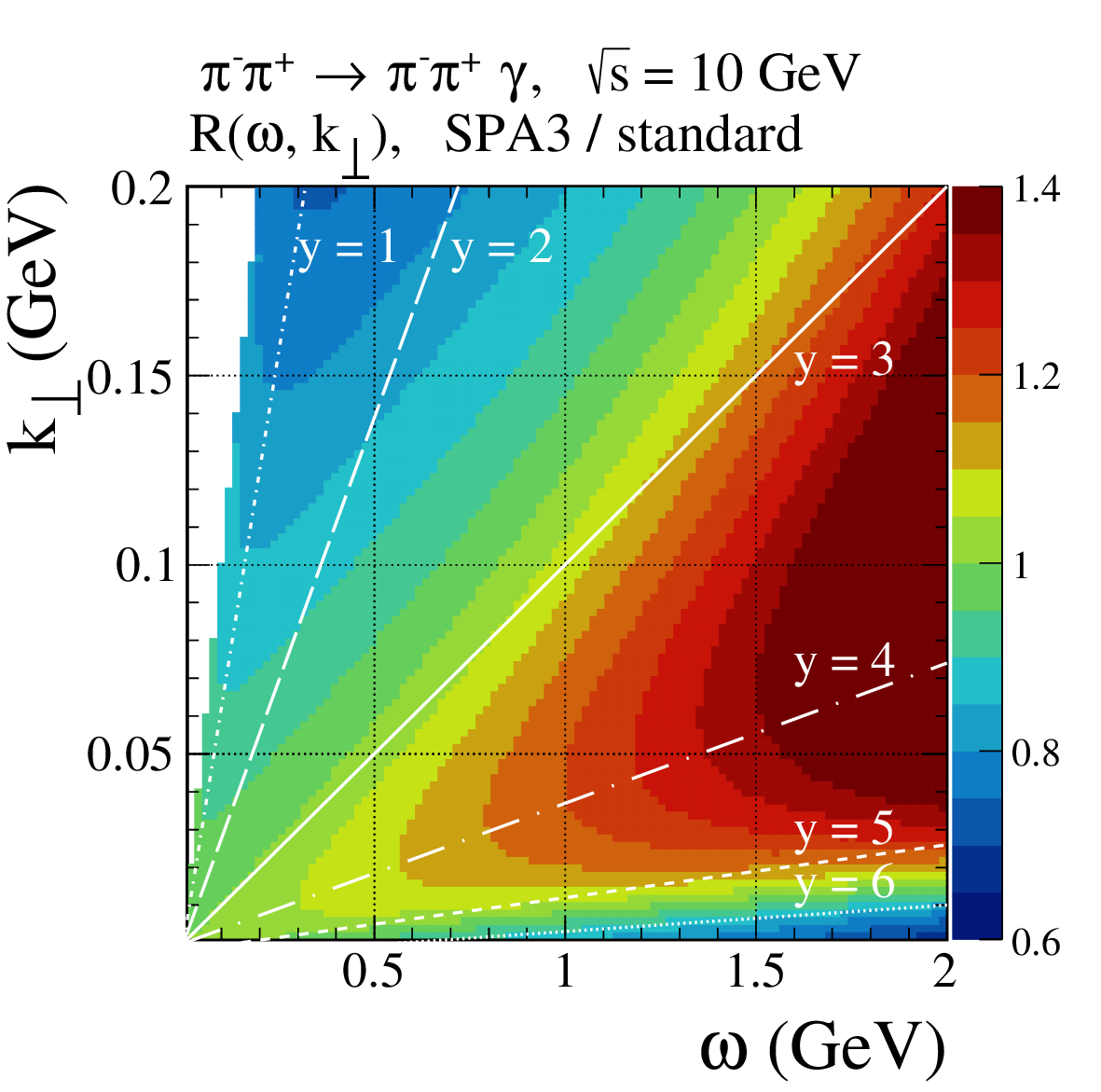}
   \includegraphics[width=0.44\textwidth]{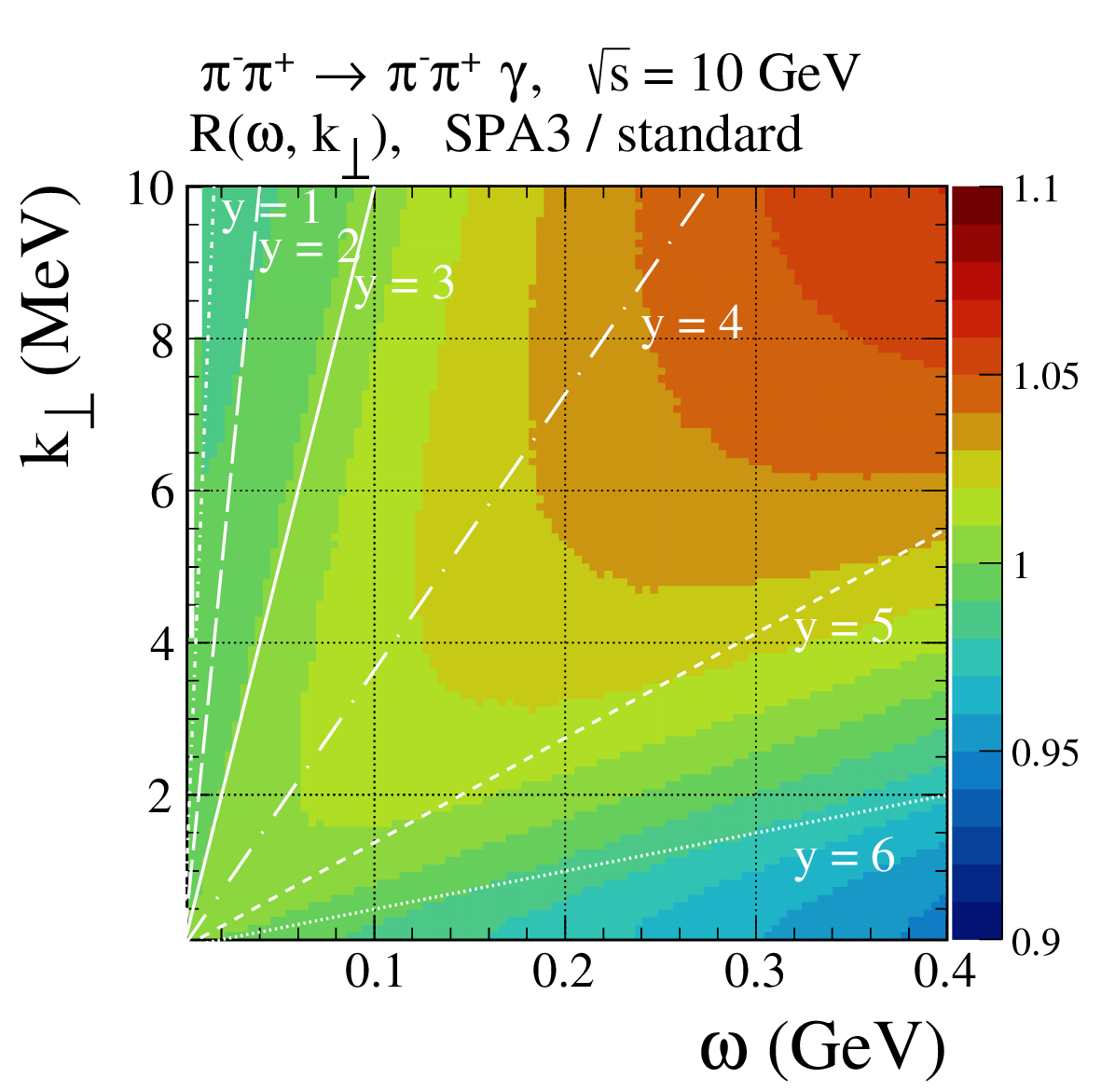}
\caption{\label{fig:ratio}
\small
The ratios ${\rm R}(\omega, k_{\perp})$ (\ref{ratio}) 
for the $\pi^{-} \pi^{+} \to \pi^{-} \pi^{+} \gamma$ reaction
for $\sqrt{s} = 10$~GeV
for the three soft-photon approximations SPA1~(\ref{4.39}),
SPA2~(\ref{4.42}), and SPA3~(\ref{4.43}).
The lines corresponding to the photon rapidities 
${\rm y} = 1, 2, \ldots, 6$ are also plotted.
The right panels show the region of small $k_{\perp}$
and small $\omega$ in greater detail.}
\end{figure}

A quantity of great interest is the ratio of the cross section calculated in one of the SPAs to the ``standard'' result.
This ratio will now be studied as a function of
$\omega = k^{0}$ and $k_{\perp}$
in the $\omega$-$k_{\perp}$ plane.
In Fig.~\ref{fig:ratio} we show, 
in two-dimensional plots,
the ratio
\begin{eqnarray}
{\rm R}(\omega, k_{\perp})=
\frac{d^{2}\sigma_{\rm SPA} / d\omega dk_{\perp}}
     {d^{2}\sigma_{\rm standard} / d\omega dk_{\perp}}\,.
\label{ratio}
\end{eqnarray}
The results for the three scenarios of the SPA amplitudes are presented.
The result on the panels~(a) corresponds to SPA1~(\ref{4.39}),
the result on the panels~(b) corresponds to SPA2~(\ref{4.42}), and
the result on the panels~(c) corresponds to SPA3~(\ref{4.43}).
We also show the lines corresponding to ${\rm y} = 1, 2, \ldots, 6$.

\begin{figure}[!ht]
\includegraphics[width=0.45\textwidth]{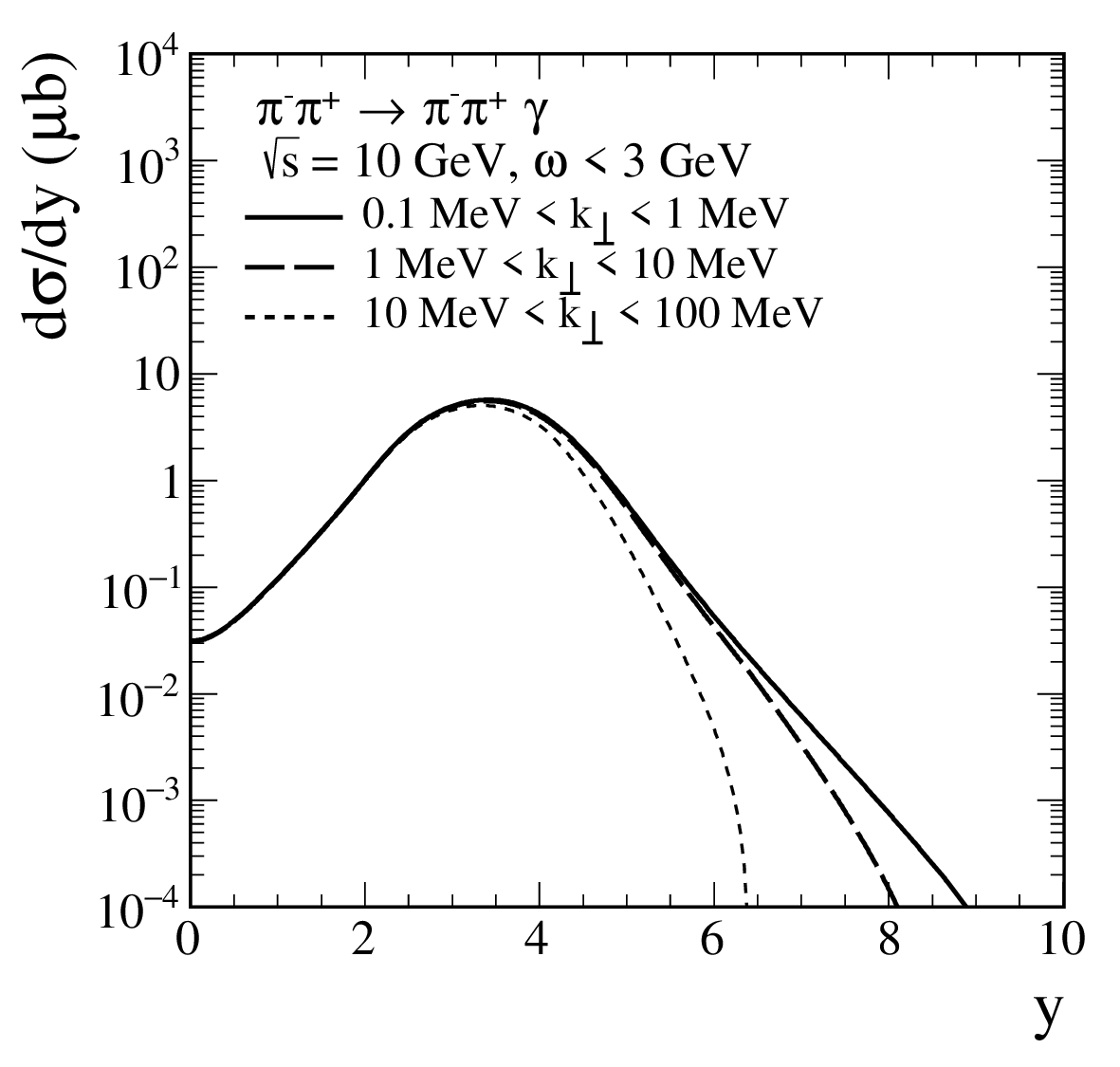}
\includegraphics[width=0.45\textwidth]{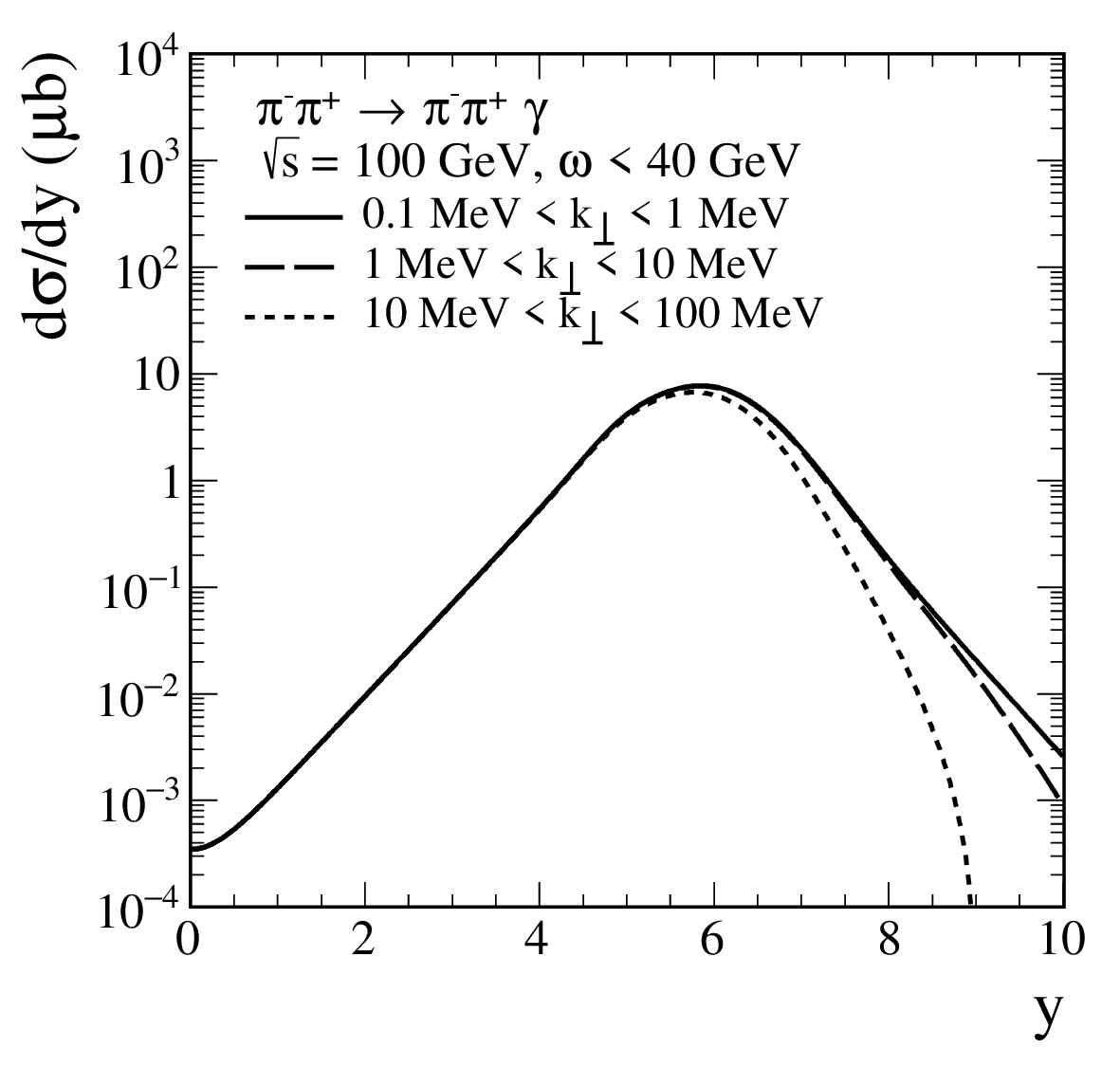}
\caption{\label{fig:y}
\small
The distributions in rapidity of the photon
in the $\pi^{-} \pi^{+} \to \pi^{-} \pi^{+} \gamma$ reaction 
calculated for $\sqrt{s} = 10$~GeV, $\omega < 3$~GeV
(left panel) and for $\sqrt{s} = 100$~GeV, $\omega < 40$~GeV (right panel) 
for different $k_{\perp}$ intervals.
Plotted are the results only for positive $\rm{y}$
since the distributions are symmetric under
$\rm{y} \to -\rm{y}$.}
\end{figure}

Now we discuss results separately for the three $k_{\perp}$ intervals
of photon transverse momenta:
$0.1\; {\rm MeV} < k_{\perp} < 1\; {\rm MeV}$,
$1\; {\rm MeV} < k_{\perp} < 10\; {\rm MeV}$,
$10\; {\rm MeV} < k_{\perp} < 100\; {\rm MeV}$. 
We do so due to
difficulties in the numerical evaluation of integrals.
In Fig.~\ref{fig:y} we show the distributions in ${\rm y}$
for the standard results [see Eq.~(\ref{4.35})] including only 
the pomeron exchange.
Calculations were done for $\sqrt{s} = 10$~GeV and 100~GeV.
When going from $\sqrt{s} = 10$~GeV to $\sqrt{s} = 100$~GeV
the maximum of the ${\rm y}$ distribution shifts from
${\rm y_{max}} \simeq 3.4$ to ${\rm y_{max}} \simeq 5.8$.

\begin{figure}[!ht]
\includegraphics[width=0.45\textwidth]{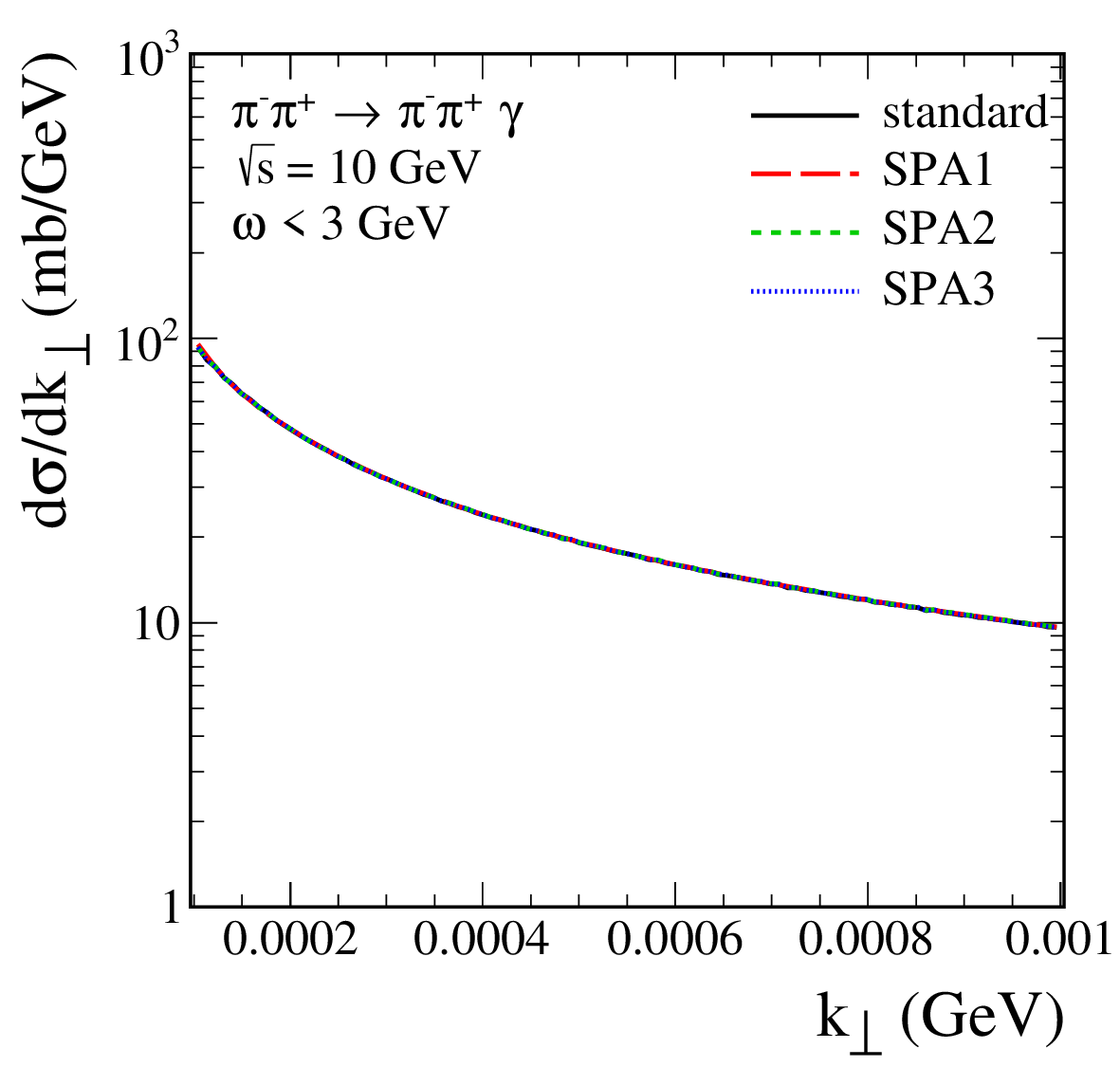}
\includegraphics[width=0.45\textwidth]{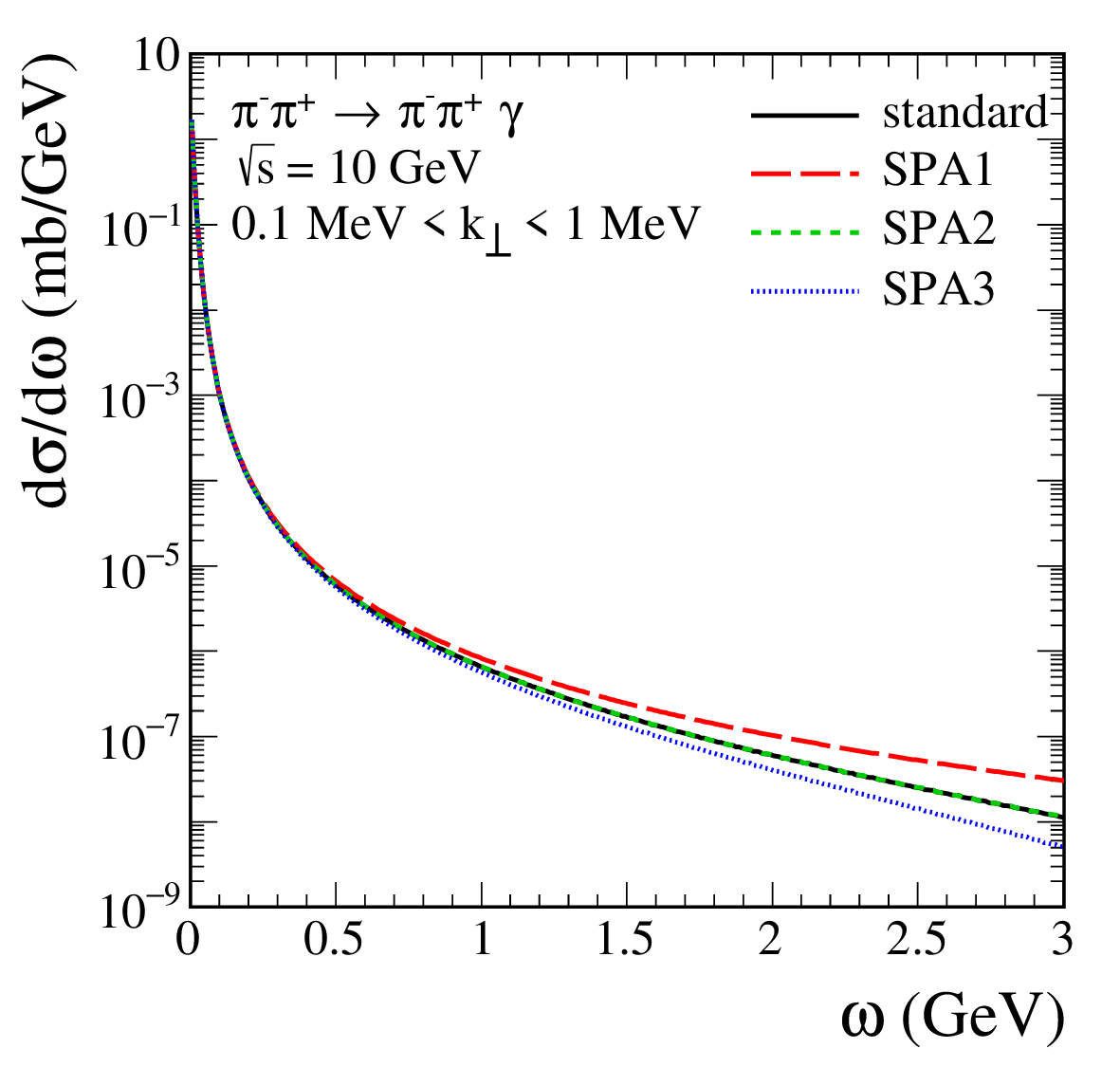}
\includegraphics[width=0.45\textwidth]{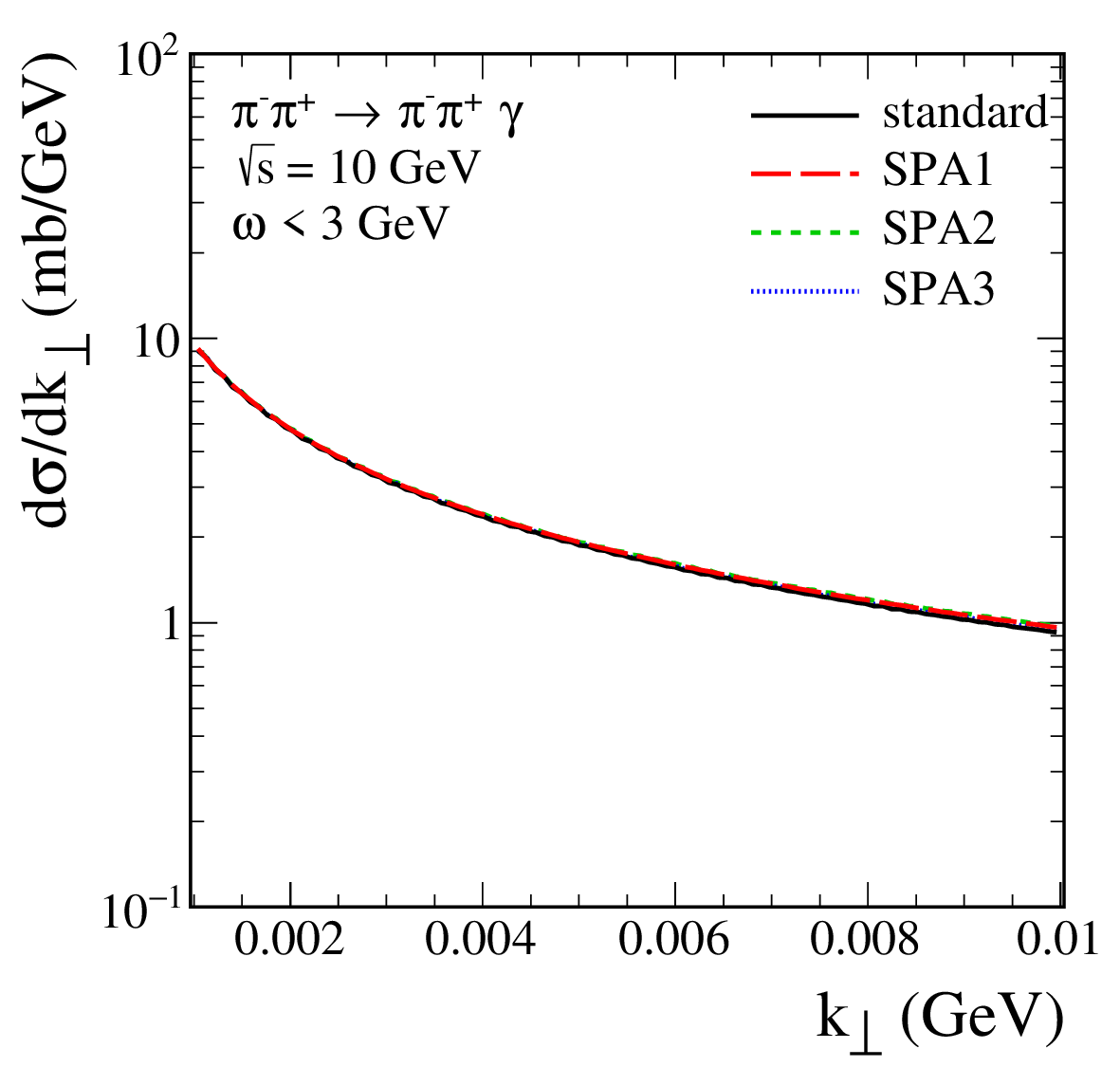}
\includegraphics[width=0.45\textwidth]{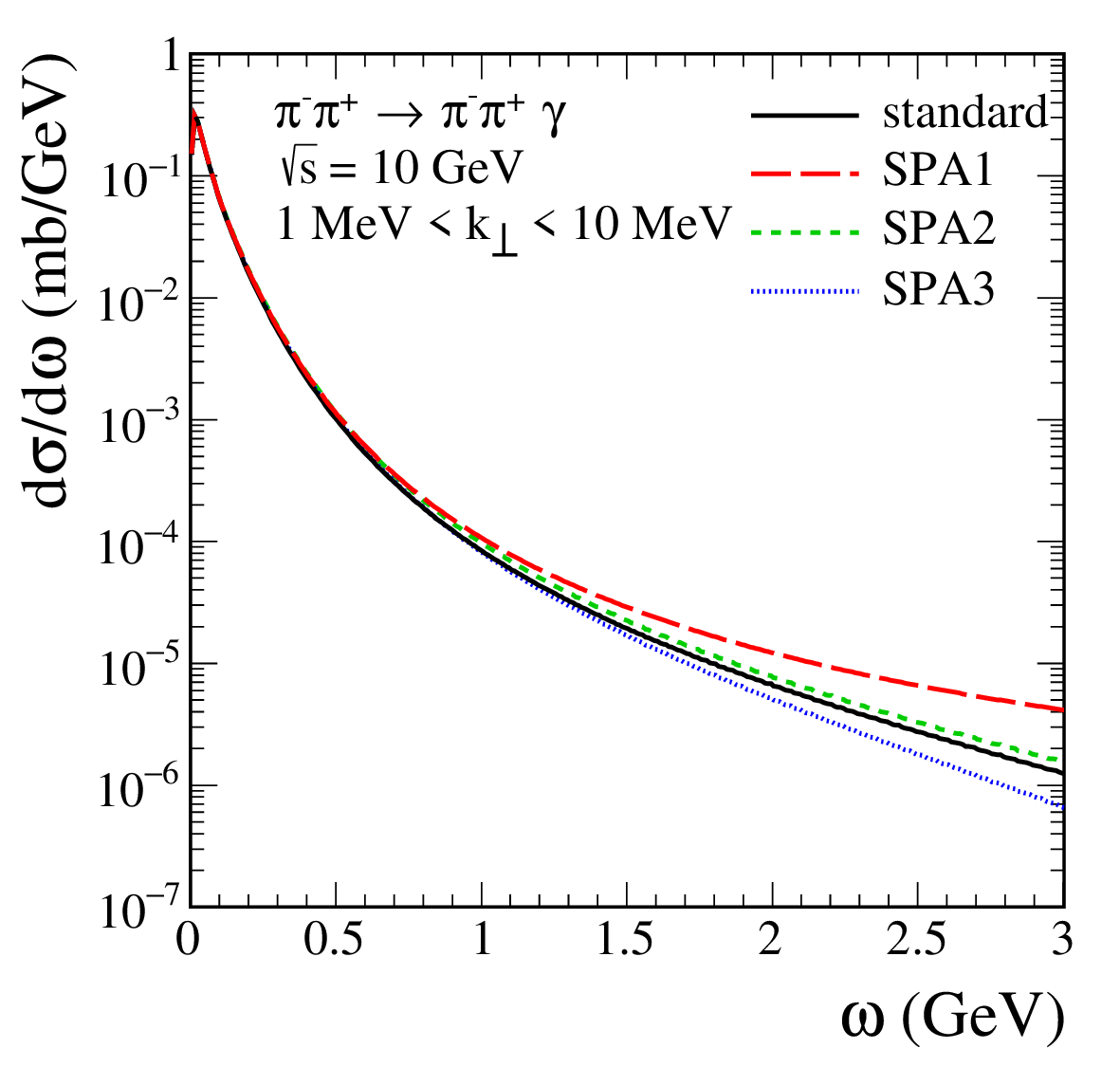}
\includegraphics[width=0.45\textwidth]{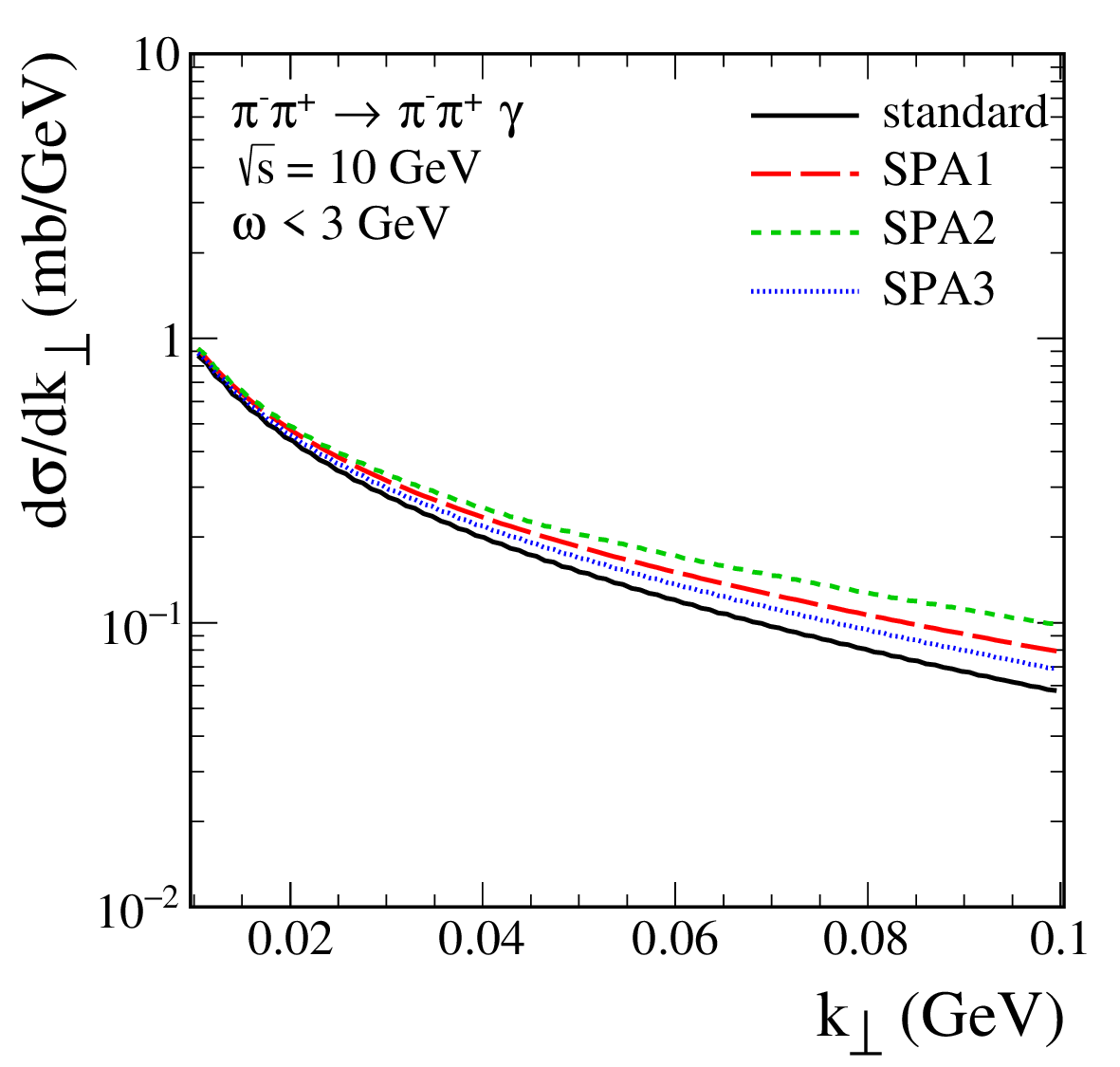}
\includegraphics[width=0.45\textwidth]{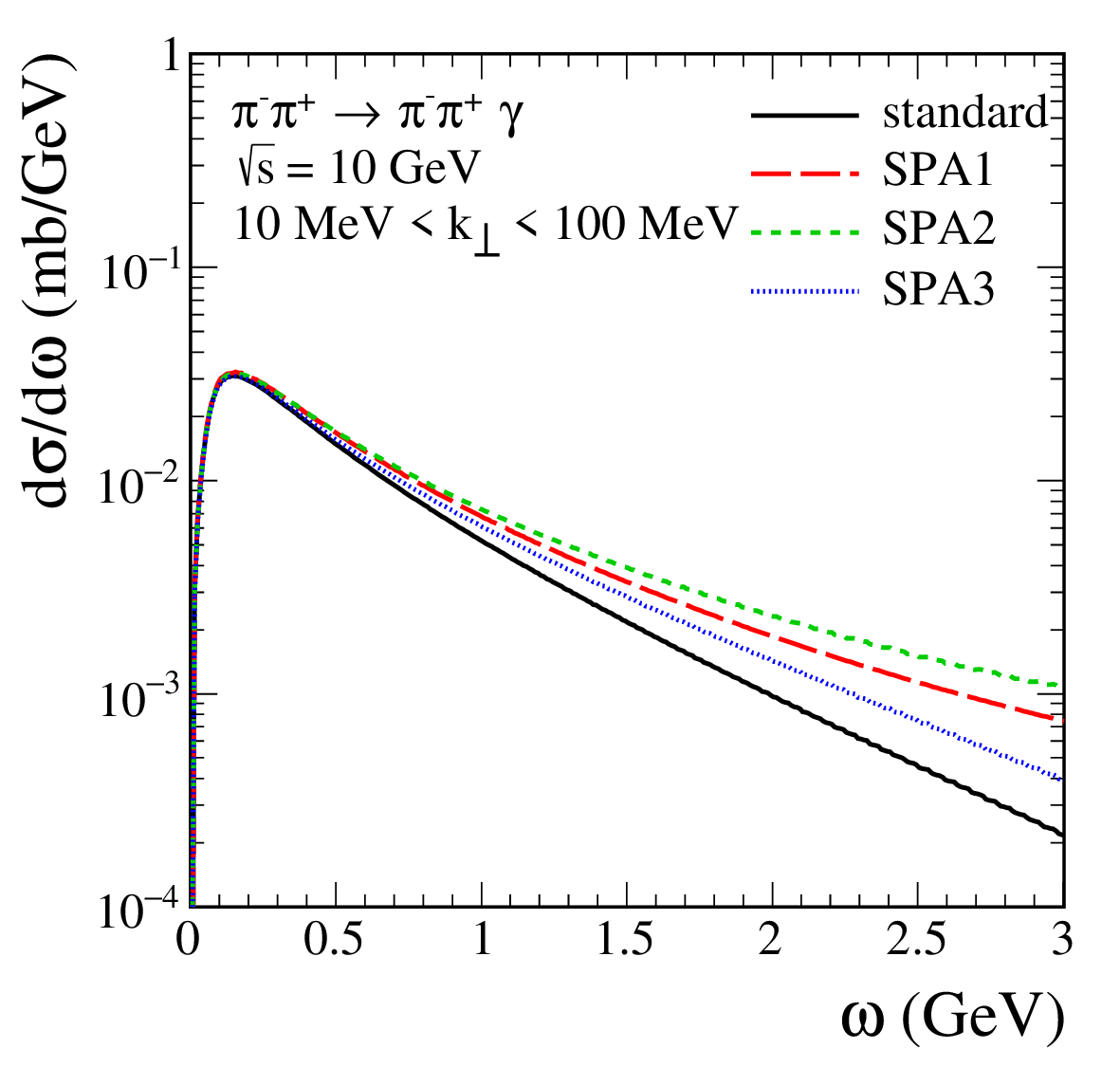}
\caption{\label{fig:10GeV}
\small
The differential distributions 
in transverse momentum of the photon
and in the energy of the photon
for the $\pi^{-} \pi^{+} \to \pi^{-} \pi^{+} \gamma$ reaction.
The calculations were done for $\sqrt{s} = 10$~GeV 
and $\omega < 3$~GeV.
The black solid line corresponds to the standard result,
the red long-dashed line corresponds to SPA1~(\ref{4.39}),
the green dashed line corresponds to SPA2~(\ref{4.42}), and
the blue dotted line corresponds to SPA3~(\ref{4.43}).}
\end{figure}

In Fig.~\ref{fig:10GeV} we present the distributions 
in $k_{\perp}$ and $\omega$
for the reaction $\pi^{-}\pi^{+} \to \pi^{-}\pi^{+} \gamma$
calculated for $\sqrt{s} = 10$~GeV 
including only the pomeron exchange.
Results are shown for three $k_{\perp}$ intervals
for our model and for the various SPAs.
From the semi-logarithmic plots of Fig.~\ref{fig:10GeV}
we see that the three SPAs follow the general trend
of our standard results quite well 
for $k_{\perp} \lesssim 20$~MeV and for $\omega \lesssim 1$~GeV.
But let us now have a closer look at these kinematic regions
at a linear scale.

\begin{figure}[!ht]
\includegraphics[width=0.45\textwidth]{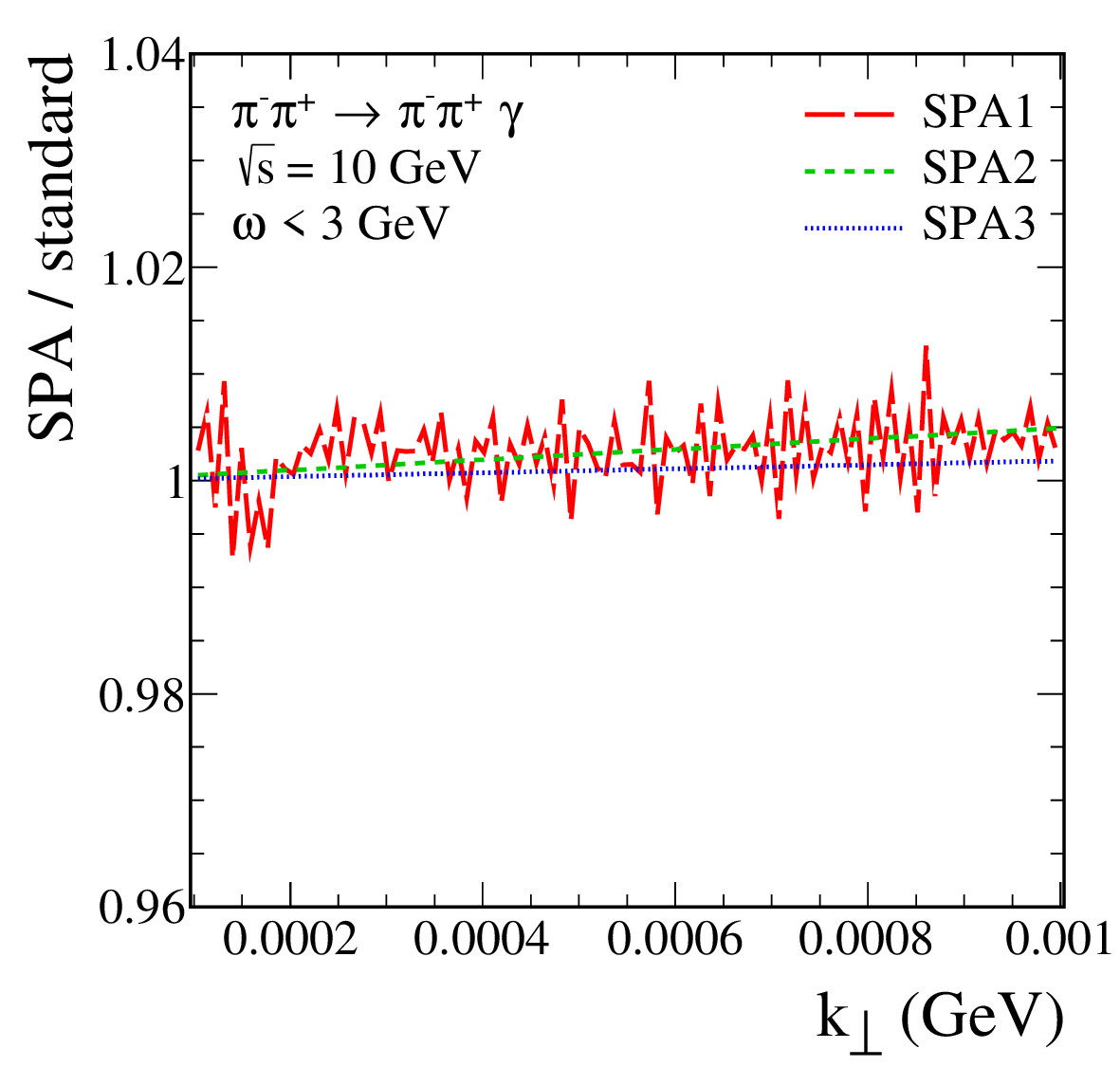}
\includegraphics[width=0.45\textwidth]{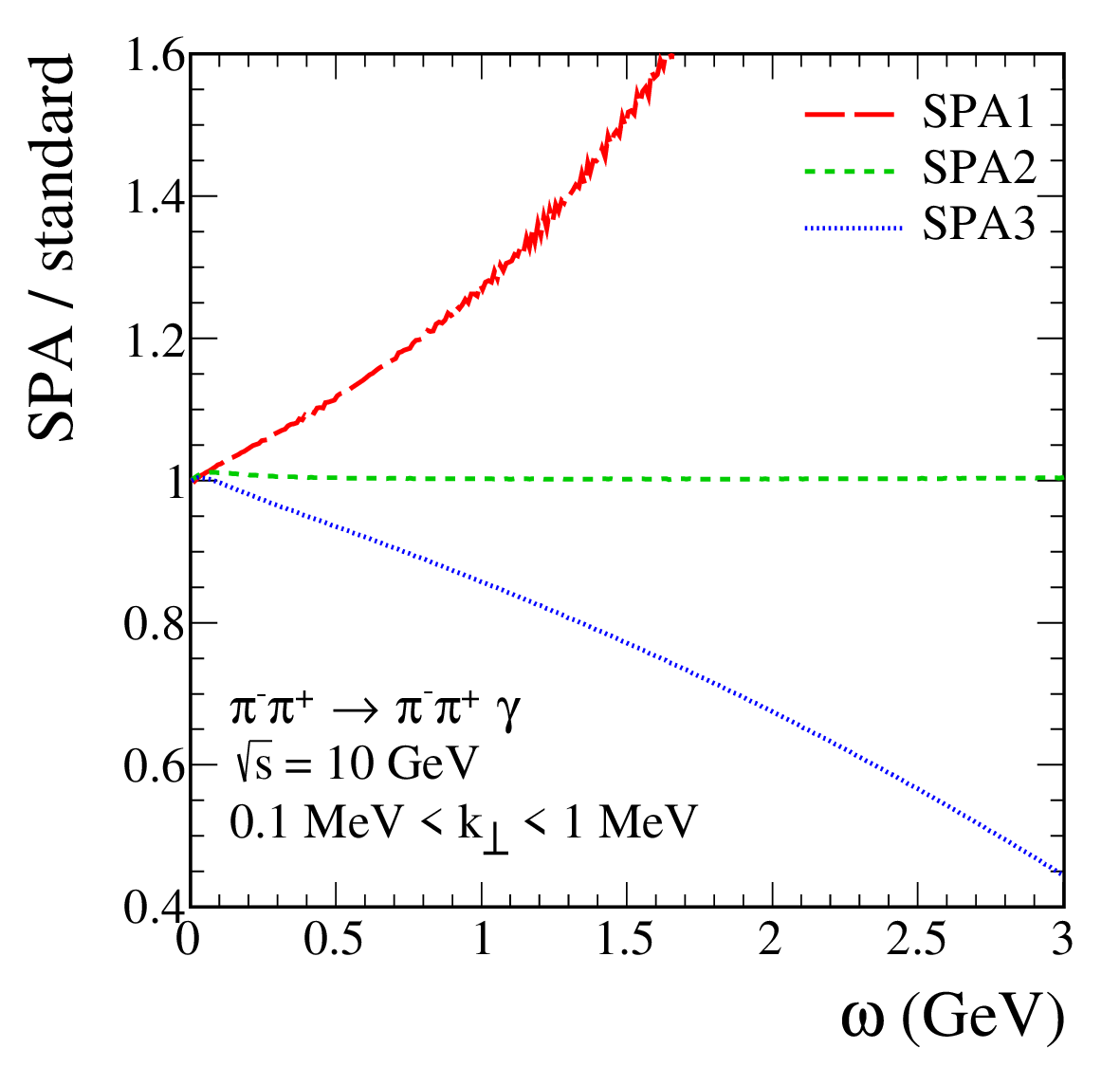}
\includegraphics[width=0.45\textwidth]{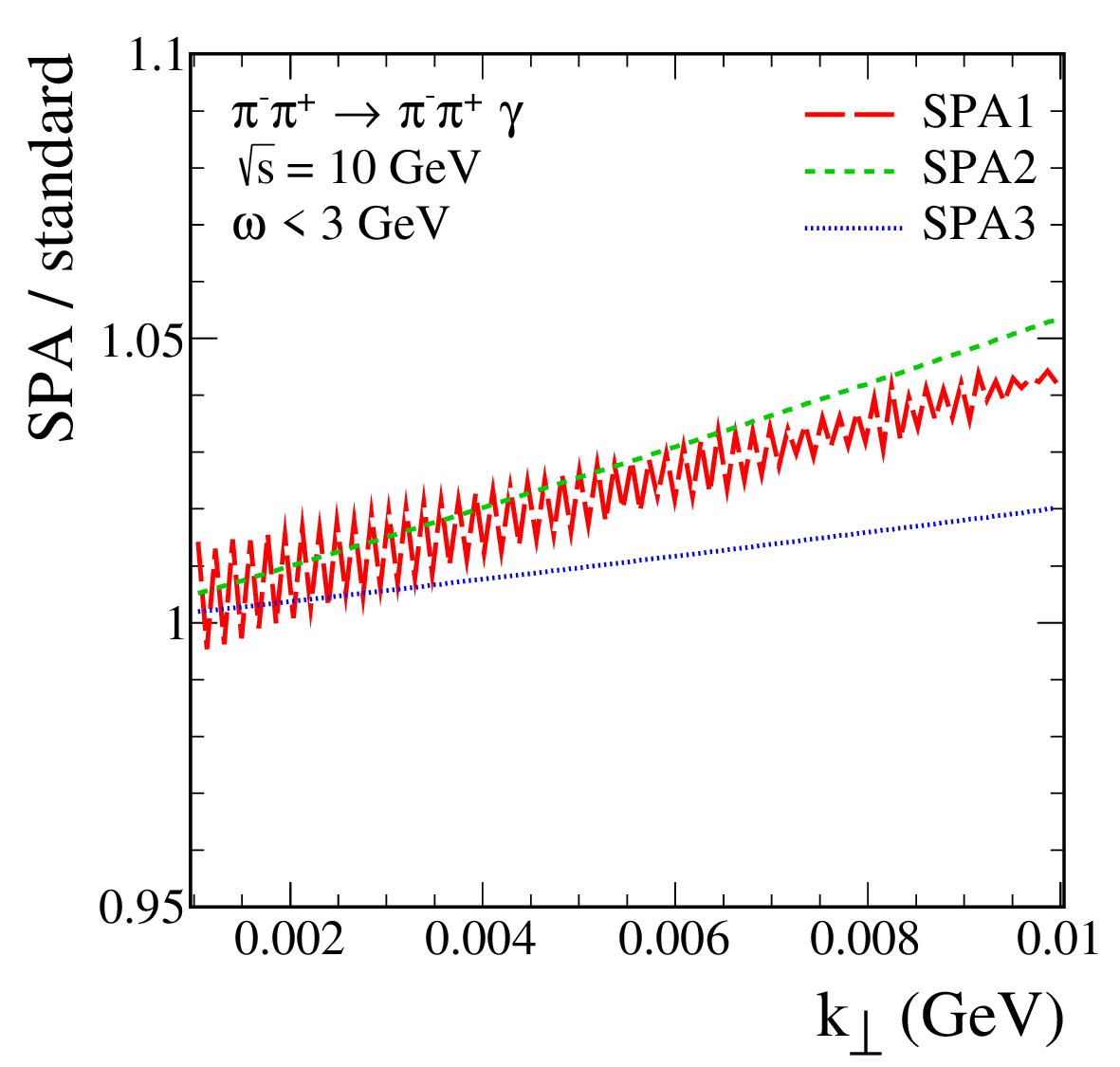}
\includegraphics[width=0.45\textwidth]{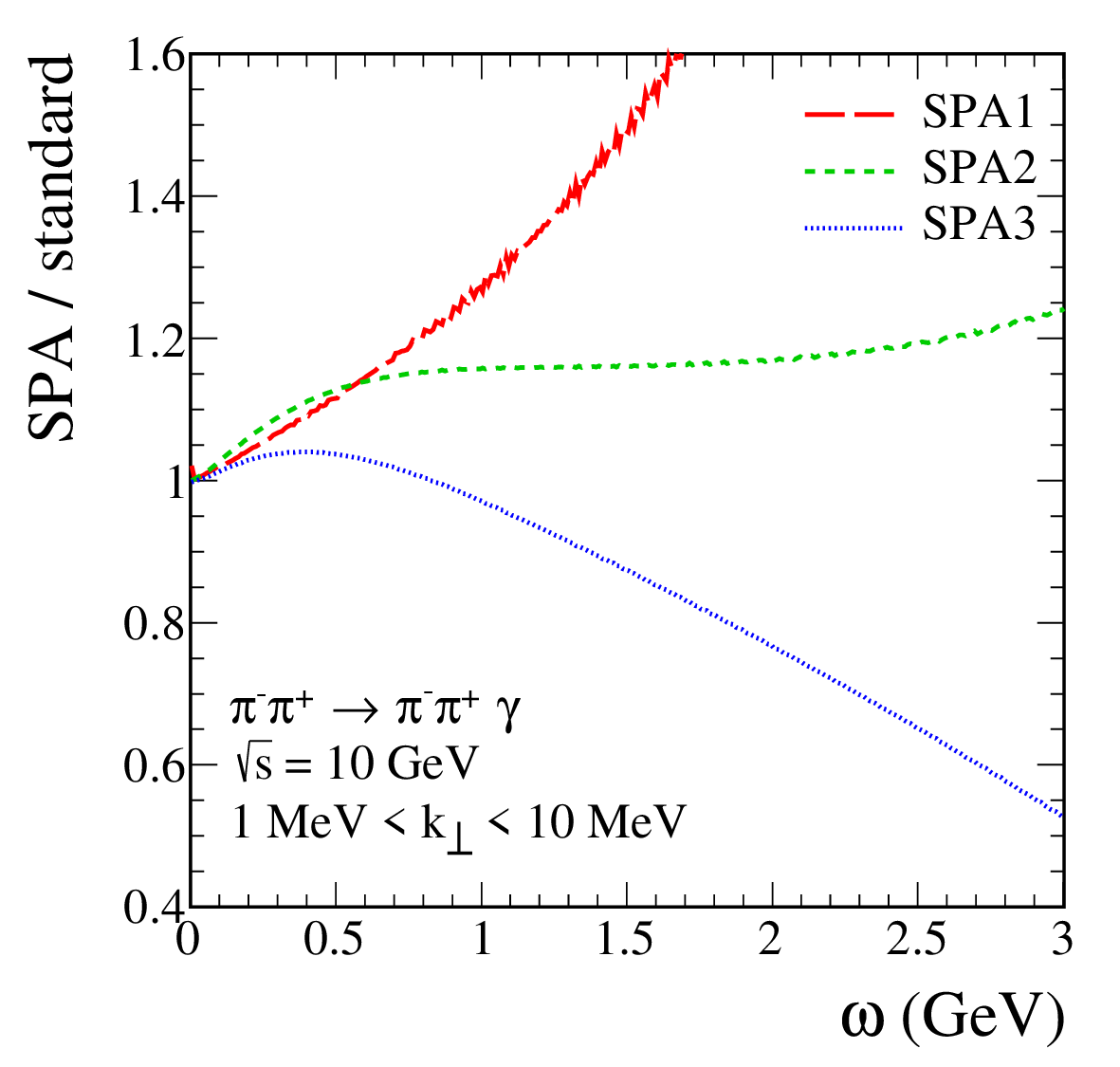}
\includegraphics[width=0.45\textwidth]{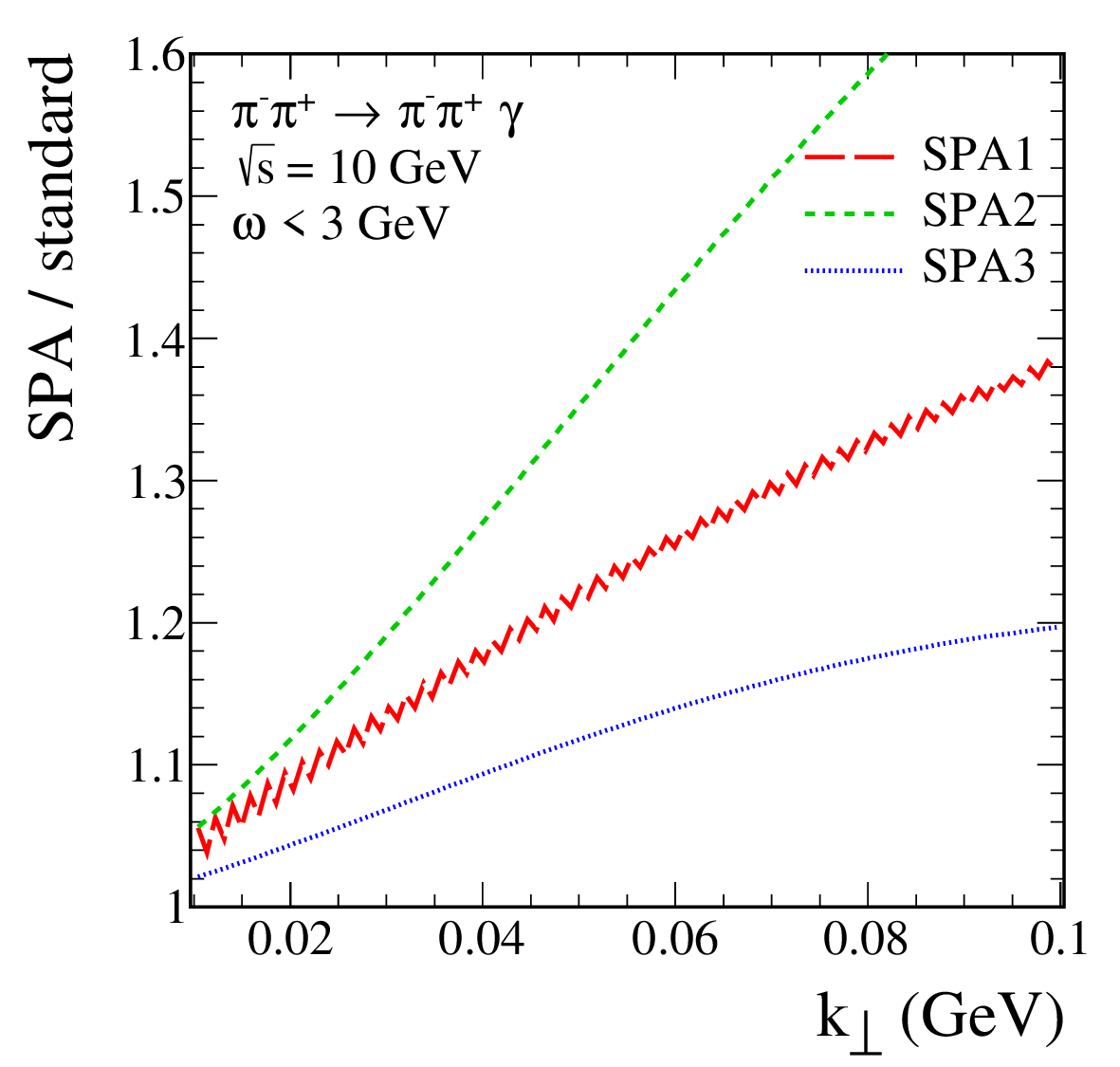}
\includegraphics[width=0.45\textwidth]{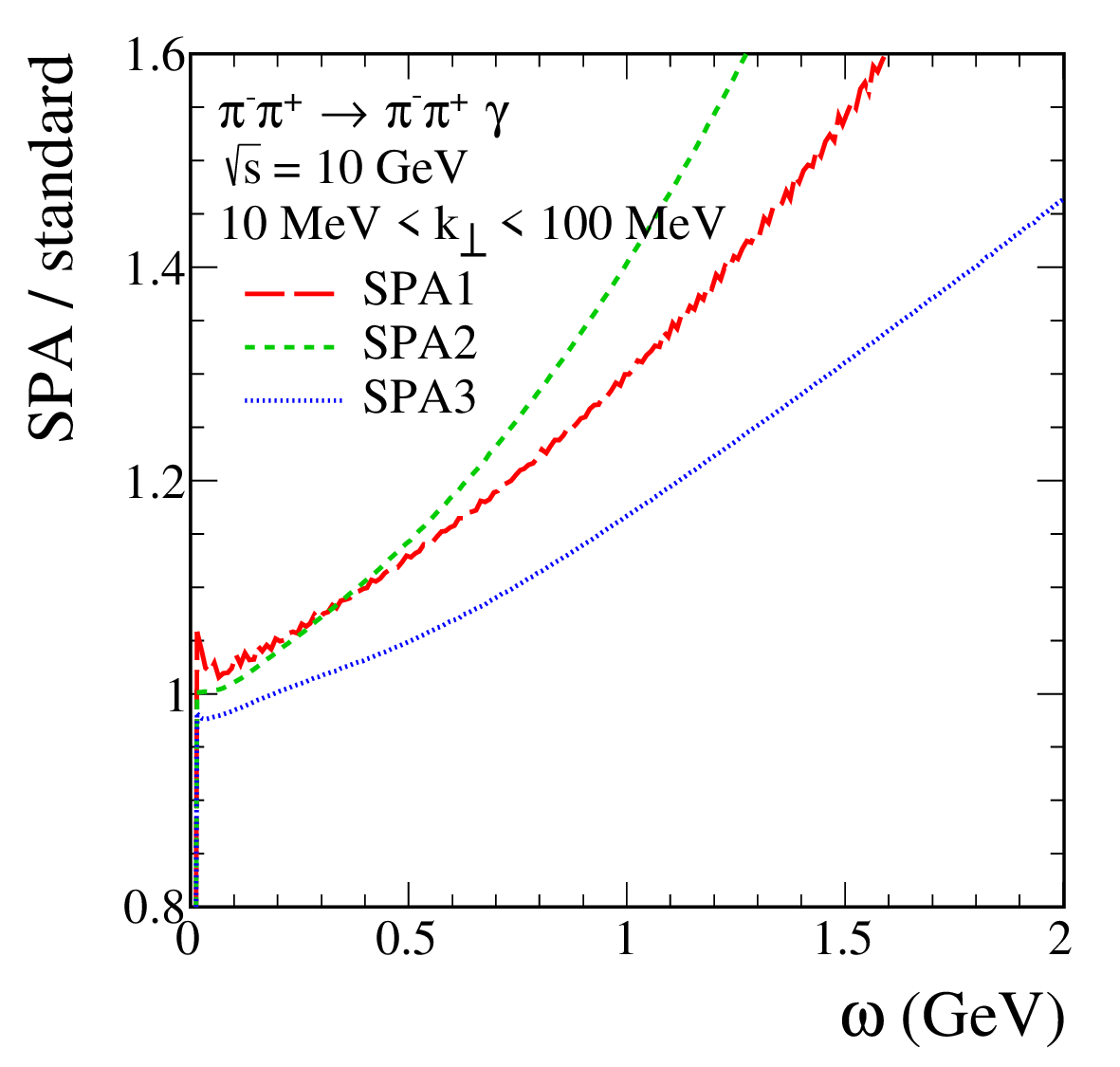}
\caption{\label{fig:ratios_10GeV}
\small
The ratios $\sigma_{\rm SPA}/\sigma_{\rm standard}$
given by (\ref{ratio_kt}) and (\ref{ratio_omega}), respectively.
The oscillations in the SPA1 results 
are of numerical origin.}
\end{figure}

Figure~\ref{fig:ratios_10GeV} shows the ratios 
of the SPAs to the standard cross section:
\begin{eqnarray}
&&\frac{d\sigma_{\rm SPA}/dk_{\perp}}{d\sigma_{\rm standard}/dk_{\perp}}\,
\label{ratio_kt} \,, \\
&&\frac{d\sigma_{\rm SPA}/d\omega}{d\sigma_{\rm standard}/d\omega}\,,
\label{ratio_omega}
\end{eqnarray}
as functions of $k_{\perp}$ and $\omega$, respectively.
The rapid fluctuations of the ratio as a function of $k_{\perp}$ are due
to different organization of integration in the two codes:
one for the full three-body phase space 
(standard approach, SPA2, SPA3) 
and one for the two-body phase space supplemented 
by additional integration over photon three momentum (SPA1).
The SPAs which we consider deviate from the standard results
only at the percent level for 
$0.1\; {\rm MeV} < k_{\perp} < 1\; {\rm MeV}$
but at the 10$\,\%$ to 50$\,\%$ level for $k_{\perp} \cong 50$~MeV;
see the left panels of Fig.~\ref{fig:ratios_10GeV}.
From the right panels of Fig.~\ref{fig:ratios_10GeV} we see
that the deviations of the SPAs from the standard results
are up to around 50$\,\%$ for $\omega < 1.5$~GeV.
We also note that the discrepancies of the SPAs to the standard results
typically increase rapidly with growing $k_{\perp}$ and $\omega$.
For the SPA1 approximation we have on purpose set $k = 0$
in the energy-momentum conserving delta function in
(\ref{4.37}), since this corresponds to a frequently used
procedure in the literature. 
Thus, the SPA1 approach does not respect the upper kinematic limit
for $\omega$.
But this is no problem for us since we are interested here only
in soft-photon production.
But we note that the accuracy of the SPA1
can be significantly improved and the region of its applicability 
can be extended by keeping
the correct energy-momentum conservation as in the SPA2 and SPA3.

\begin{figure}[!ht]
\includegraphics[width=0.45\textwidth]{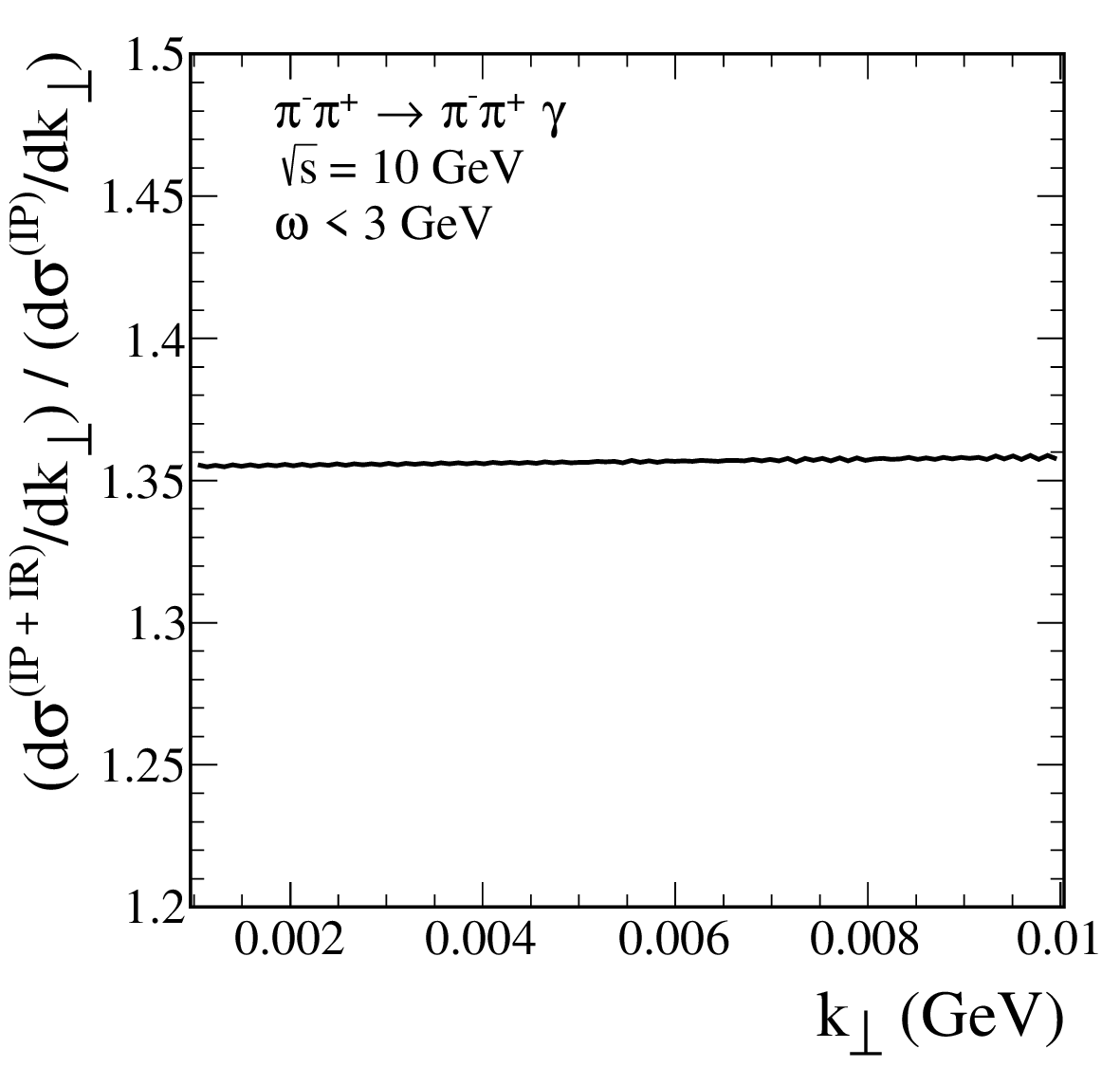}
\includegraphics[width=0.45\textwidth]{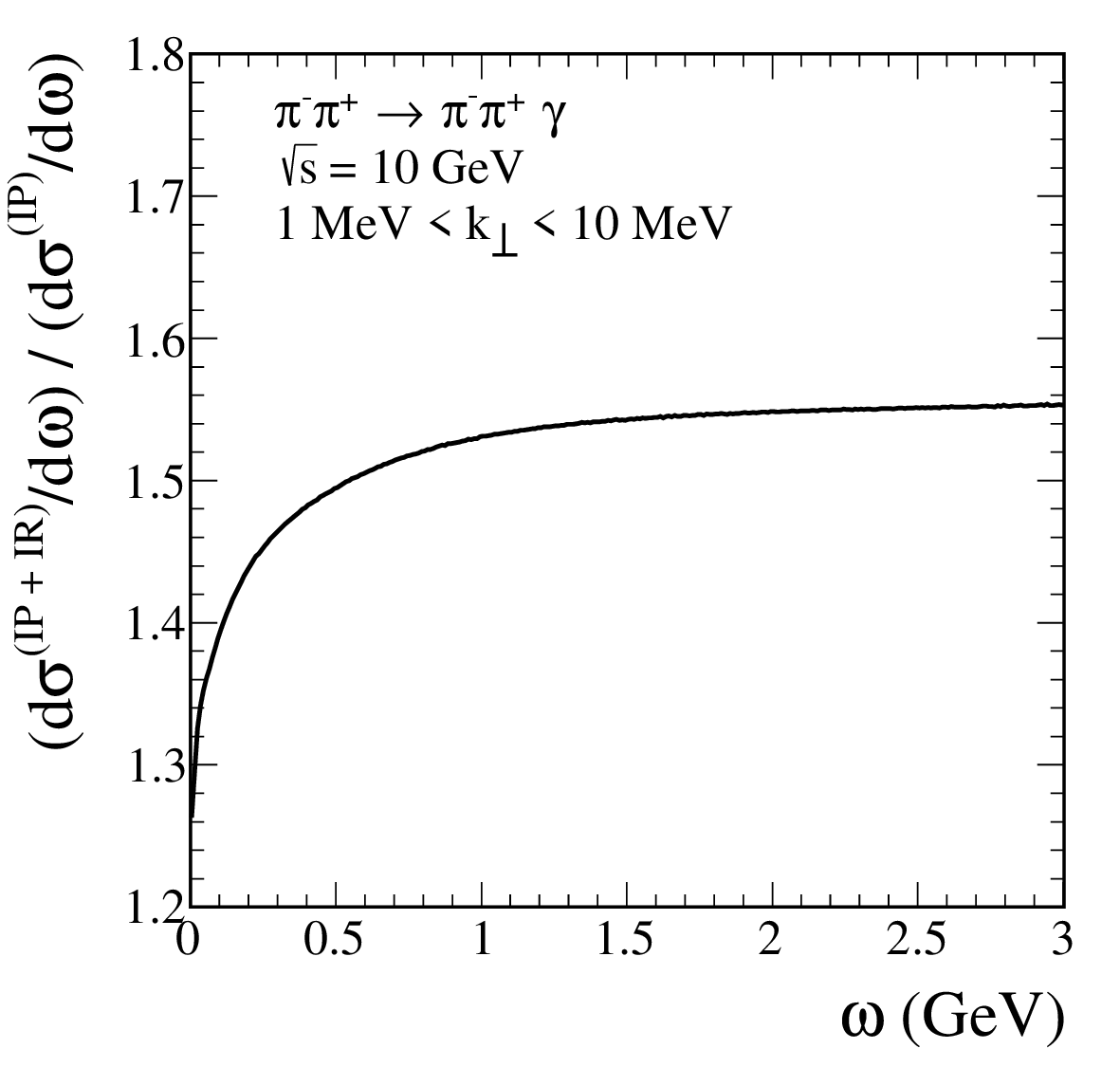}
\includegraphics[width=0.45\textwidth]{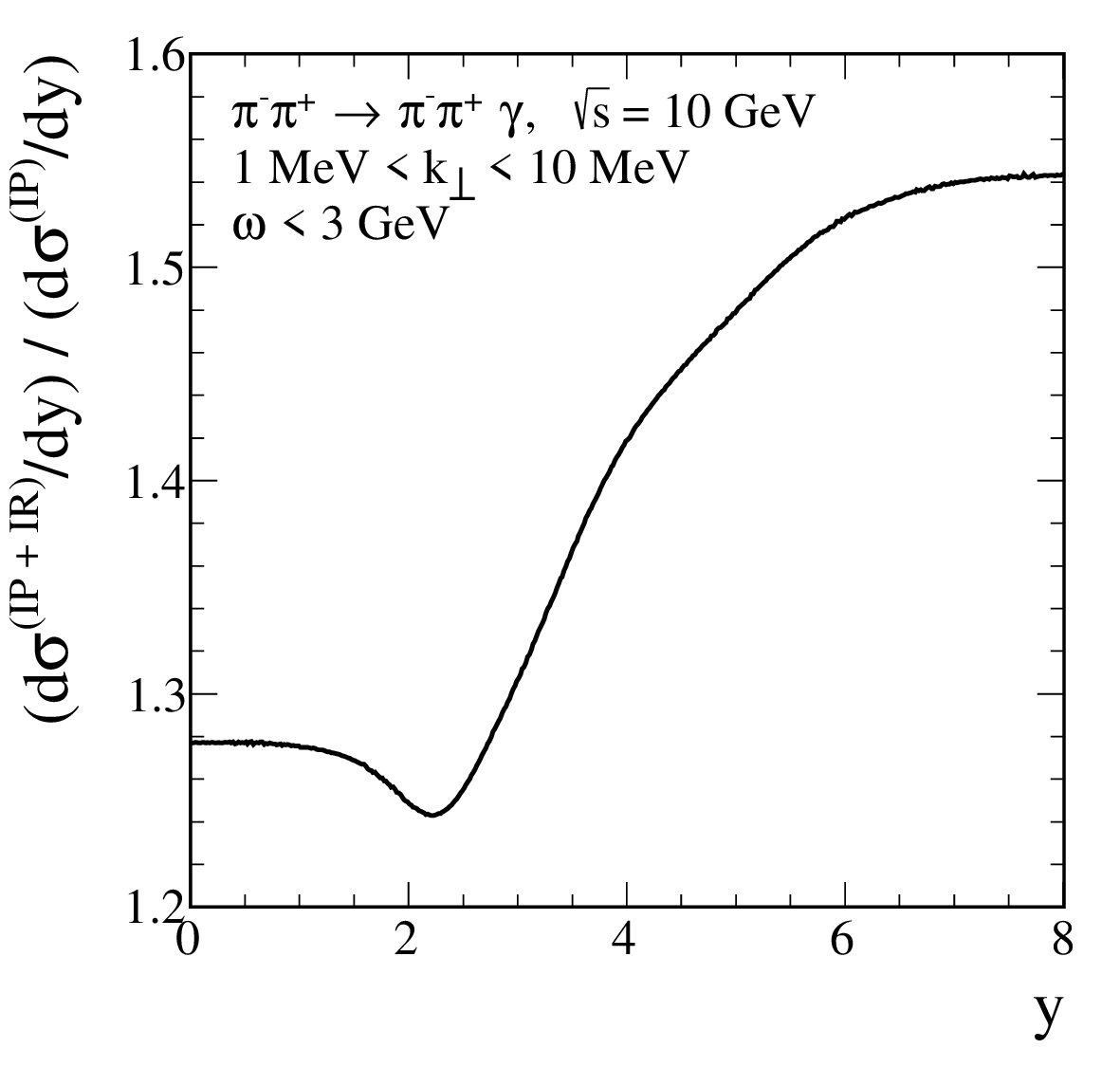}
\caption{\label{fig:regge}
\small
The ratios
$\sigma_{\rm standard}^{(\Pom + \Reg)}/\sigma_{\rm standard}^{(\Pom)}$
in the $\pi^{-} \pi^{+} \to \pi^{-} \pi^{+} \gamma$ reaction 
calculated for $\sqrt{s} = 10$~GeV,
$1\; {\rm MeV} < k_{\perp} < 10\; {\rm MeV}$, 
and $\omega < 3$~GeV.}
\end{figure}

Now we wish to illustrate the effect of inclusion of 
reggeon exchanges ($\rho_{\Reg}$ and $f_{2 \Reg}$) 
in addition to the pomeron exchange.
In Fig.~\ref{fig:regge} we show the ratio
$\sigma_{\rm standard}^{(\Pom + \Reg)}/\sigma_{\rm standard}^{(\Pom)}$ 
for our model as a function of $k_{\perp}$, $\omega$, and $\rm{y}$.
Inclusion of the subleading reggeon exchanges 
in the calculations leads to 
a sizable increase of the cross section.
We get for the ratio of the total cross sections
with the cuts
$1\; {\rm MeV} < k_{\perp} < 10\; {\rm MeV}$
and $\omega < 3$~GeV
\begin{eqnarray}
\frac{\sigma_{\rm standard}^{(\Pom + \Reg)}}
     {\sigma_{\rm standard}^{(\Pom)}} = 
\frac{29.50\; \mu{\rm b}}{21.76\; \mu{\rm b}} \simeq 1.36\,,
\label{ratio_xsection}
\end{eqnarray}
that is, an about 36$\,\%$ increase due to the reggeon exchanges.
From the ratios of differential distributions 
in $\omega$ and in $\rm{y}$
we see that these ratios vary from 1.25 to 1.55 
depending on kinematics.

Now we turn to the results at c.m. energy $\sqrt{s} = 100$~GeV.
Here we include in the calculations only the pomeron-exchange contributions. 
As we see already from Fig.~\ref{fig:XS}
the nonleading exchanges are negligible there.

\begin{figure}[!ht]
(a)\includegraphics[width=0.45\textwidth]{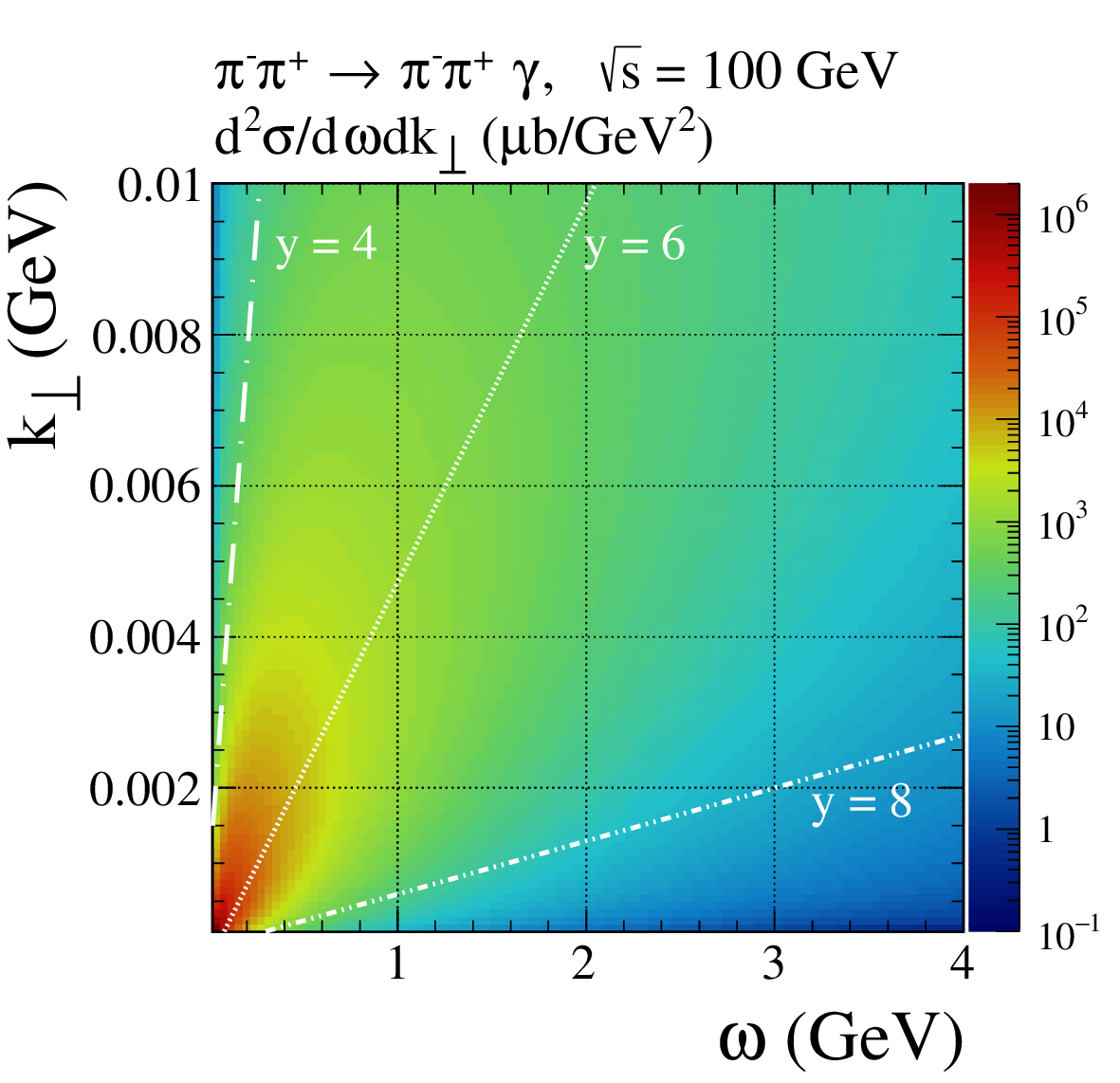}\\
(b)\includegraphics[width=0.45\textwidth]{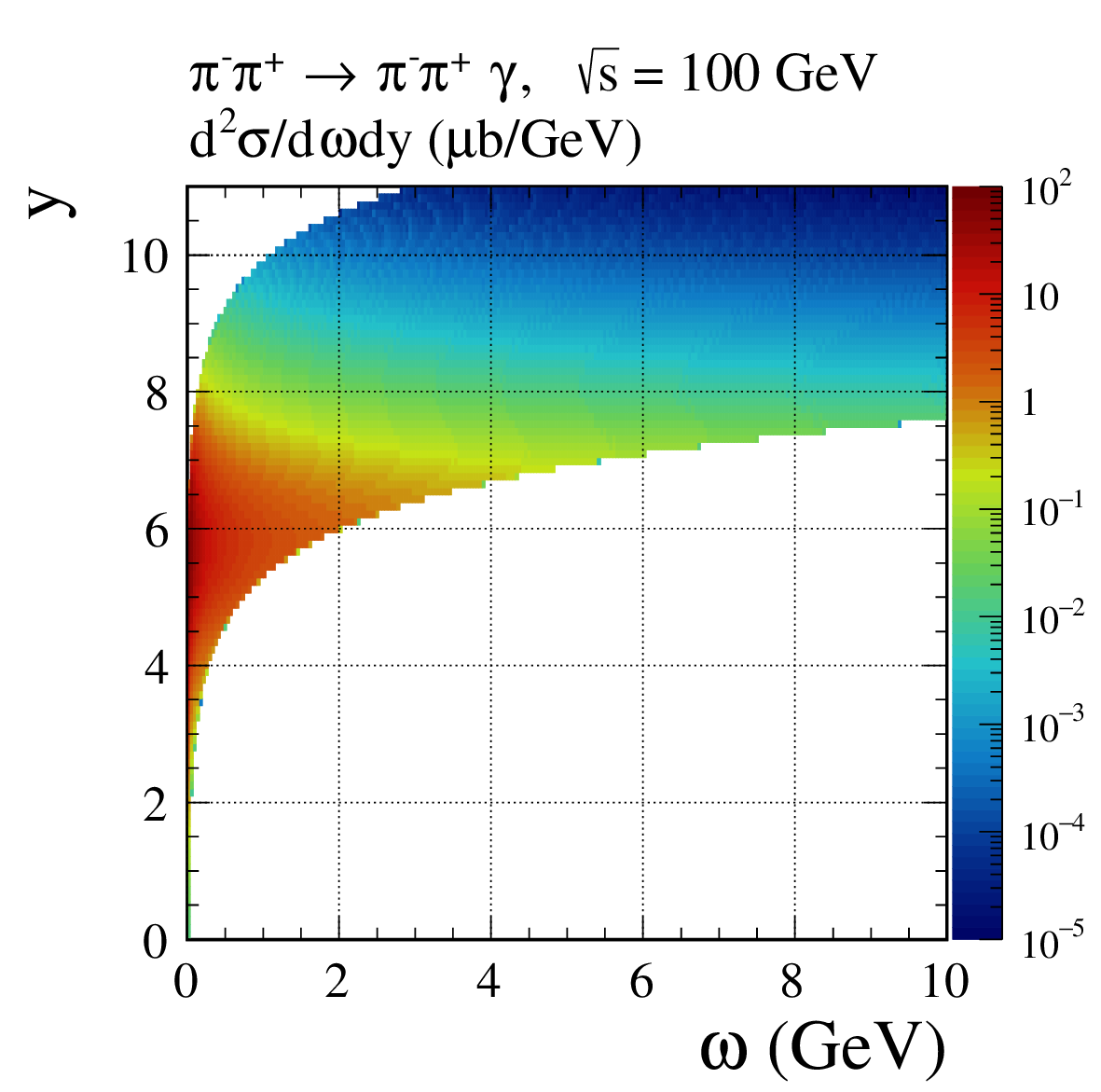}
(c)\includegraphics[width=0.45\textwidth]{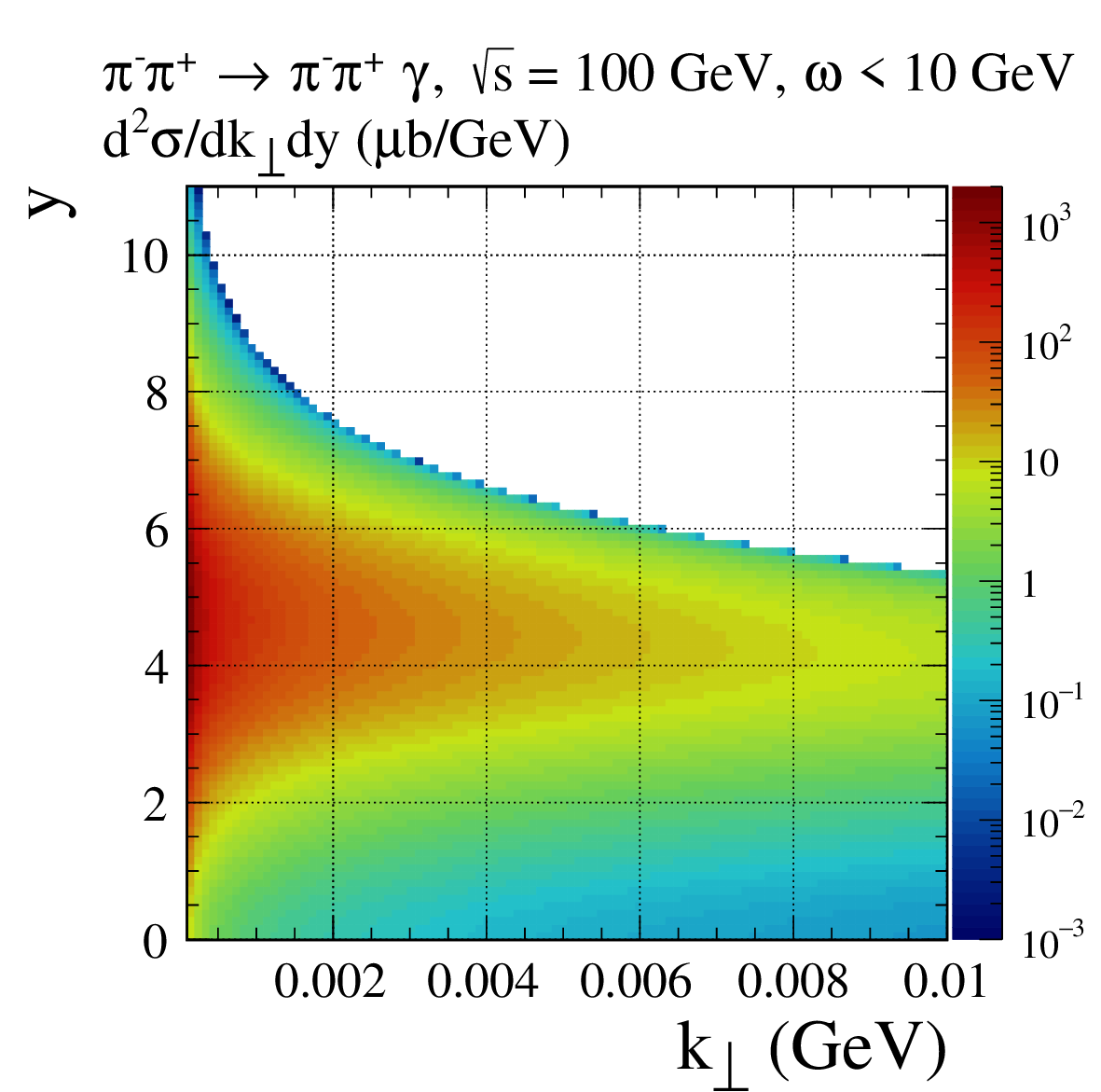}
\caption{\label{fig:2dim_exact_100}
\small
The two-dimensional distributions
in ($\omega$, $k_{\perp}$), ($\omega$, $\rm{y}$), and
($k_{\perp}$, $\rm{y}$),
for the reaction $\pi^{-} \pi^{+} \to \pi^{-} \pi^{+} \gamma$ at $\sqrt{s} = 100$~GeV.
This is the same as in Fig.~\ref{fig:2dim_exact} 
but for $\sqrt{s} = 100$~GeV, $\omega < 10$~GeV,
and $|\rm{y}| < 11$.
The lines plotted in the panel~(a) correspond 
to the photon rapidities ${\rm y} = 4, 6, 8$.}
\end{figure}

In Fig.~\ref{fig:2dim_exact_100} we show the distributions
in \mbox{($\omega$, $k_{\perp}$)}, \mbox{($\omega$, $\rm{y}$)}, 
and \mbox{($k_{\perp}$, $\rm{y}$)}, for our standard results.
Here we consider only c.m. photon energies $\omega < 10$~GeV.
The constraint (\ref{4.23d}), setting $k^{2} = 0$,
is then always well satisfied.
That is, we are in the Regge regime for all relevant amplitudes.
These distributions are the analogs of those shown
in Fig.~\ref{fig:2dim_exact} for $\sqrt{s} = 10$~GeV.

\begin{figure}[!ht]
(a)\includegraphics[width=0.45\textwidth]{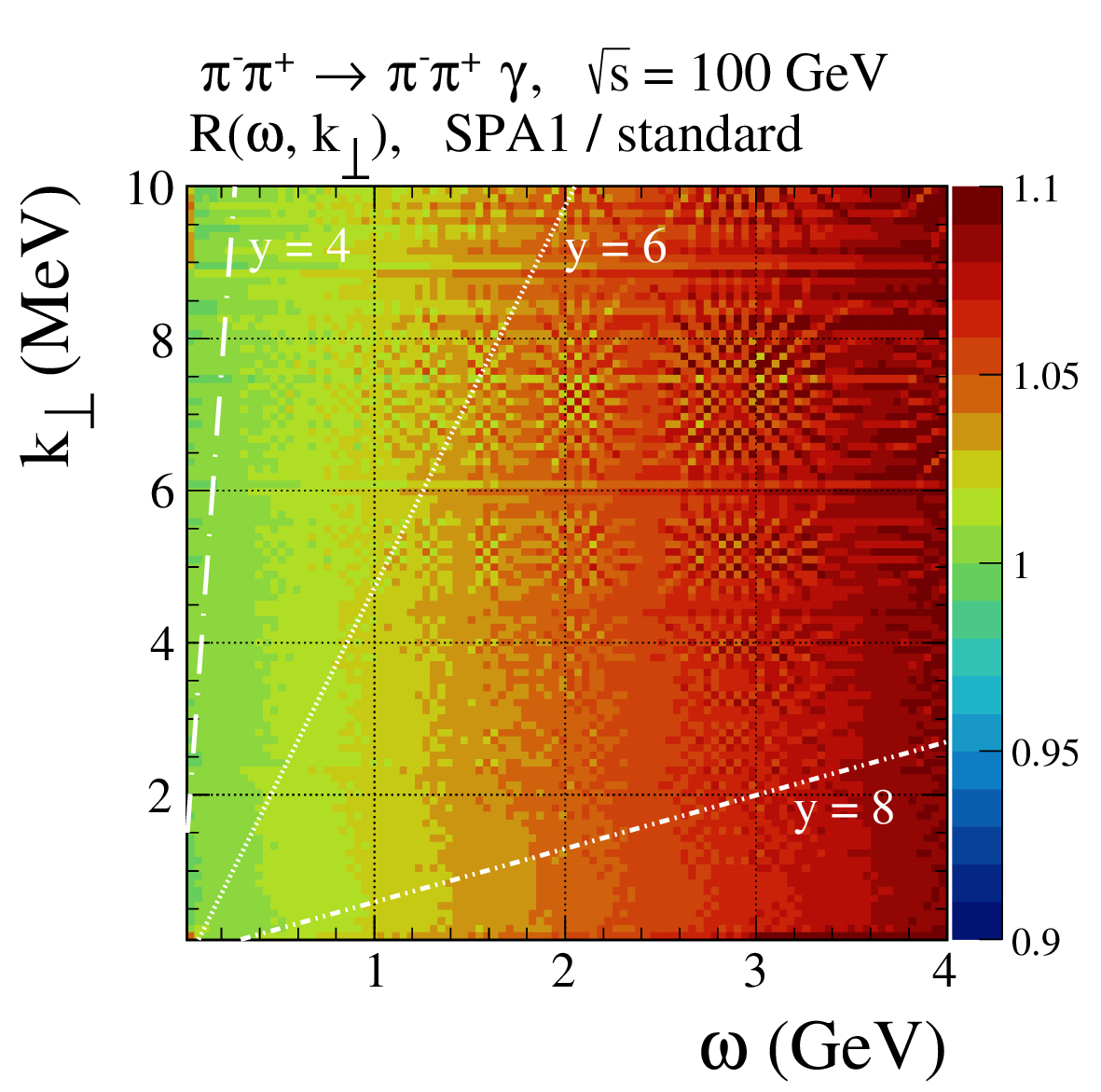}
(b)\includegraphics[width=0.45\textwidth]{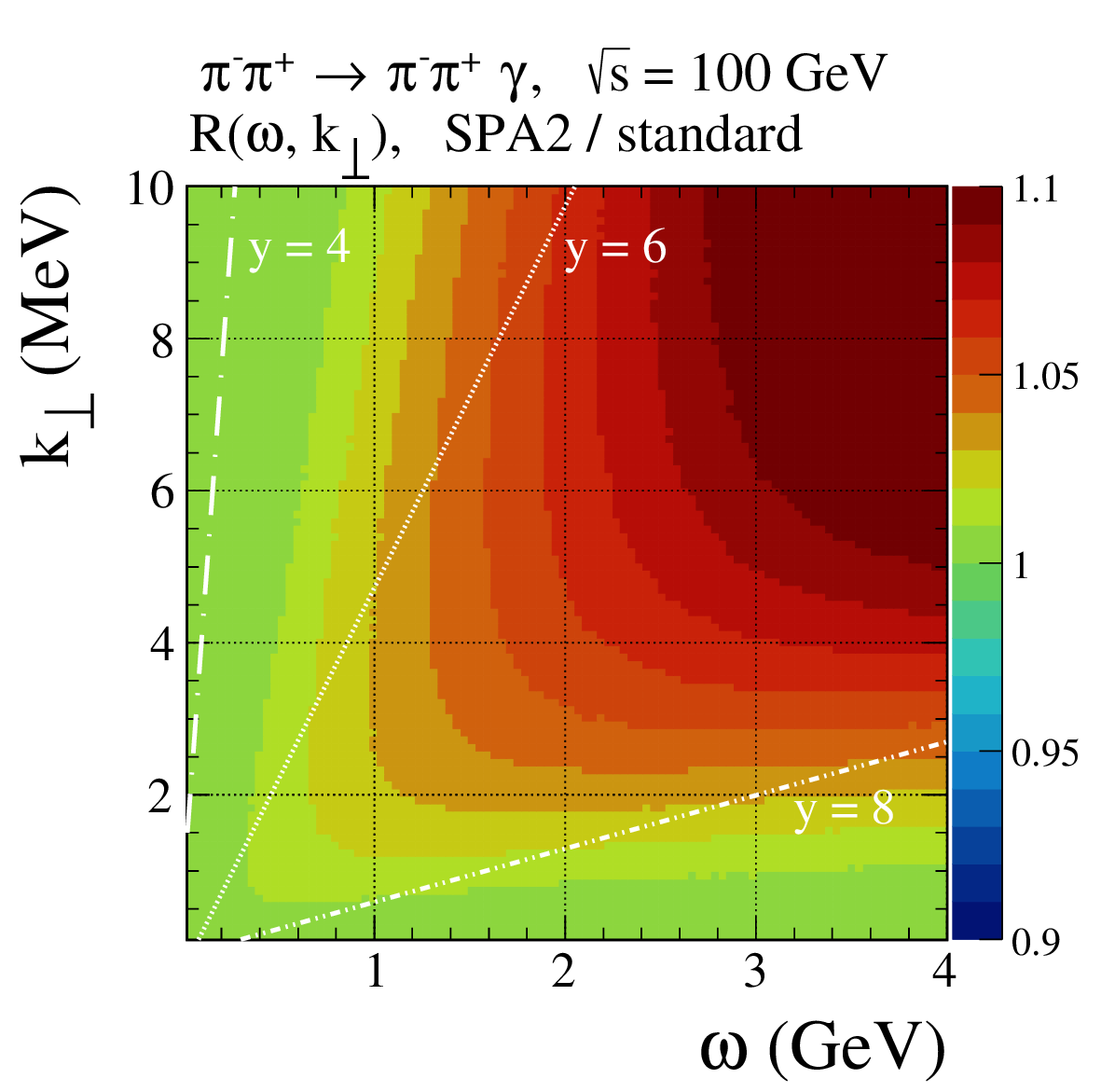}
(c)\includegraphics[width=0.45\textwidth]{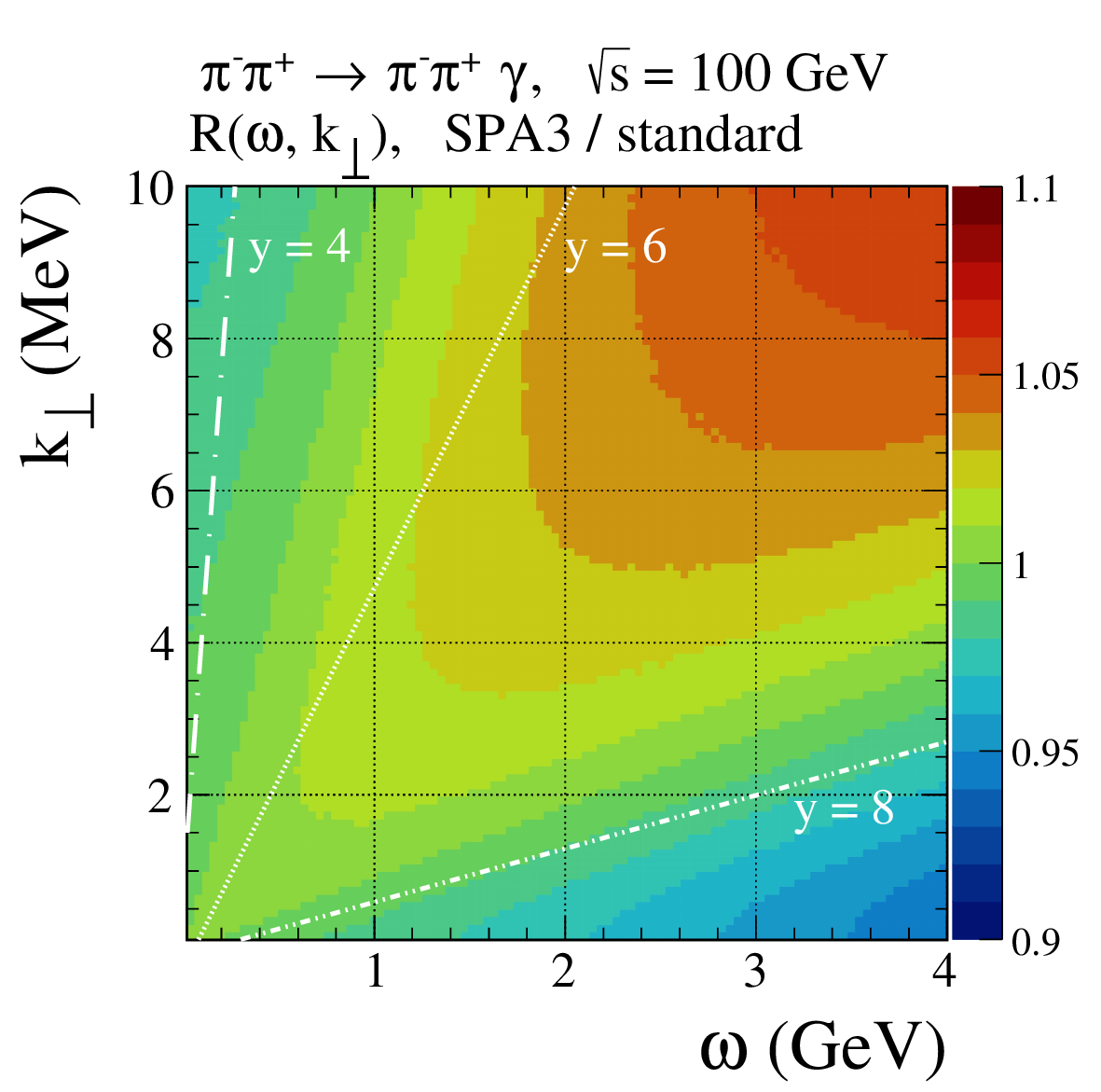}
\caption{\label{fig:2Dratio_100}
\small
The ratios ${\rm R}(\omega,k_{\perp})$ (\ref{ratio}) 
for the $\pi^{-} \pi^{+} \to \pi^{-} \pi^{+} \gamma$ reaction
for $\sqrt{s} = 100$~GeV for the three SPAs.
The lines corresponding to the photon rapidities 
${\rm y} = 4, 6, 8$ are also plotted.}
\end{figure}

Figure~\ref{fig:2Dratio_100} shows the ratios 
${\rm R}(\omega,k_{\perp})$ (\ref{ratio}) 
for the reaction $\pi^{-} \pi^{+} \to \pi^{-} \pi^{+} \gamma$
at $\sqrt{s} = 100$~GeV for the approximations
SPA1~(\ref{4.39}), SPA2~(\ref{4.42}) and SPA3~(\ref{4.43}).

\begin{figure}[!ht]
\includegraphics[width=0.45\textwidth]{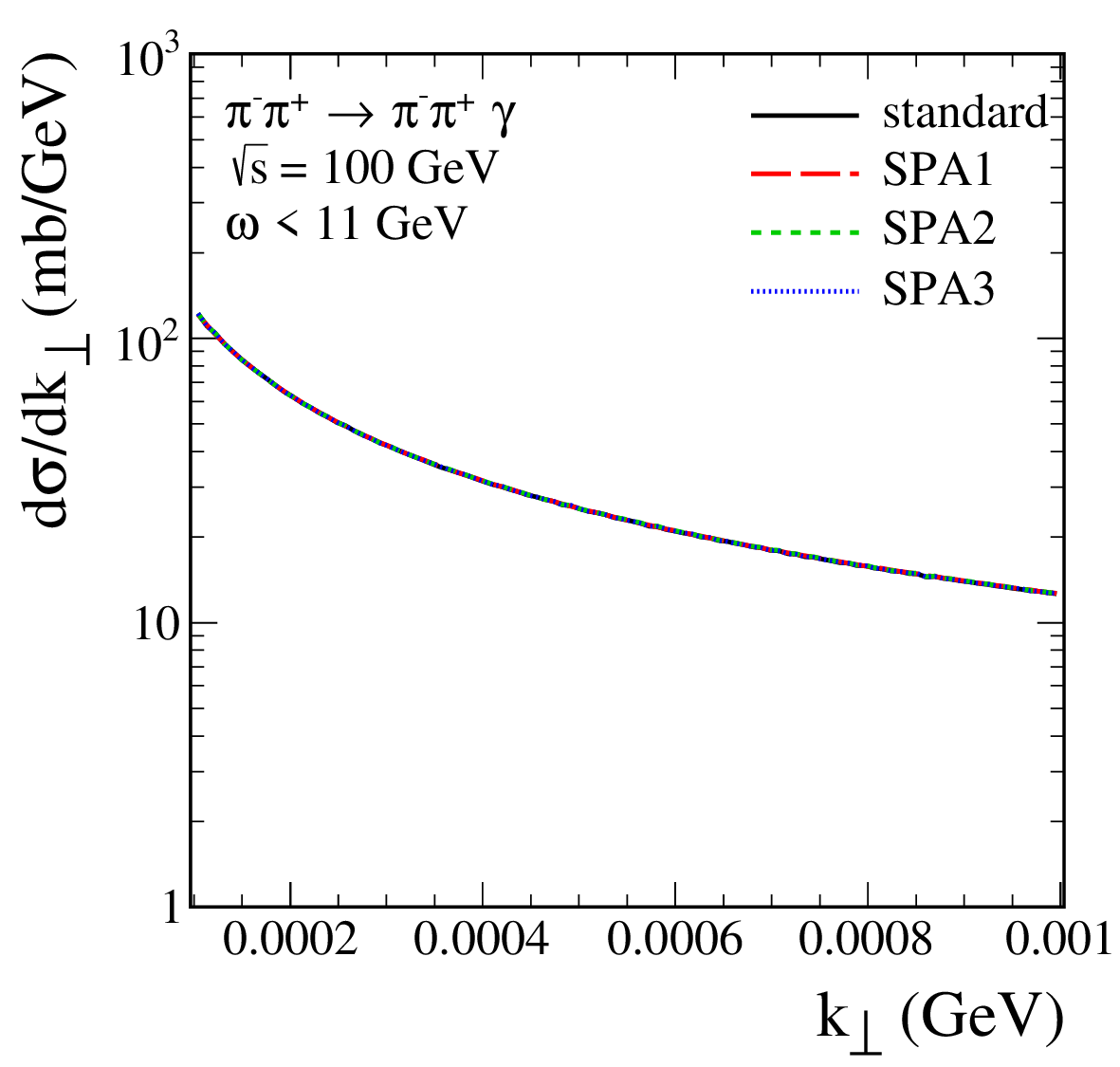}
\includegraphics[width=0.45\textwidth]{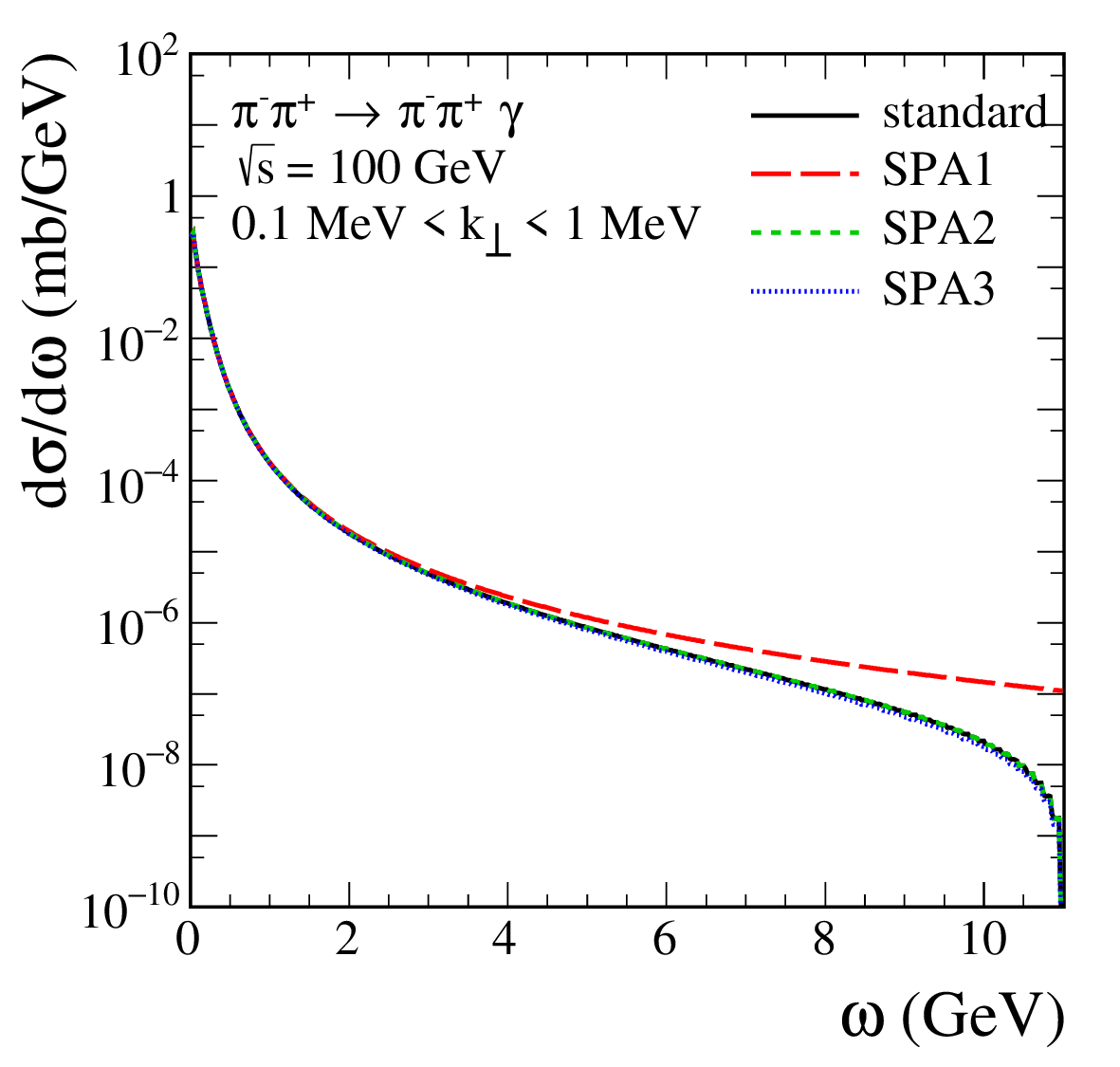}
\includegraphics[width=0.45\textwidth]{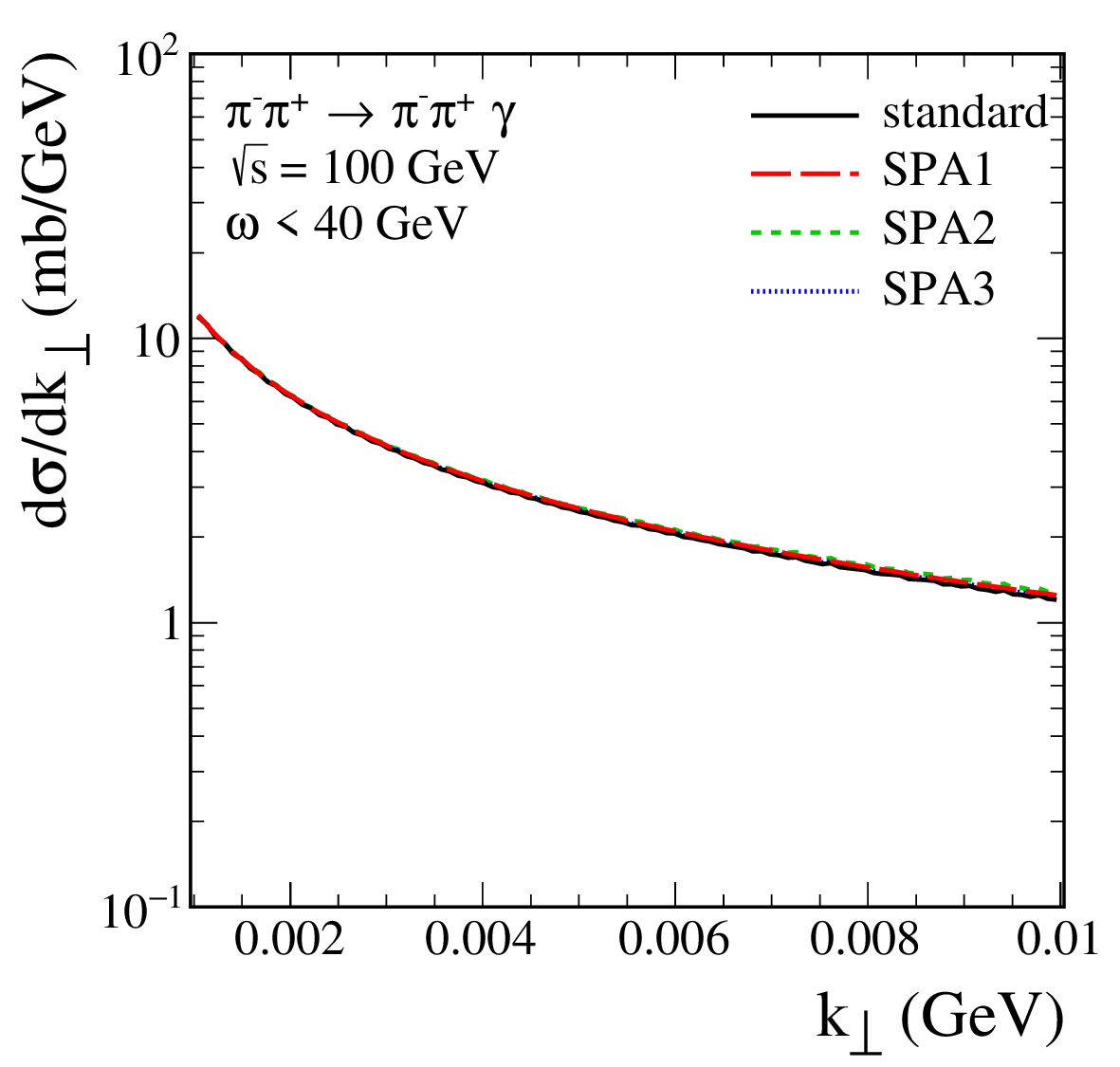}
\includegraphics[width=0.45\textwidth]{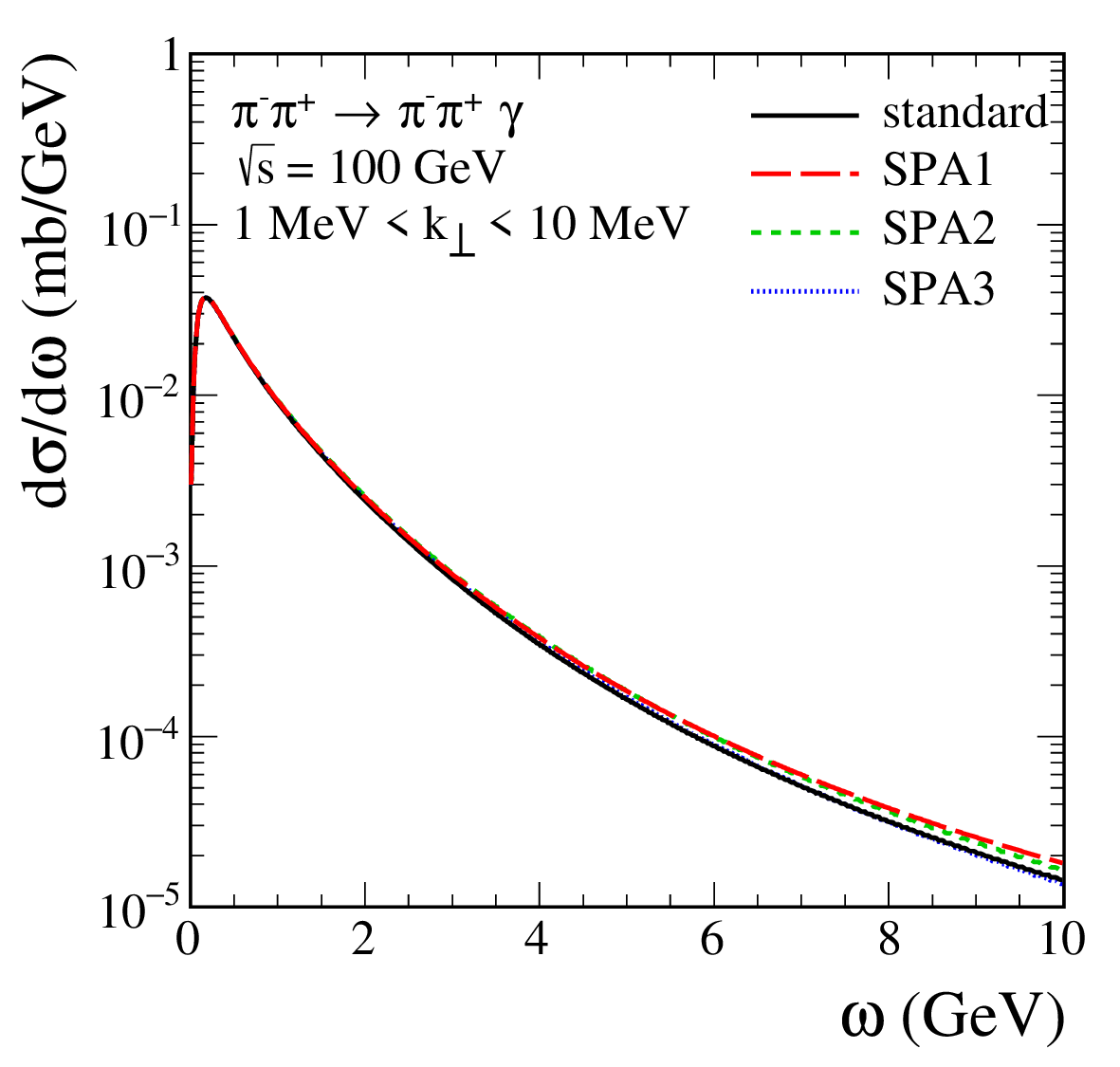}
\includegraphics[width=0.45\textwidth]{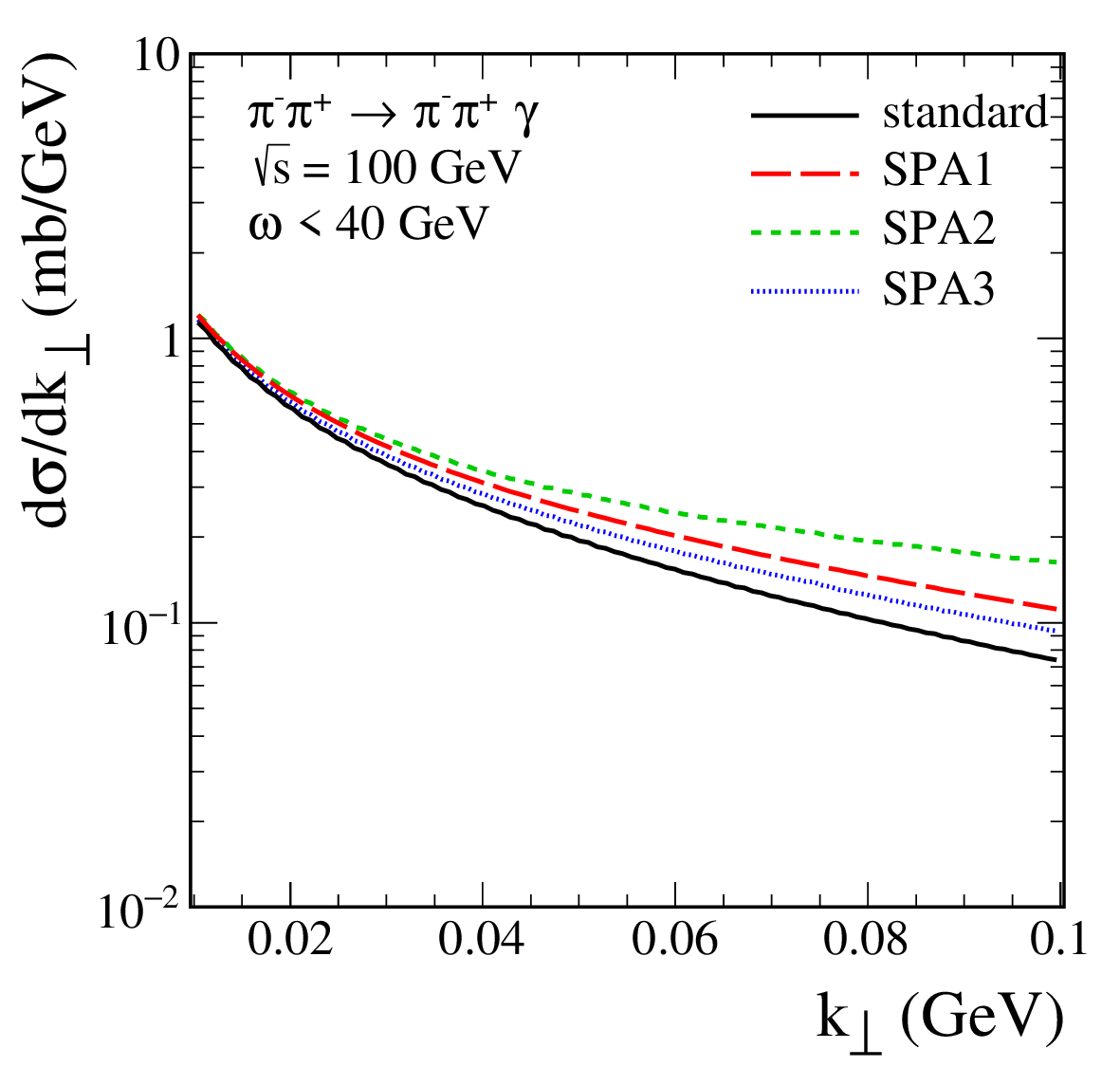}
\includegraphics[width=0.45\textwidth]{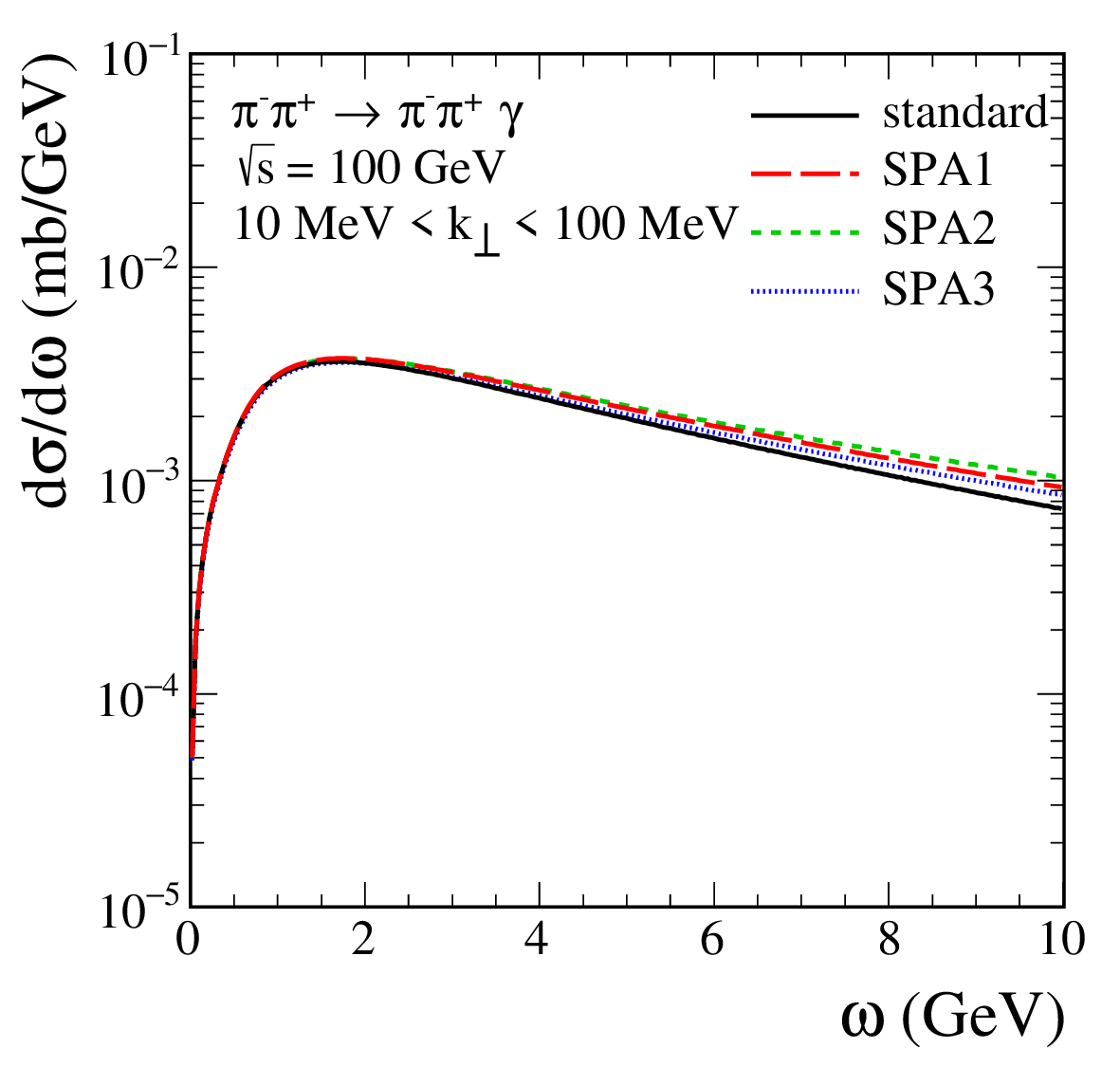}
\caption{\label{fig:100GeV}
\small
The same as in Fig.~\ref{fig:10GeV} but for $\sqrt{s} = 100$~GeV.
Shown are results for three $k_{\perp}$ intervals
and with cuts on $\omega$ specified in the figure legends.}
\end{figure}

\begin{figure}[!ht]
\includegraphics[width=0.45\textwidth]{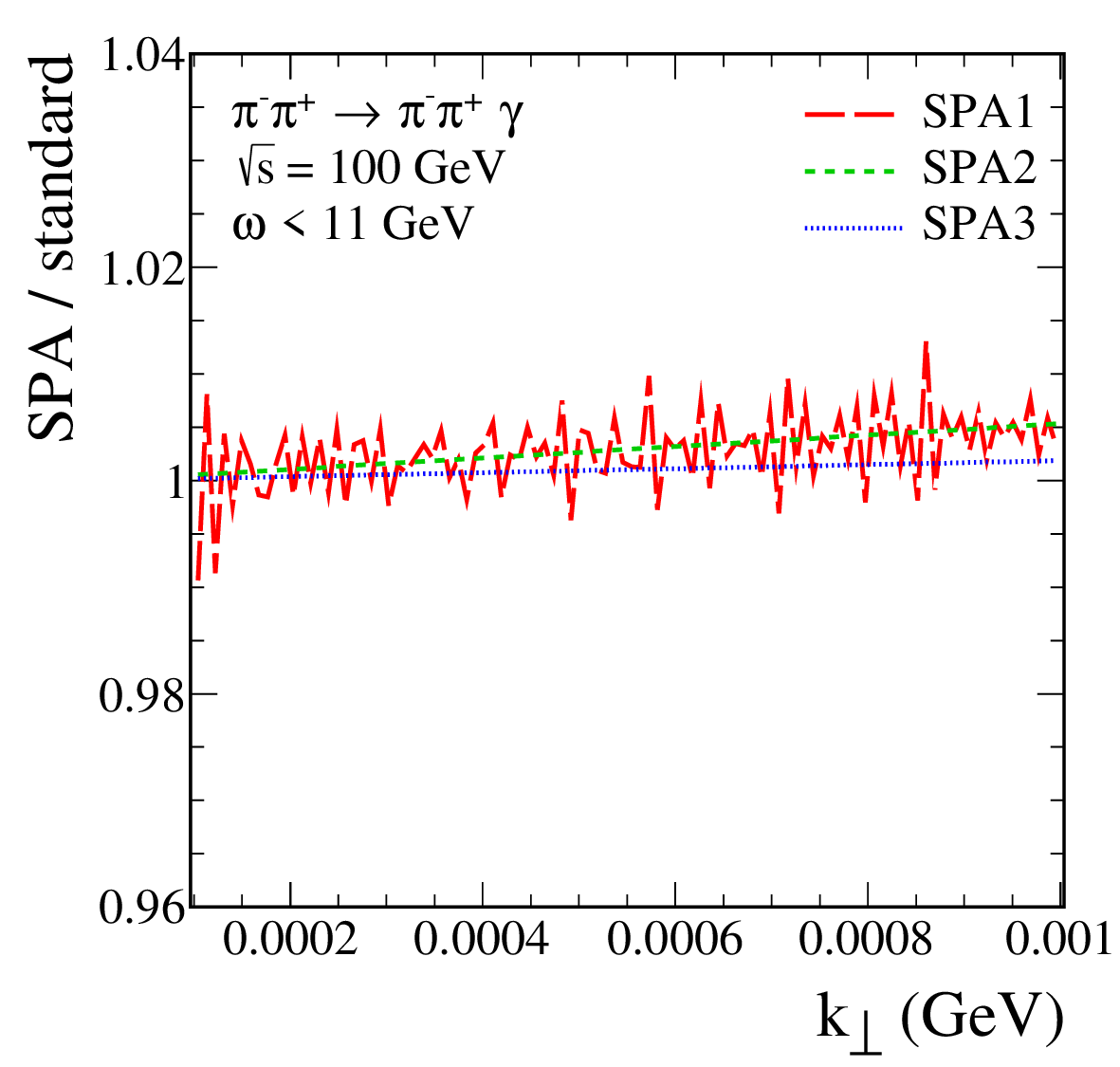}
\includegraphics[width=0.45\textwidth]{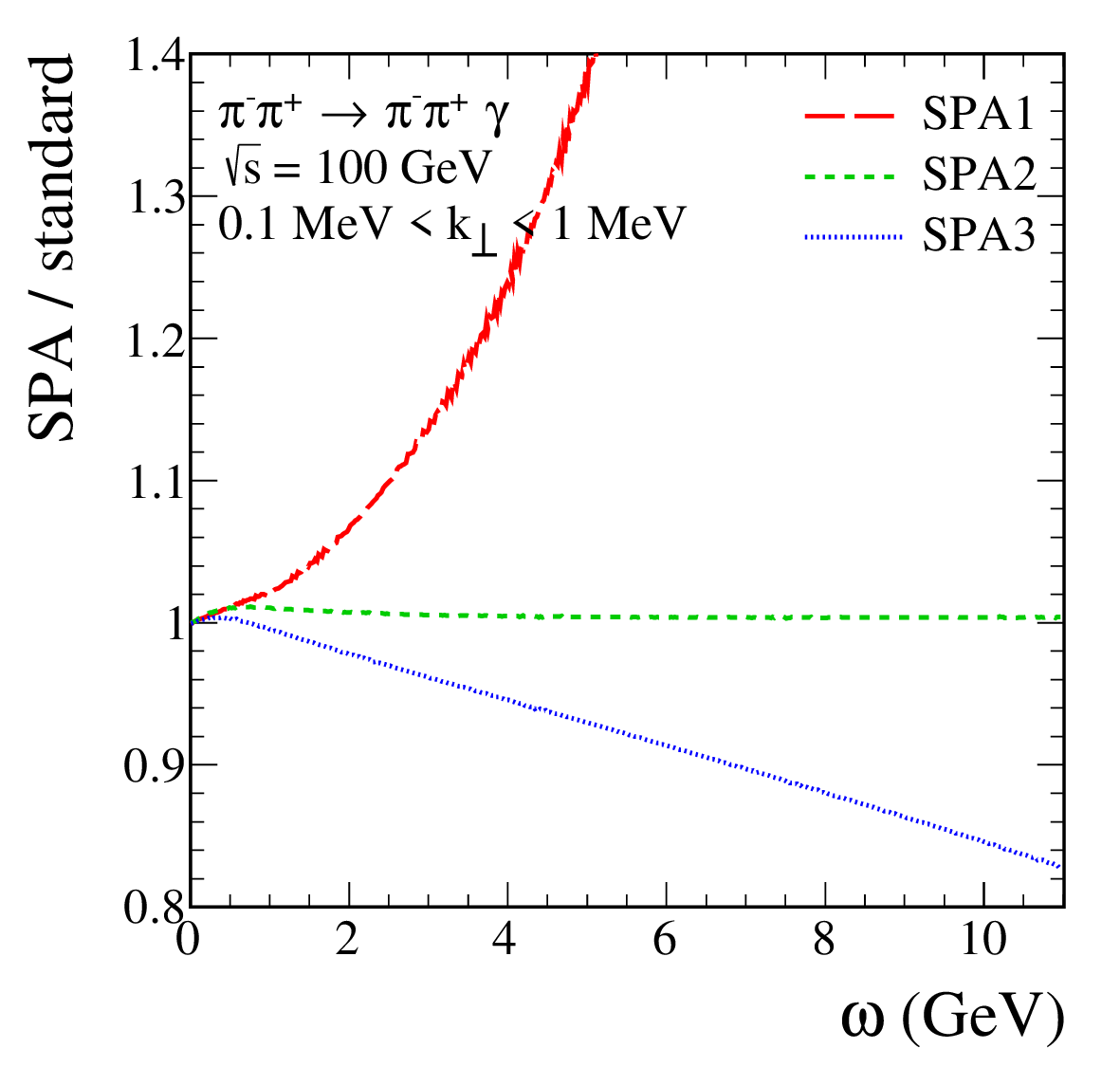}
\includegraphics[width=0.45\textwidth]{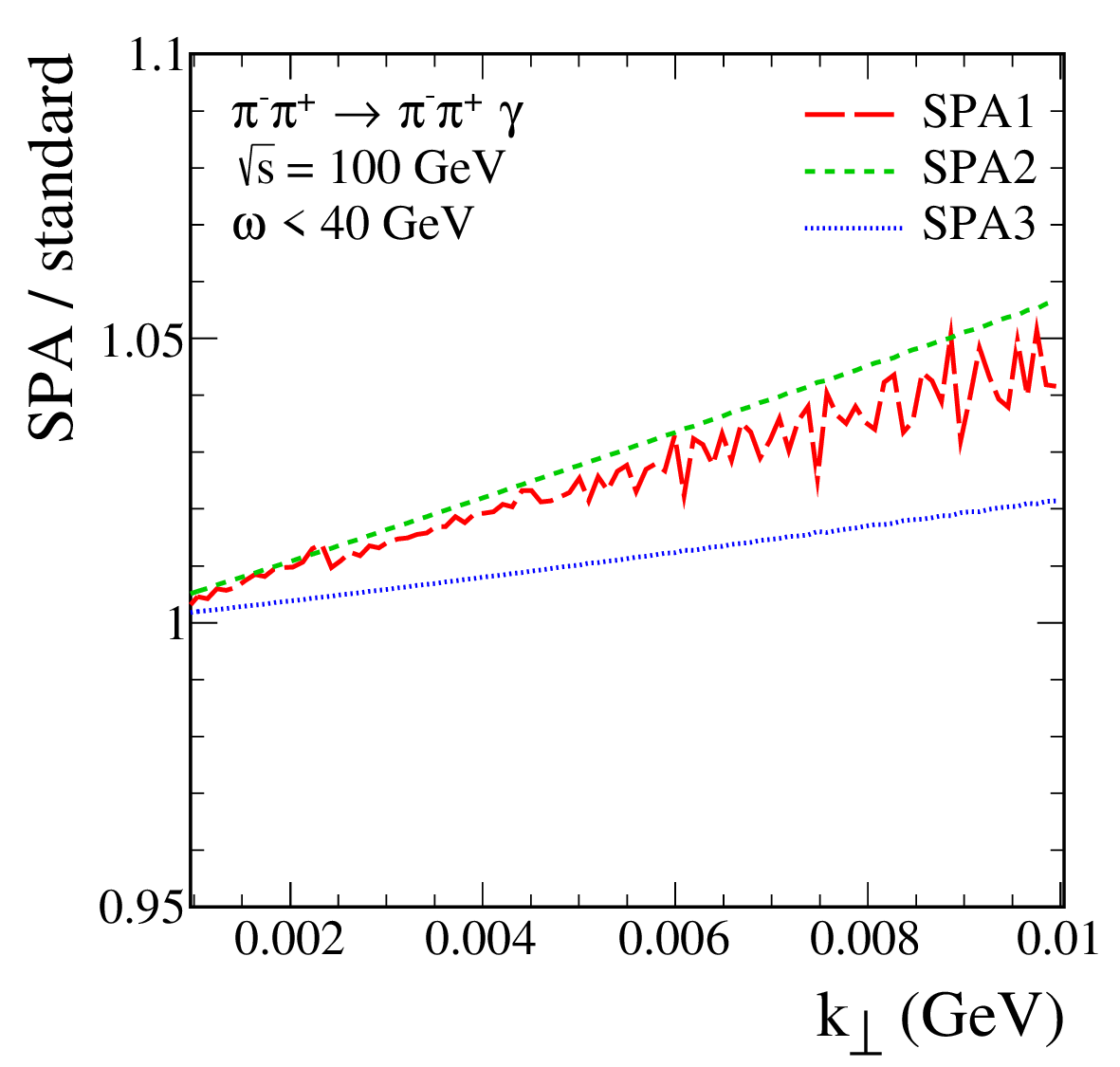}
\includegraphics[width=0.45\textwidth]{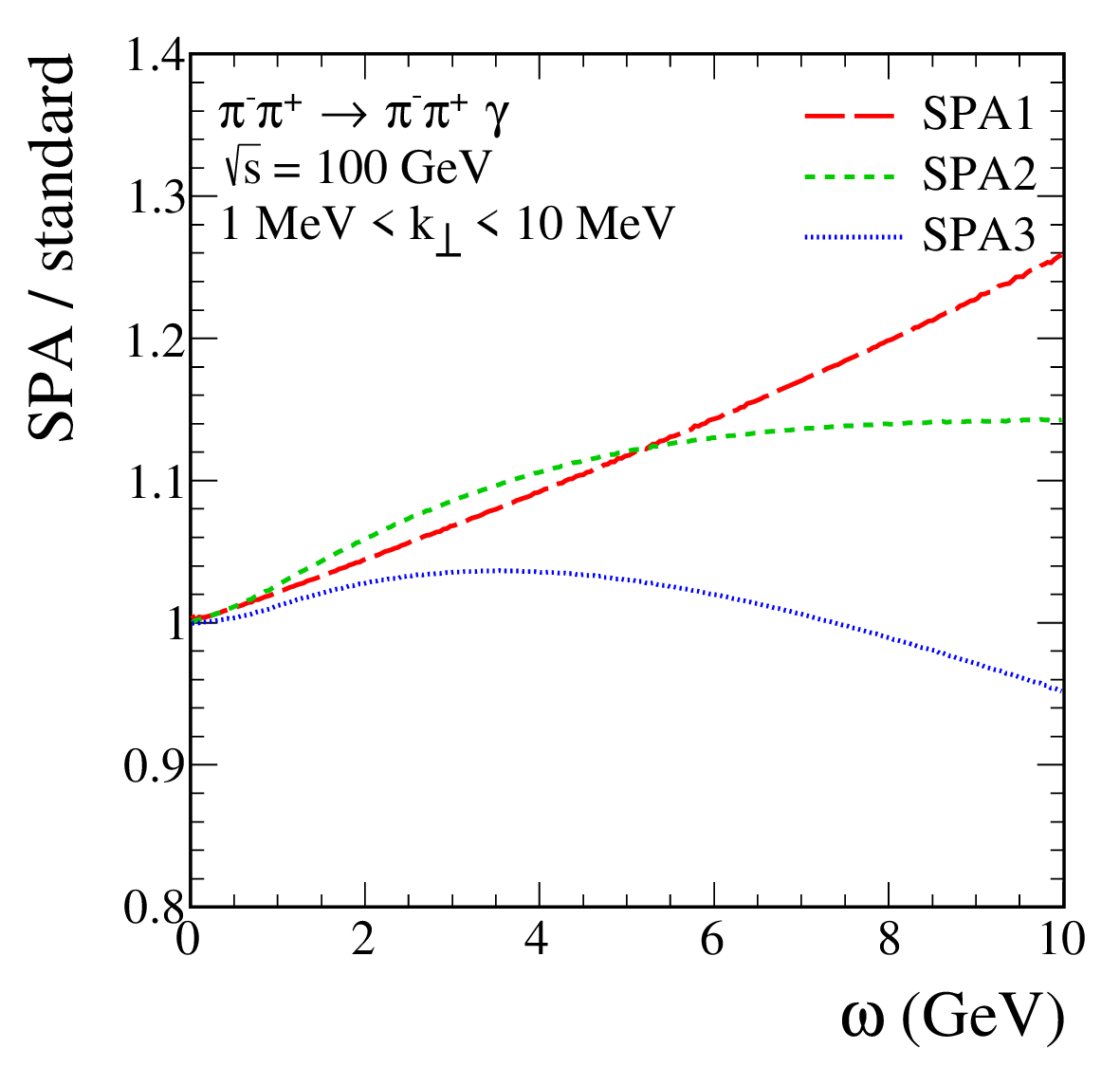}
\includegraphics[width=0.45\textwidth]{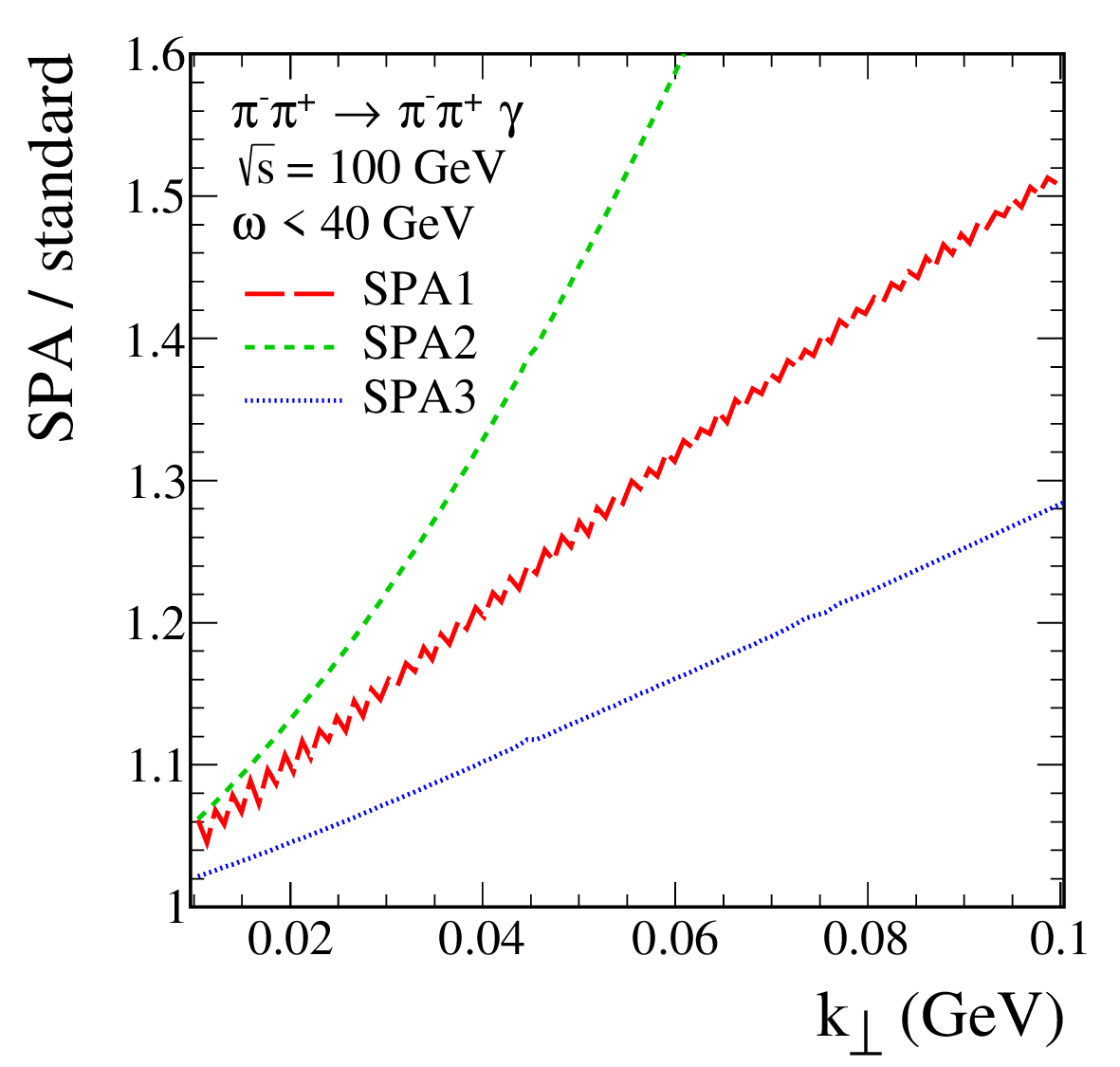}
\includegraphics[width=0.45\textwidth]{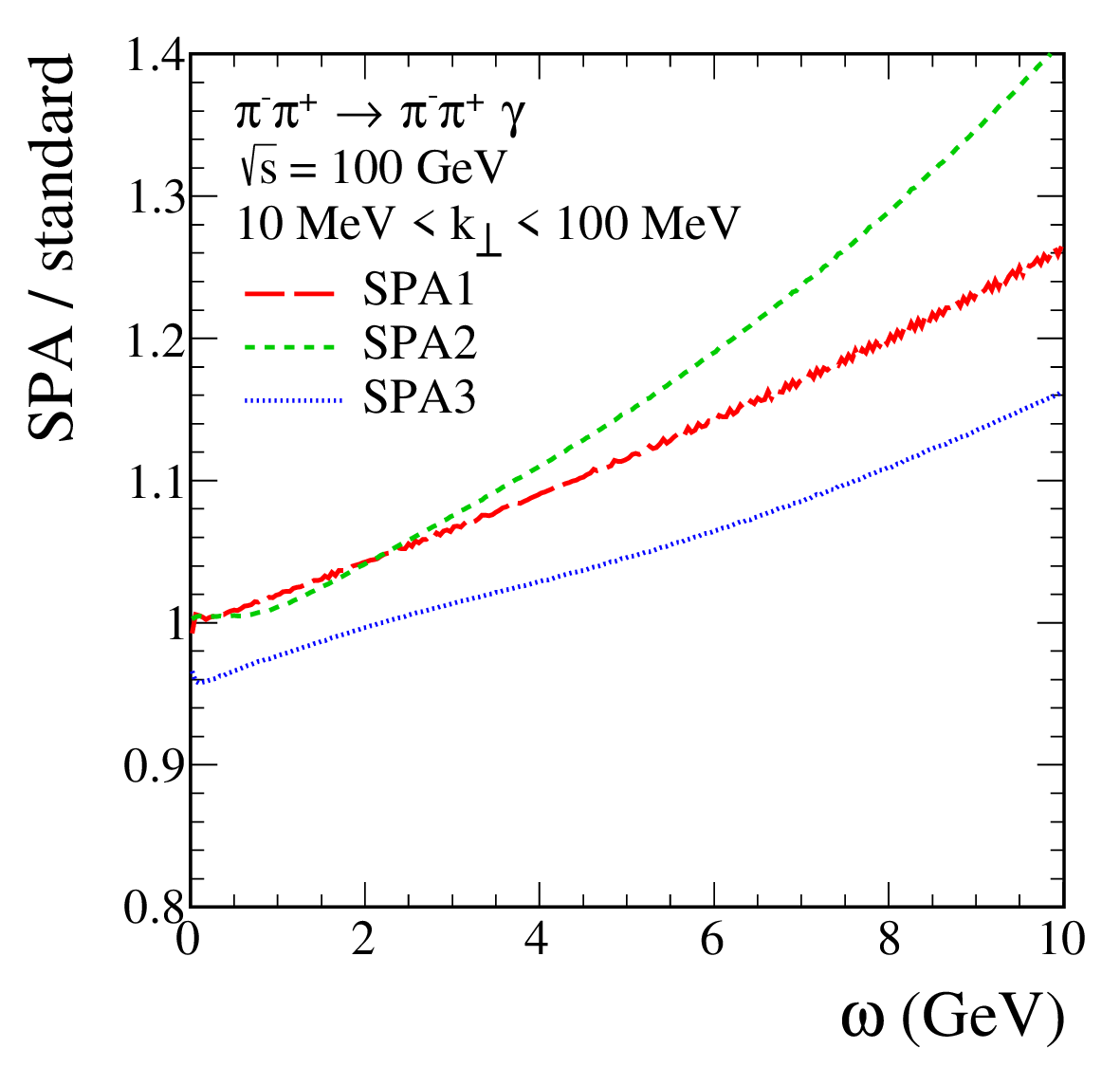}
\caption{\label{fig:ratios_100GeV}
\small
The same as in Fig.~\ref{fig:ratios_10GeV} but for $\sqrt{s} = 100$~GeV.
Shown are results for three $k_{\perp}$ intervals 
and with cuts on $\omega$ specified in the figure legends.
The oscillations in the SPA1 results 
are of numerical origin.}
\end{figure}

In Figs.~\ref{fig:100GeV} and \ref{fig:ratios_100GeV} 
we show the results for $\sqrt{s} = 100$~GeV which are analogs
of those shown in Figs.~\ref{fig:10GeV} and \ref{fig:ratios_10GeV}
for $\sqrt{s} = 10$~GeV. 
The calculations were done with cuts on $\omega$ specified
in the figure legends.
In all cases the constraint on $\omega$ from
(\ref{4.23d}) is well satisfied.
We see that at $\sqrt{s} = 100$~GeV
the three SPAs are all close to our standard results
in the region of small $k_{\perp}$ and $\omega$.
For $0.1\; {\rm MeV} < k_{\perp} < 1\; {\rm MeV}$
the SPA1 result deviates strongly from the standard result
for $\omega \gtrsim 4$~GeV;
see the upper most right panel of Fig.~\ref{fig:100GeV}.
This is due to the incorrect energy-momentum $\delta$ function
used, on purpose, there; see (\ref{4.38})--(\ref{4.41}).
Figure~\ref{fig:ratios_100GeV} shows 
that for $k_{\perp} \lesssim 10\; {\rm MeV}$
the deviations of the SPAs from the standard results are 
only at the percent level.
For the $\omega$ distributions these differences are up to around
10$\,\%$ for $\omega \lesssim 3$~GeV.

We also note that in Fig.~\ref{fig:ratios_100GeV}
the SPA results are in most cases above the standard results
(ratio > 1) but in some cases also below (ratio < 1).
Thus, the ratios SPA/standard depend strongly on the kinematics.

As for $\sqrt{s} = 10$~GeV, 
the rapid oscillations of the ratios for SPA1
in Fig.~\ref{fig:ratios_100GeV} are a numerical artefact
caused by different integration procedures in two different codes.

\begin{figure}[!ht]
\includegraphics[width=0.48\textwidth]{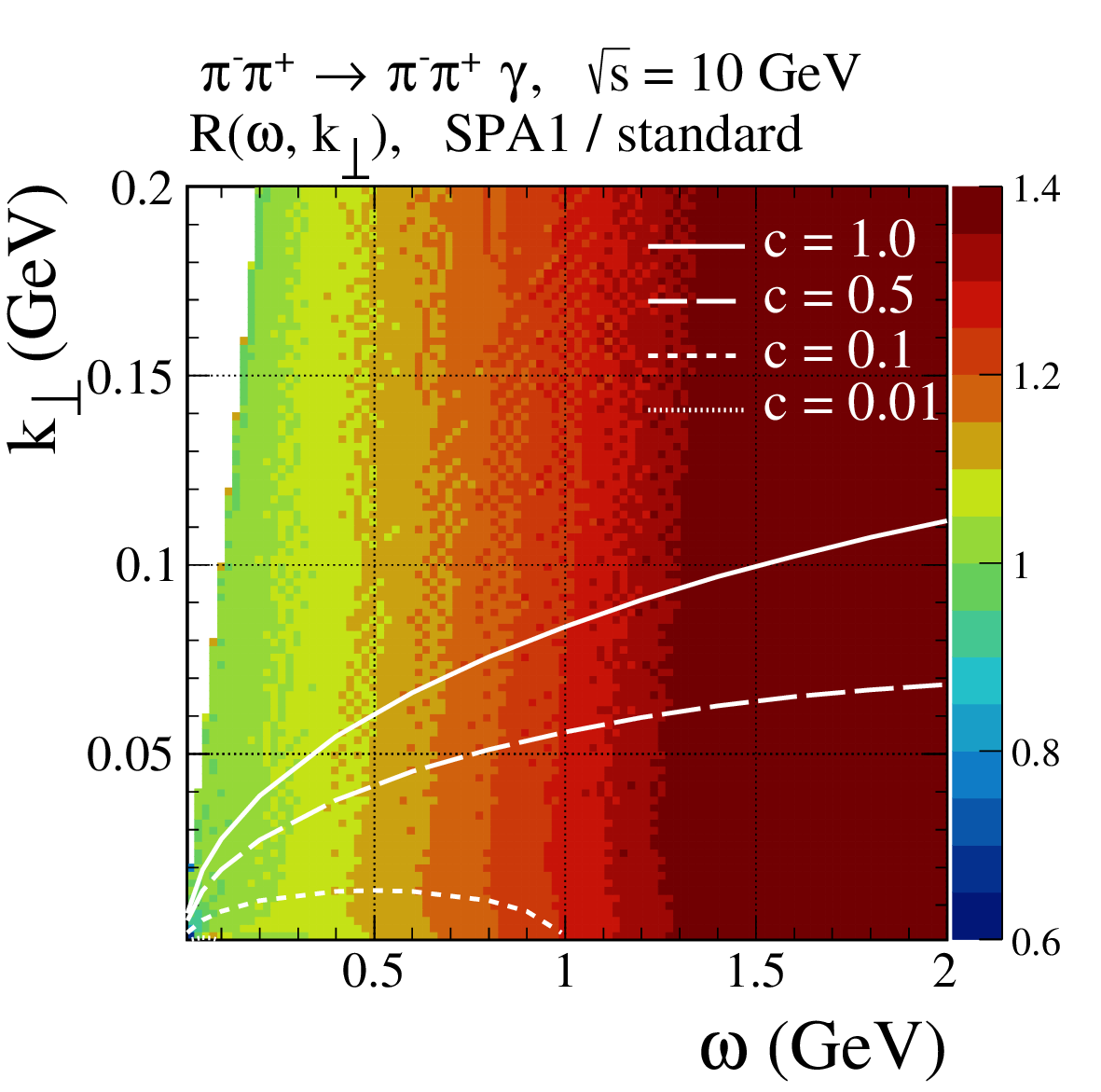}\\
\includegraphics[width=0.48\textwidth]{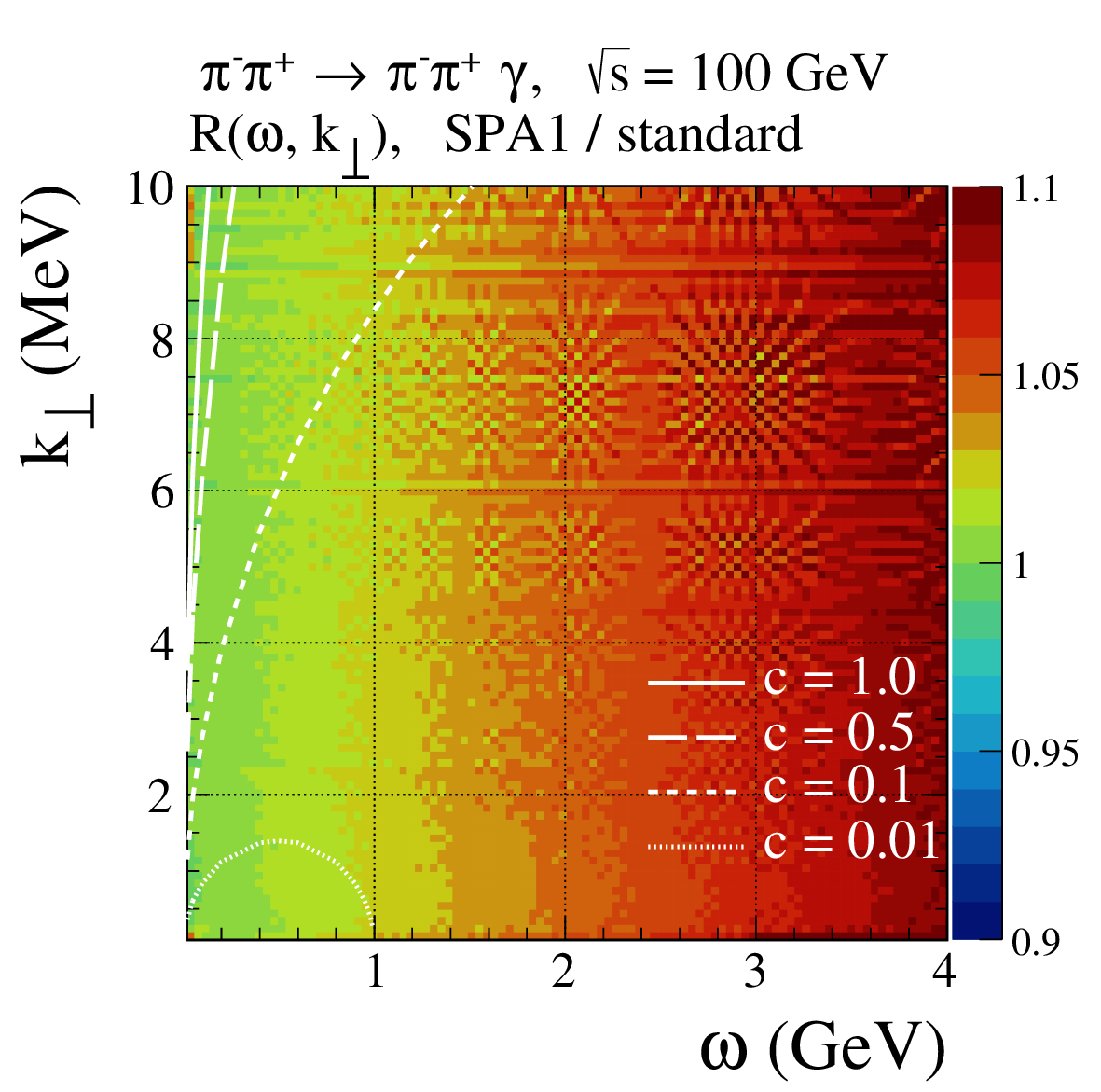}
\includegraphics[width=0.48\textwidth]{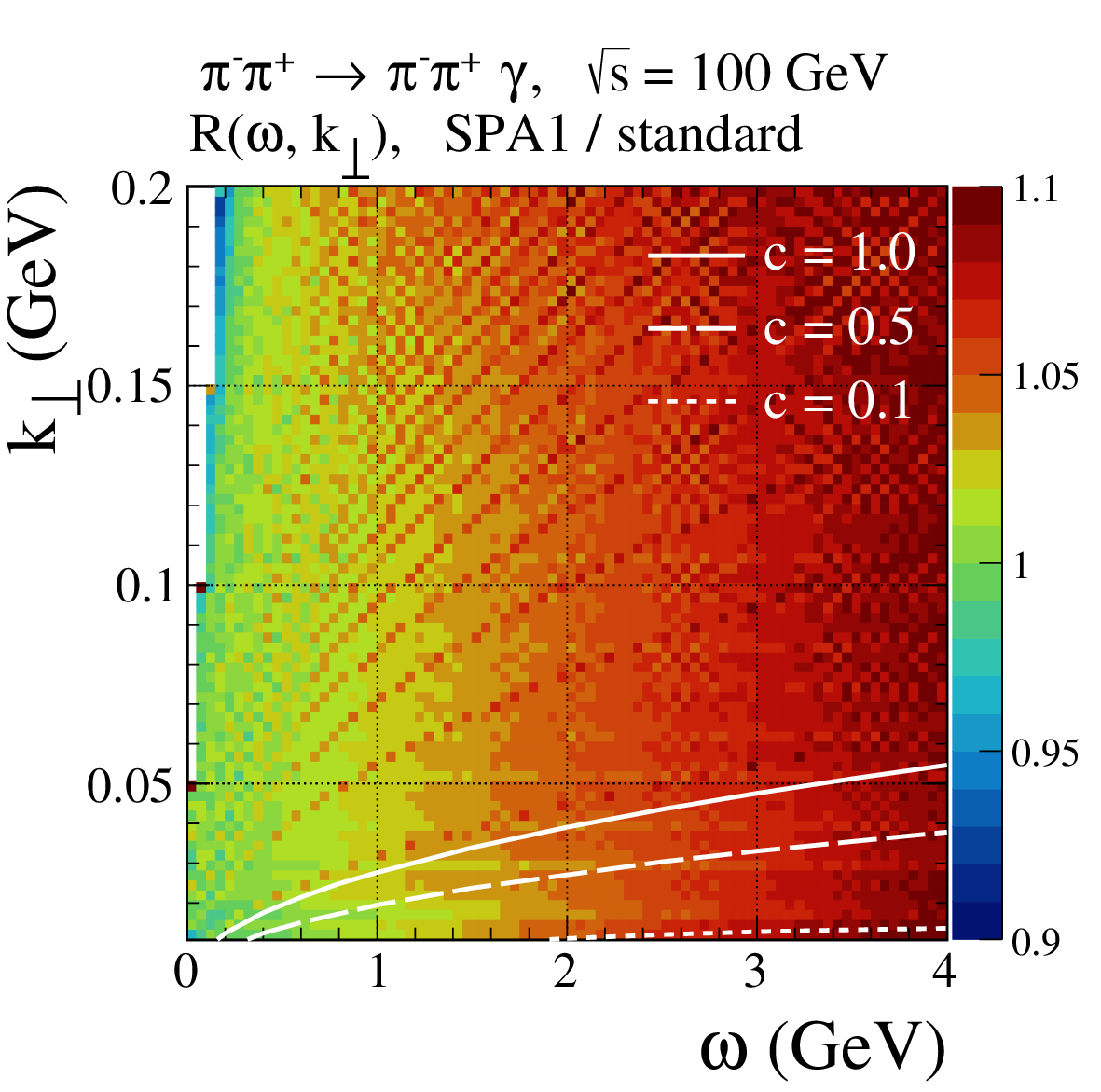}
\caption{\label{fig:ratios_GR}
\small
The ratio SPA1/standard (\ref{ratio})
for $\sqrt{s} = 10$~GeV (top panel) and for $\sqrt{s} = 100$~GeV (bottom panels).
The lines corresponding to $c = 1.0, 0.5, 0.1$, and 0.01 in 
(\ref{5.10}) are plotted.
All other curves in the bottom right panel are numerical artefacts.}
\end{figure}

At this point we can discuss the qualitative accuracy estimates
of the SPA1 approximation as given a long time ago in \cite{Gribov:1966hs}.
For the scattering reaction this estimate 
is given following Eq.~(19) of \cite{Gribov:1966hs}
and reads for our case
\begin{eqnarray}
&&p_{a} \cdot k \approx p_{1} \cdot k \ll m_{\pi}^{2}\,,
\label{5.6}\\
&&p_{b} \cdot k \approx p_{2} \cdot k \ll m_{\pi}^{2}\,.
\label{5.7}
\end{eqnarray}
In the c.m. system, choosing $\bpa$ in $z$ direction,
$p_{a} \cdot k$ ($p_{b} \cdot k$) can only become small
for the longitudinal component $k_{L}$ of $k$ being
positive (negative). 
We have then
\begin{eqnarray}
p_{a} \cdot k = p_{a}^{0} \omega - | \bpa | k_{L} 
&=& 
p_{a}^{0} \omega - | \bpa | \sqrt{\omega^{2} - k_{\perp}^{2}} \nonumber \\
&=&
\frac{2 m_{\pi}^{2} \omega^{2} + \frac{1}{2}(s - 4 m_{\pi}^{2}) \,k_{\perp}^{2}}
{\sqrt{s} \omega + \sqrt{s - 4 m_{\pi}^{2}} \sqrt{\omega^{2} - k_{\perp}^{2}}}
\quad {\rm for} \; k_{L} > 0 \,,
\label{5.8}
\end{eqnarray}
and
\begin{eqnarray}
p_{b} \cdot k =
\frac{2 m_{\pi}^{2} \omega^{2} + \frac{1}{2}(s - 4 m_{\pi}^{2}) \,k_{\perp}^{2}}
{\sqrt{s} \omega + \sqrt{s - 4 m_{\pi}^{2}} \sqrt{\omega^{2} - k_{\perp}^{2}}}
\quad {\rm for} \; k_{L} < 0 \,.
\label{5.9}
\end{eqnarray}
In Fig.~\ref{fig:ratios_GR} we show again the ratio
SPA1/standard in the $\omega$-$k_{\perp}$ plane together with the lines
\begin{eqnarray}
\frac{2 m_{\pi}^{2} \omega^{2} + \frac{1}{2}(s - 4 m_{\pi}^{2}) \,k_{\perp}^{2}}
{\sqrt{s} \omega + \sqrt{s - 4 m_{\pi}^{2}} \sqrt{\omega^{2} - k_{\perp}^{2}}}
= c \, m_{\pi}^{2} \,, \quad c = 1.0, \;0.5, \;0.1, \;0.01 \,.
\label{5.10}
\end{eqnarray}
%

If the accuracy estimate (\ref{5.6}), (\ref{5.7}) would be valid,
say with $p_{a} \cdot k \leqslant c m_{\pi}^{2}$
and $p_{b} \cdot k \leqslant c m_{\pi}^{2}$ and 
$c = 0.1$ or 0.01, the area below the corresponding curves 
in Fig.~\ref{fig:ratios_GR} should be coloured green,
indicating that there the SPA1 is a good representation
of our standard result. But we see from Fig.~\ref{fig:ratios_GR}
that the green areas, which we obtained from explicit calculations,
seem to have only very little to do with the qualitative conditions
(\ref{5.6}), (\ref{5.7}).
The areas below the curves of constant $c$ have regions where the SPA1
is not a good representation of the standard results.
On the other hand, there are large green areas above these curves
where the SPA1 gives a good representation of the standard results.
Of course, all these statements refer to a comparison of SPA1
to our standard results and things could be different for possible
other calculations including, for instance, ``anomalous'' terms.

\begin{figure}[!ht]
(a)\includegraphics[width=0.45\textwidth]{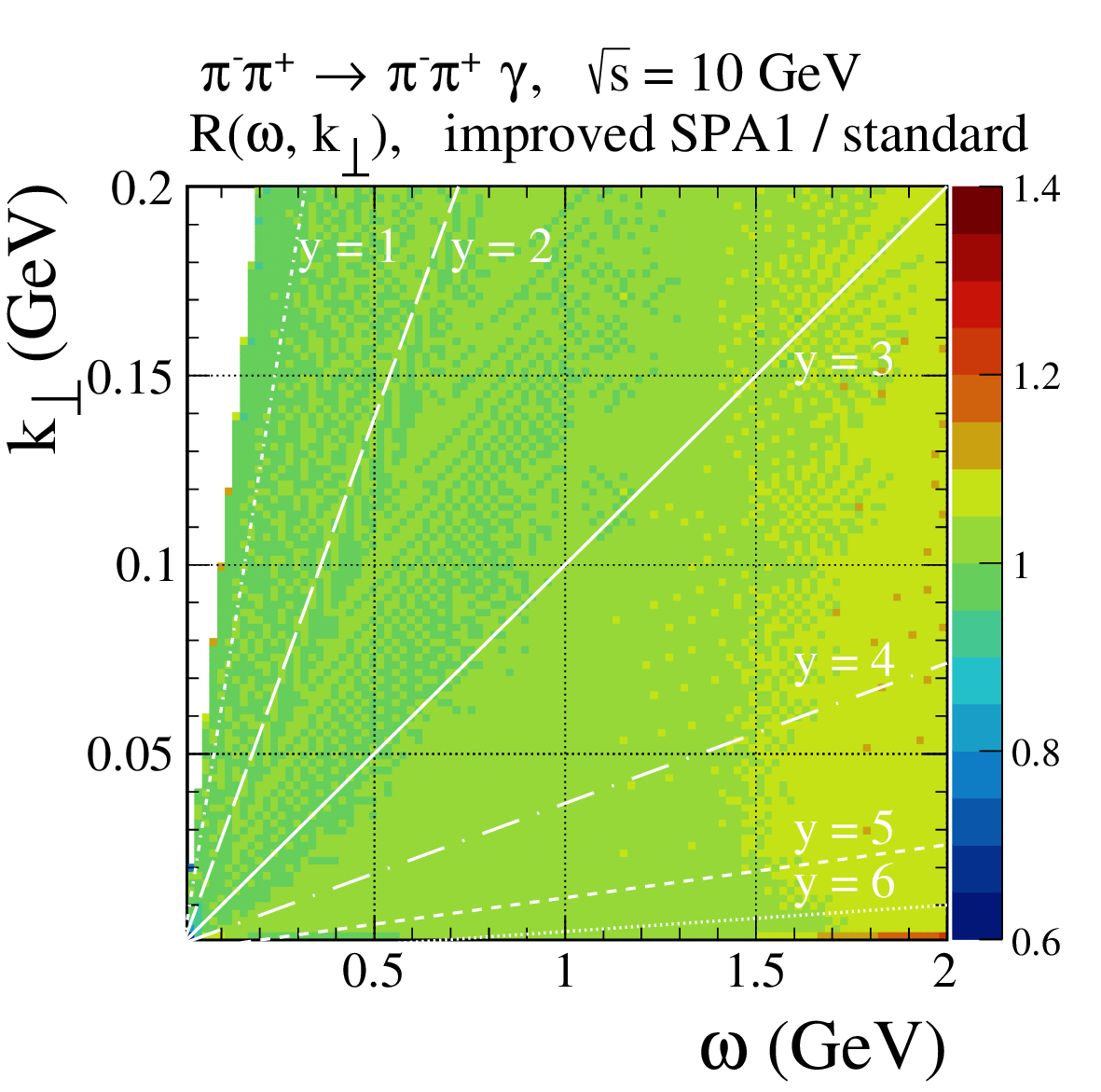}
(b)\includegraphics[width=0.45\textwidth]{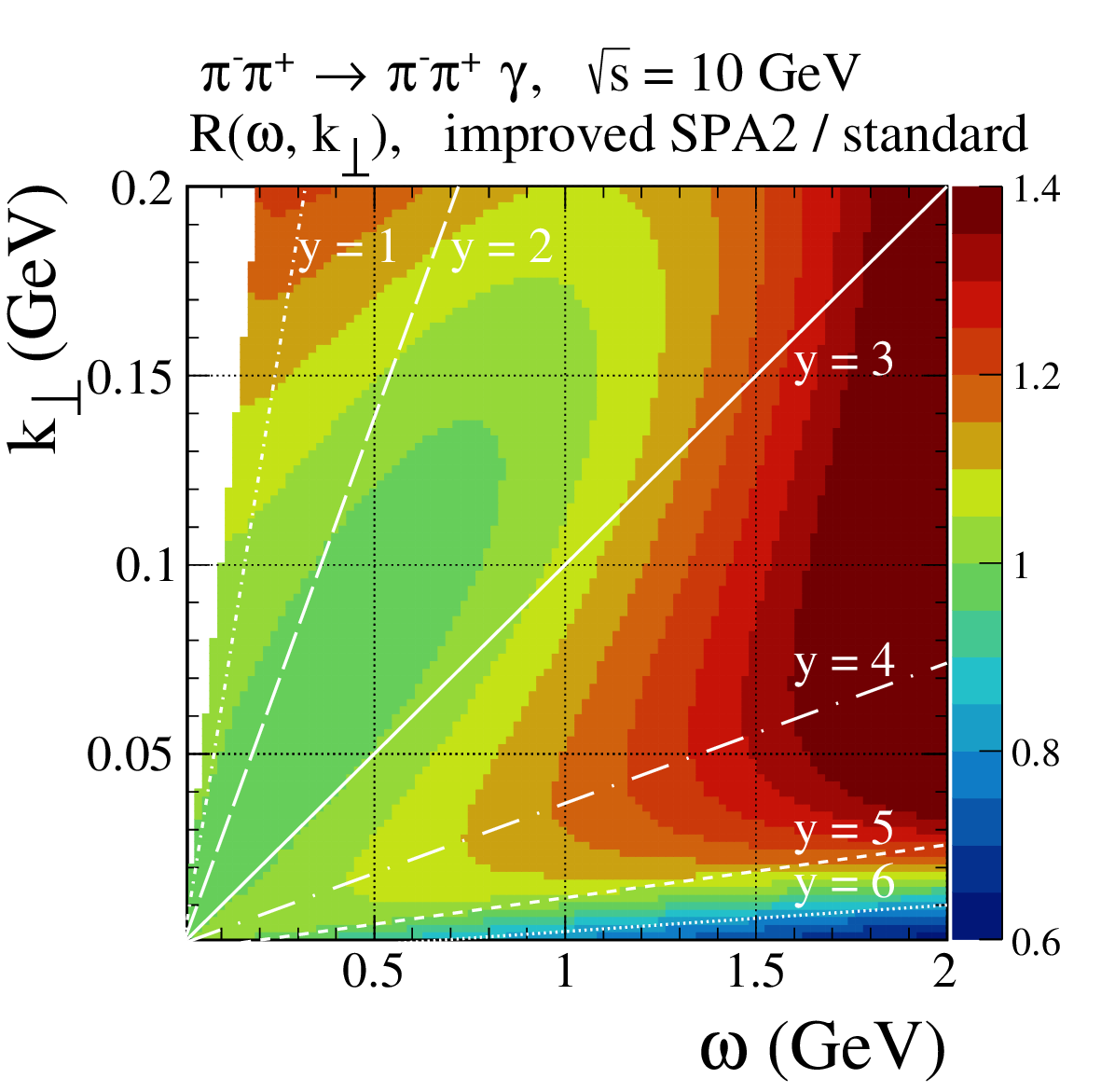}
(c)\includegraphics[width=0.45\textwidth]{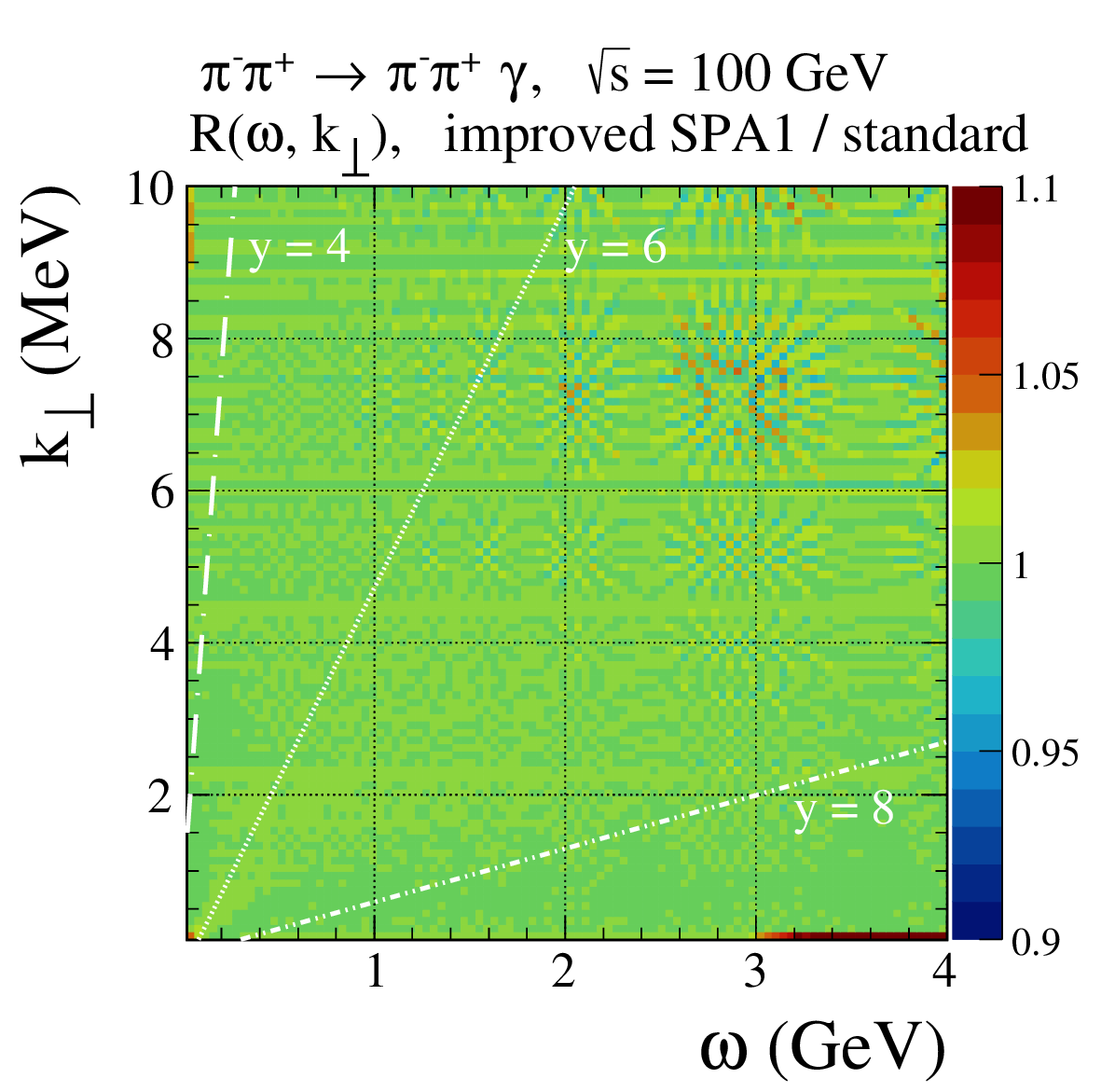}
(d)\includegraphics[width=0.45\textwidth]{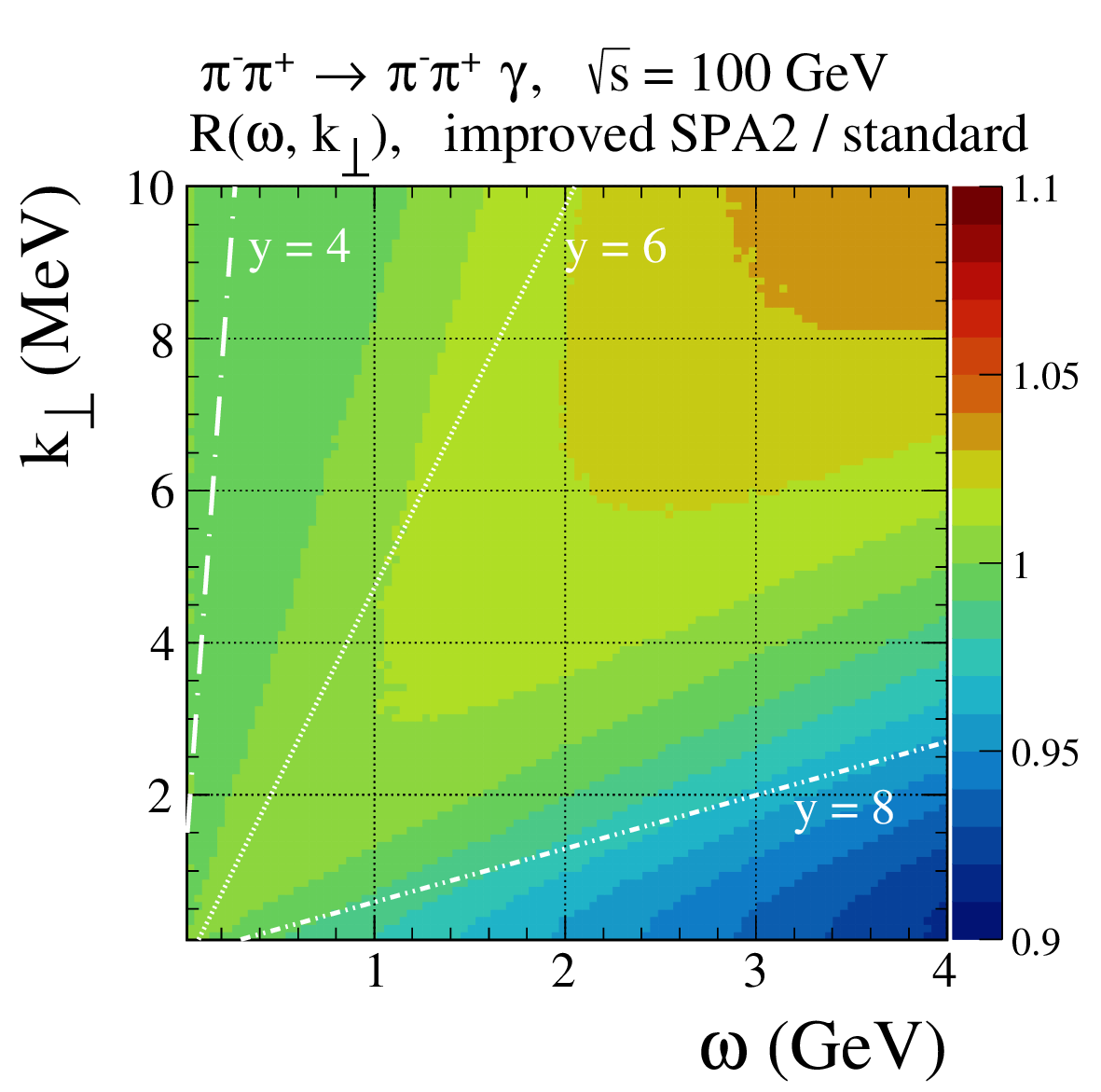}
(e)\includegraphics[width=0.45\textwidth]{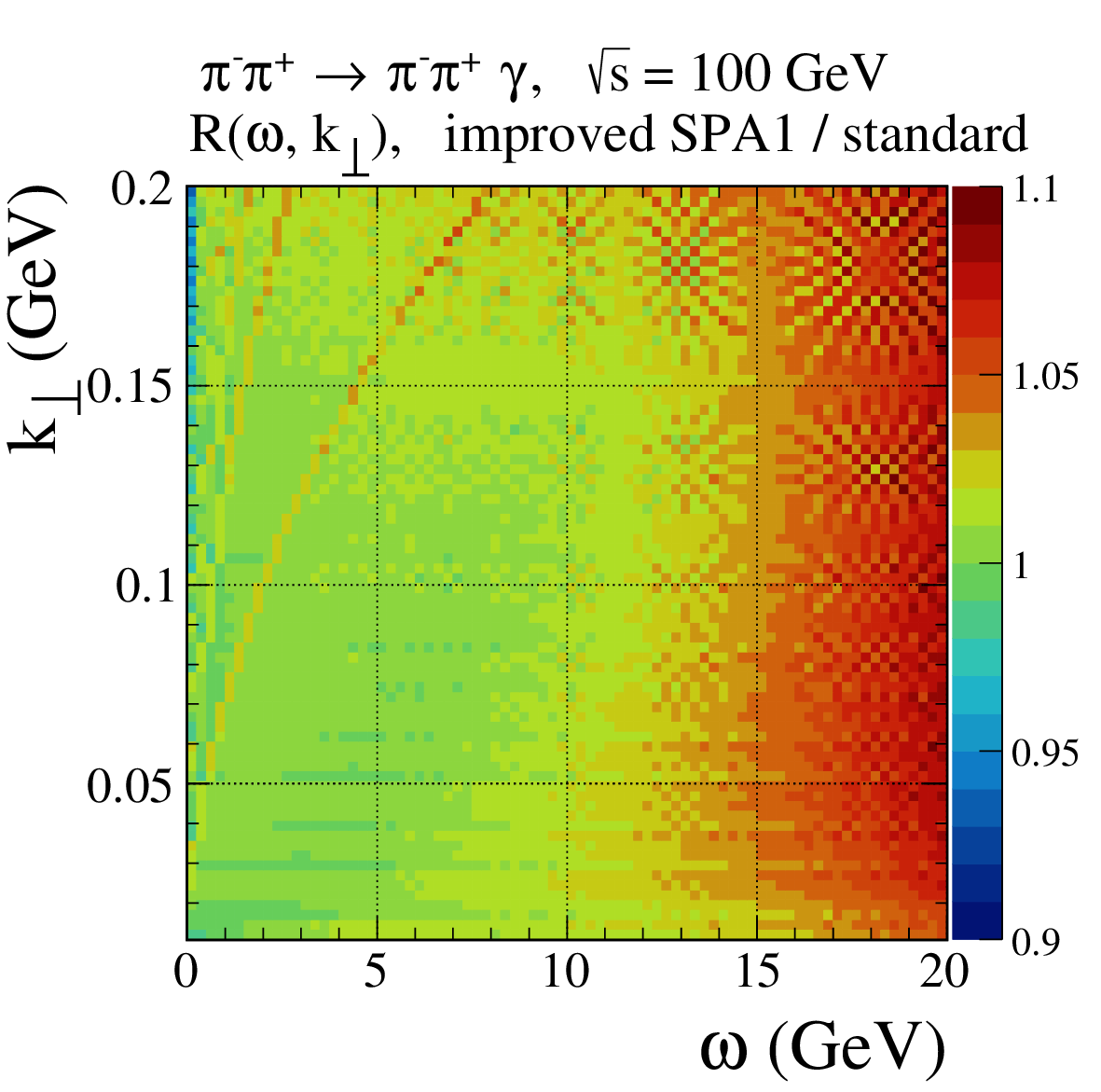}
\caption{\label{fig:2Dratio_SPAimproved}
\small
The ratios ${\rm R}(\omega,k_{\perp})$ (\ref{ratio}) 
for the $\pi^{-} \pi^{+} \to \pi^{-} \pi^{+} \gamma$ reaction 
for our ``improved SPA'' scenarios
for $\sqrt{s} = 10$~GeV and 100~GeV.}
\end{figure}

Figure~\ref{fig:2Dratio_SPAimproved} shows the ratios 
${\rm R}(\omega,k_{\perp})$
for the two ``improved SPA'' scenarios
calculated for $\sqrt{s} = 10$~GeV and 100~GeV.
Comparing these results to the corresponding 
results from Figs.~\ref{fig:ratio} and \ref{fig:2Dratio_100}
we observe that the ``improved SPAs''
greatly reduce the discrepancies to our standard results.
This is particularly the case for the improved SPA1 case.
One can see from Figs.~\ref{fig:2Dratio_SPAimproved}~(a) and (e)
that the improved SPA1 is a good approximation 
with the accuracy up to 10$\,\%$
for $k_{\perp} \lesssim 0.2$~GeV and $\omega \lesssim 2$~GeV
for $\sqrt{s} = 10$~GeV, and
for $k_{\perp} \lesssim 0.2$~GeV and $\omega \lesssim 20$~GeV
for $\sqrt{s} = 100$~GeV.

\section{Conclusions}
\label{sec:6}

In this paper we have studied elastic pion-pion scattering
without and with photon radiation.
In Sec.~\ref{sec:2} we~have given a detailed analysis,
from a QFT point of view, 
of the reactions $\pi^{-} \pi^{0} \to \pi^{-} \pi^{0}$
and $\pi^{-} \pi^{0} \to \pi^{-} \pi^{0} \gamma$.
We have used this analysis in Sec.~\ref{sec:3} to derive
the expansion of the amplitude for $\pi^{-} \pi^{0} \to \pi^{-} \pi^{0} \gamma$
in powers of~$\omega$, the photon energy in the overall center-of-mass system, 
for $\omega \to 0$.
The term of order $\omega^{-1}$ in the Laurent expansion (\ref{3.30}), specialized for $k^{2} = 0$,
agrees with the version of the soft-photon theorem
due to S.~Weinberg \cite{Weinberg:1965nx}.
We have given the term of order $\omega^{0}$
of this Laurent expansion.
This expansion (\ref{3.30}) should not be confused
with the expressions for the radiative amplitude given in F.E.~Low's version of the soft-photon theorem \cite{Low:1958sn}.
The latter gives an approximate expression for the radiative amplitude at a given phase-space point
and not an expansion of the amplitude around
the phase-space point of zero radiation as given in (\ref{3.30}).
All this is discussed in detail in \cite{Nachtmann_talk, Lebiedowicz:2023ell} where also the relation
of Low's formula, (1.7) of \cite{Low:1958sn},
and our formula (\ref{3.30}) is given.
We emphasize that our result (\ref{3.30}) is a strict
consequence of QFT. Therefore, absolutely no model dependence
is contained there.
As a non trivial check of this general result
we have considered the Laurent expansion
of our tensor-pomeron-model amplitude for ${\cal M}_{\lambda}$;
see (\ref{4.19}), (\ref{4.21})--(\ref{4.23}). 
The terms of order $\omega^{-1}$ and $\omega^{0}$
are found exactly as expected from (\ref{3.25}).

In Sec.~\ref{sec:4} we have calculated the amplitudes for
$\pi \pi \to \pi \pi$ and $\pi \pi \to \pi \pi \gamma$
in the tensor-pomeron model.
The diagrams for the latter process where the photon
is emitted from the external pion lines
[Fig.~\ref{fig:pipigam}~(a), (b), (d), (e)] are determined 
completely by the (off-shell) $\pi \pi \to \pi \pi$ scattering amplitude.
The amplitudes corresponding to the diagrams of Fig.~\ref{fig:pipigam}~(c)
and Fig.~\ref{fig:pipigam}~(f), the ``structure terms'',
have to satisfy gauge-invariance constraints involving
the previous amplitudes.
We have given a solution of these constraints which involves again
only the (off-shell) $\pi \pi \to \pi \pi$ scattering amplitude.
But we have emphasized that this solution is not unique
(as is well known in the literature, see e.g. \cite{Bern:2014vva})
and there ``anomalous'' terms in the $\pi \pi \to \pi \pi \gamma$
amplitudes, not directly related to the $\pi \pi \to \pi \pi$ amplitude,
could come up. We considered then as ``standard'', or ``exact'' model, 
our $\pi \pi \to \pi \pi \gamma$ amplitudes without 
such ``anomalous'' terms.
Clearly, in Sec.~\ref{sec:4} we used a model.
We summarize here our main model assumptions.
\begin{enumerate}
\item[(1)] We used the tensor-pomeron model, both for the on-shell
and off-shell $\pi \pi \to \pi \pi$ amplitudes.
For the high-energy reactions which are our main interest we needed
the effective pomeron propagator 
and the ${\rm pomeron} \,\pi \pi$ vertex.
These quantities were taken from \cite{Ewerz:2013kda}
where they were derived from comparison of theory to data,
in particular for nucleon-(anti)nucleon and pion-nucleon scattering.
\item[(2)] We used the standard pion propagator and $\gamma \pi \pi$ vertex;
see (\ref{4.11}).
Possible off-shell form factors in the pomeron- and photon-pion
vertices and the pion propagator are set to one.
\item[(3)] To determine the structure terms
[Figs.~\ref{fig:pimpi0_pimpi0gamma_diagrams}~(c),
\ref{fig:pipigam}~(c), \ref{fig:pipigam}~(f)]
we used the simplest solutions of the respective gauge-invariance relations;
see (\ref{4.18}), (\ref{4.19}), (\ref{4.30d}).
Throughout our paper we denote as ``anomalous'' terms
possible additional structure terms,
which then have to satisfy the gauge-invariance relations by themselves.
In our model we excluded such ``anomalous'' terms
in the radiative amplitudes.

We consider point (3) as our main model assumption.
\end{enumerate}
We have defined three soft-photon approximations
to our above ``exact model'': SPA1, SPA2, and SPA3;
see Sec.~\ref{sec:4.2}. 
In the SPA1 the photon momentum $k$ was, 
on purpose, omitted
in the energy-momentum conserving $\delta^{(4)}(.)$ 
function in the evaluation of the cross section.
In the SPA2 and SPA3 the correct energy-momentum conservation was required.

In Sec.~\ref{sec:5} we have presented quantitative calculations 
for the elastic $\pi \pi$ scattering without and with photon radiation 
within the tensor-pomeron model.
We have shown results for our ``standard model'' and for
the three SPAs for two different collision energies
$\sqrt{s} = 10$~GeV and 100~GeV.
We have shown, for instance, the results of our model
for the two-dimensional distributions in
photon transverse momentum $k_{\perp}$ and rapidity~${\rm y}$
[Figs.~\ref{fig:2dim_exact}~(c) and \ref{fig:2dim_exact_100}~(c)].
For $\sqrt{s} = 10$~GeV the distribution is largest for
$k_{\perp} \lesssim 0.1$~GeV and $2 \lesssim {\rm y} \lesssim 5$.
For $\sqrt{s} = 100$~GeV the distribution is largest for
$k_{\perp} \lesssim 0.004$~GeV and $3 \lesssim {\rm y} \lesssim 7$.
These are the results of our calculations in the framework
of our ``standard'' model where we have listed the assumptions
in (1), (2), (3) above.
We note that the distributions in $k_{\perp}$ and ${\rm y}$
give very small values for ${\rm y} \approx 0$, that is,
in the mid-rapidity region.
Clearly, this is then a region where ``anomalous'' contributions
to the radiative amplitude could be large compared to our ``standard''.
But we note that such anomalous contributions cannot come directly
from the high-energy exchange object, the pomeron~$\Pom$.
Charge-conjugation ($C$) invariance of the strong
and electromagnetic interactions forbids a $\Pom \Pom \gamma$ vertex.
The pomeron has $C = +1$ and the photon $C = -1$.
But in $\pi \pi$ scattering there can be central exclusive production (CEP)
of single photons by the fusion reactions
$\Pom + \rho_{\Reg} \to \gamma$ and $f_{2 \Reg} + \rho_{\Reg} \to \gamma$.
In the terminology used in our paper these would be called
``anomalous'' photon contributions even if their
origin is quite conventional. These photons, indeed, can be expected to populate preferentially the mid-rapidity region;
see~\cite{Lebiedowicz:2013xlb}.

Another main purpose of our paper was a study of the various
soft-photon approximations (SPAs).
How close or far away are they from our standard results?
As expected, the SPAs are good approximations to the standard
results for low $k_{\perp}$ and low $\omega$.
To be concrete: this means $k_{\perp} \lesssim 10$~MeV and $\omega \lesssim 50$~MeV
for $\sqrt{s} = 10$~GeV (see Fig.~\ref{fig:ratio})
and $k_{\perp} \lesssim 10$~MeV and $\omega \lesssim 0.5$~GeV
for $\sqrt{s} = 100$~GeV (see Fig.~\ref{fig:2Dratio_100}). 
For larger values of $k_{\perp}$ and/or $\omega$
the discrepancies between the standard and SPA results 
increase rapidly.
But these discrepancies also depend on the detailed kinematics.
The ``improved SPA'' approaches
with the variable $\tilde{s}$, defined in (\ref{4.73}), 
in the $\pi \pi \to \pi \pi$ amplitudes
greatly reduce the discrepancies to our standard results,
especially in the case of SPA1 
(see Fig.~\ref{fig:2Dratio_SPAimproved}).
For these numerical studies we have considered only the leading
exchange at high energies, the pomeron.
This should be a very good approximation for $\sqrt{s} = 100$~GeV.
For $\sqrt{s} = 10$~GeV we have also considered the subleading
reggeon exchanges and we found that they increase the cross sections
for $\pi^{-} \pi^{+} \to \pi^{-} \pi^{+} \gamma$
by about 20$\,\%$ to 40$\,\%$.

As already mentioned in the Introduction there are
plans for a new detector for the LHC, ALICE~3.
One physics aim for this new initiative is an experimental
study of soft-photon emission in hadronic reactions.
What can we say in this context from
our investigation of $\pi \pi$ scattering 
without and with photon radiation?
From the theory side we have a good model for the basic process
$\pi \pi \to \pi \pi$.
This allowed us to construct our standard amplitude
for $\pi \pi \to \pi \pi \gamma$ but we have excluded
anomalous terms, as described above.
Suppose now that we have experimental measurements
at all photon energies $\omega$.
Then we could study, as an example, the ratio
\begin{eqnarray}
R_{\rm exp}(\omega) = 
\frac{d\sigma_{\rm exp}/d\omega}{d\sigma_{\rm standard}/d\omega}\,.
\label{6.1}
\end{eqnarray}
From the results of our present paper we know that the terms
of order $1/\omega$ and $\omega^{0}$
in the expansion of the standard amplitude are strict
results from QFT without approximations,
given the on-shell $\pi\pi \to \pi\pi$ amplitudes.
Therefore, if QFT describes experiment we must have
(see Appendix~\ref{sec:appendixB} for a detailed discussion)
\begin{eqnarray}
\lim_{\omega \to 0} R_{\rm exp}(\omega) = 1\,, \quad
\lim_{\omega \to 0} \frac{dR_{\rm exp}(\omega)}{d\omega} = 0\,.
\label{6.2}
\end{eqnarray}
A violation of these relations would mean a terrible crisis for QFT!
For higher $\omega$ a value $R_{\rm exp}(\omega) \neq 1$ would mean
that there are soft photons from ``anomalous'' terms
(in the sense defined above) present in experiment.
From our point of view the origin of such ``anomalous'' terms
should be searched for in nonperturbative QCD.
One will have to first consider carefully all conventional
sources of ``anomalous'' photons like
photons from central exclusive production reactions (see above)
and then more unconventional sources;
see for instance \cite{Wong:2014pY} and \cite{Nachtmann:1983uz,Botz:1994bg,
Nachtmann:ELFE,
Nachtmann:Lectures,Nachtmann:2014qta}.
Let us note that for very small $\omega$ one has to take care
of infrared divergences and multiple soft photon emission.
But these effects can be calculated with the methods originally
developed by Bloch and Nordsieck \cite{Bloch:1937pw}.

What can we do if we do not have a good model for the amplitude
of the basic process, e.g. for multi-particle production?
Typically one has then the experimental or theoretical
distributions of particles and one uses the analog
of our SPA1 approximation (\ref{4.38})--(\ref{4.41})
instead of $d\sigma_{\rm standard}/d\omega$ in (\ref{6.1}):
\begin{eqnarray}
\tilde{R}_{\rm exp}(\omega) = 
\frac{d\sigma_{\rm exp}/d\omega}{d\sigma_{\rm SPA1}/d\omega}\,.
\label{6.3}
\end{eqnarray}
Then, the firm prediction from QFT is only
\begin{eqnarray}
\lim_{\omega \to 0} \tilde{R}_{\rm exp}(\omega) = 1\,.
\label{6.4}
\end{eqnarray}
Note that the ratios $R(\omega)$ for SPA1 shown in the right panels
of Fig.~\ref{fig:ratios_10GeV} and Fig.~\ref{fig:ratios_100GeV}
do \underline{not} satisfy
\begin{eqnarray}
\lim_{\omega \to 0} \frac{dR(\omega)}{d\omega} = 0\,,
\label{6.5}
\end{eqnarray}
and this must be expected to be the case in general.
If then $\tilde{R}_{\rm exp}(\omega)$ turns out $\neq 1$
for larger $\omega$ the conclusions for ``anomalous'' terms
in the photon-emission process will not be so straightforward,
since it will depend on an estimate of the accuracy
of the SPA used.
For our $\pi \pi$ scattering reaction these accuracies
can be read off, as function of the kinematic region considered,
from the figures shown in Sec.~\ref{sec:5}.
But, in general, such accuracy estimates are a difficult task.

In the future we plan to study proton-proton
elastic scattering and central exclusive production (CEP) reactions
like $pp \to p \pi^{+} \pi^{-} p$ without and with
soft photon production using the methods which we have developed here
for the $\pi \pi$ scattering case. We hope that with
the planned ALICE~3 detector at the LHC our theoretical studies
of soft photon emission in exclusive reactions
will find their experimental counterparts.
The goals will be to establish if QFT has a crisis there
in the sense of a violation of relations of the type (\ref{6.2})
and if ``anomalous'' soft photons, compatible with QFT, are present.

\acknowledgments

We thank Johanna Stachel, Peter Braun-Munzinger, Carlo Ewerz,
and Stefan Fl{\"o}rchinger for very useful
discussions and for providing us information on relevant
literature.
We thank Charles Gale and Heinrich Leutwyler 
for correspondence on the topics of our paper.
This work is partially supported by
the Polish National Science Centre under Grant
No. 2018/31/B/ST2/03537
and by the Center for Innovation and Transfer of Natural Sciences 
and Engineering Knowledge in Rzesz\'ow (Poland).

\appendix

\section{Some remarks on the literature concerning Low's theorem}
\label{sec:appendixA}

Here we compare our findings concerning the Laurent series
for the soft-photon expansion from Sec.~\ref{sec:3}
to results from a number of papers from the literature.
We find it surprising that so many versions of 
``Low's theorem'' can be found in the literature.
This, clearly, poses the question if they are all equivalent.
A genuine theorem of QFT should give a unique result.
We think that a clarification of this question
is important especially for experimentalist trying
to check this theorem. They should know precisely
what they are supposed to check.
In this spirit, as a service to experimentalists,
and in order to answer to various points raised 
by the referee of our paper,
we shall in the following compare results from the literature
to our findings.

For these comparisons we shall mainly restrict
ourselves to the simple pion-pion scattering reactions
(\ref{2.1}) and (\ref{2.2}).

We start by recalling our Eqs.~(\ref{3.25}) and (\ref{3.26}) which
we specialise here for real photons, $k^{2} = 0$.
We get then for $\pi^{-} \pi^{0} \to \pi^{-} \pi^{0} \gamma$,
dropping gauge terms proportional to~$k_{\lambda}$,
\begin{eqnarray}
{\cal M}_{\lambda}
&=& e 
\big[ \frac{p_{a \lambda}}{(p_{a} \cdot k)} -\frac{p_{1 \lambda}}{(p_{1} \cdot k)} \big]
{\cal M}^{(0)}(s_{L},t,m_{\pi}^{2},m_{\pi}^{2},m_{\pi}^{2},m_{\pi}^{2})
\nonumber \\
&&+ e 
\big\lbrace 
-\frac{1}{(p_{1} \cdot k)^{2}}
\big[ p_{1 \lambda} (l_{1} \cdot k) - l_{1 \lambda} (p_{1} \cdot k)  \big]
+ 2 \big[ - \frac{(p_{b} \cdot k)}{(p_{a} \cdot k)} p_{a\lambda} + p_{b\lambda} \big]
\frac{\partial}{\partial s_{L}}  \nonumber \\
&&
- 2 \big[ (p_{a} - p_{1},k) - (p_{a} \cdot l_{1}) \big] 
\big[ \frac{p_{a\lambda}}{(p_{a} \cdot k)} - \frac{p_{1\lambda}}{(p_{1} \cdot k)} \big] 
\frac{\partial}{\partial t} 
\big\rbrace
{\cal M}^{(0)}(s_{L},t,m_{\pi}^{2},m_{\pi}^{2},m_{\pi}^{2},m_{\pi}^{2}) \nonumber \\
&&
+ {\cal O}(\omega)\,.
\label{A1}
\end{eqnarray}

Now we have a look at Gribov's paper \cite{Gribov:1966hs}.
As far as we can see the emphasis there is on the question
of determining the kinematic region of validity of the $\omega^{-1}$ term
in Low's theorem.
The $\omega^{0}$ term is only mentioned in the context
of the cancellation of the off-mass-shell effects.
This happens also in our calculation
when adding the amplitudes ${\cal M}_{\lambda}^{(a)}$,
${\cal M}_{\lambda}^{(b)}$, and ${\cal M}_{\lambda}^{(c)}$ in (\ref{3.25}).
The question where the $\omega^{-1}$ term gives a reliable result
is discussed in detail in our Sec.~\ref{sec:5} 
where we also compare to \cite{Gribov:1966hs}.

Next we study Lipatov's paper \cite{Lipatov:1988ii}.
From the many interesting considerations presented in
this paper we are only concerned with the form of Low's theorem
given there for the photon case.
From Eq.~(11) of \cite{Lipatov:1988ii} 
we get for our process (\ref{2.2}), using our notation,
\begin{eqnarray}
\left.{\cal M}_{\lambda}\right|_{\rm Lipatov}
&=& e 
\big[ \frac{p_{a \lambda}}{(p_{a} \cdot k)} -\frac{p'_{1 \lambda}}{(p'_{1} \cdot k)} \big]
{\cal M}^{(0)}(s_{L},t,m_{\pi}^{2},m_{\pi}^{2},m_{\pi}^{2},m_{\pi}^{2})
\nonumber \\
&&- e  (p_{a} - p_{1},k) 
\big[ \frac{p_{a\lambda}}{(p_{a} \cdot k)} - \frac{p_{1\lambda}}{(p_{1} \cdot k)} \big] 
\frac{\partial}{\partial t} 
{\cal M}^{(0)}(s_{L},t,m_{\pi}^{2},m_{\pi}^{2},m_{\pi}^{2},m_{\pi}^{2})  \nonumber \\
&&
+ {\cal O}(\omega)\,.
\label{A2}
\end{eqnarray}
This result clearly is different from our Eq.~(\ref{A1}).
There is in (\ref{A2}) no term $\partial{\cal M}^{(0)}/\partial s_{L}$
and the term proportional to $\partial{\cal M}^{(0)}/\partial t$
is different from our result.

Concerning the paper \cite{DelDuca:1990gz} we see no overlap and,
therefore, no conflict with our results.
In \cite{DelDuca:1990gz} \underline{hard} processes
with photon emission are considered.
Let $Q$ be the scale of a hard process.
Then, including photon radiation,
the region of $\omega$ mainly discussed in \cite{DelDuca:1990gz}
is $m^{2}/Q \leq \omega \lesssim m$, with $m$ being some mass scale.
But we are considering a \underline{soft} process and
we are interested in the strict limit $\omega \to 0$.

Concerning the papers which we want to discuss next
we would like to make a general remark.
In many papers the results for the soft-photon expansion
of the amplitude, say for $\pi^{-} \pi^{0} \to \pi^{-} \pi^{0} \gamma$,
contains derivatives with respect to the momenta of the basic amplitude,
here for $\pi^{-} \pi^{0} \to \pi^{-} \pi^{0}$.
Let us consider as in (\ref{2.8})
the, in general off-shell, amplitude
\begin{eqnarray}
\widetilde{\cal T}(p_{a}, p_{b}, p_{1}, p_{2}) = 
\left.{\cal T}(p_{a}, p_{b}, p_{1}, p_{2})\right|_{\rm off\; shell\; or\; on\; shell}.
\label{A3}
\end{eqnarray}
In order to calculate derivatives like
$\partial \widetilde{\cal T} / \partial p_{a}^{\mu}$
we have to consider
\begin{eqnarray}
\widetilde{\cal T}(p_{a} + \delta p_{a}, p_{b}, p_{1}, p_{2}) = 
{\cal T}(p_{a}, p_{b}, p_{1}, p_{2}) +
\delta p_{a}^{\mu} \, 
\frac{\partial \widetilde{\cal T}}{\partial p_{a}^{\mu}}(p_{a}, p_{b}, p_{1}, p_{2})
+ {\cal O}((\delta p_{a})^{2})\,. \qquad
\label{A4}
\end{eqnarray}
But clearly, with (\ref{A4}) we have to go outside the physical
region for $\widetilde{\cal T}$ which requires always
$p_{a} + p_{b} - p_{1} - p_{2} = 0$.
Thus, it is our opinion that all expressions for Low's theorem
which contain derivatives like
$\partial \widetilde{\cal T} / \partial p_{a}^{\mu}$ etc.
have to be considered as potentially problematic.

Keeping this in mind we shall now discuss the aspects
relevant for Low's theorem 
in the paper by H.~Gervais \cite{Gervais:2017yxv}.
In \cite{Gervais:2017yxv} the reactions considered
are fermion-scalar ($f$-$s$) elastic scattering and
the corresponding photon-emission process.
In our notation we have then
\begin{eqnarray}
&&f (p_{a}) + s (p_{b}) \to f (p_{1}) + s (p_{2})\,,
\label{A5} \\
&&f (p_{a}) + s (p_{b}) \to f (p_{1}') + s (p_{2}') 
+ \gamma (k, \epsilon)\,.
\label{A6}
\end{eqnarray}
The masses of $f$ and $s$ are $m_f$ and $m_s$.
Now it is correctly stated in \cite{Gervais:2017yxv}
that the momenta of $f$ and $s$
cannot all stay fixed when going from (\ref{A5}) to (\ref{A6});
see Eq.~(9) of \cite{Gervais:2017yxv}.
The four variables $\xi_{i}$, $\eta_{i}$ $(i = 1, 2)$ introduced there
correspond to our $l_{1,2}$ variables; see (\ref{3.14}).
We have only two such variables since we keep 
$p_{a}$ and $p_{b}$ the same in (\ref{A5}) and (\ref{A6})
which is, of course, legitimate.
Let $\widetilde{\cal T}(p_{a}, p_{b}, p_{1}, p_{2})$
be the elastic amplitude,
stripped from the spinors, and, in general,
for off-shell particles.
The relevant formula giving the terms of order $\omega^{-1}$ and
$\omega^{0}$ for the amplitude from (\ref{A6}) is then
Eq.~(20) of \cite{Gervais:2017yxv}. 
There, the following expressions, using our notation, occur:
\begin{eqnarray}
&&I_{1} = 
\big[ -l_{1}^{\alpha} \frac{\partial}{\partial p_{1}^{\alpha}}
       -l_{2}^{\alpha} \frac{\partial}{\partial p_{2}^{\alpha}}
       -k^{\alpha} \frac{\partial}{\partial p_{a}^{\alpha}}
\big]
\widetilde{\cal T}(p_{a}, p_{b}, p_{1}, p_{2}) \,,\nonumber \\
&&I_{2} = 
\big[ -l_{1}^{\alpha} \frac{\partial}{\partial p_{1}^{\alpha}}
       -l_{2}^{\alpha} \frac{\partial}{\partial p_{2}^{\alpha}}
       +k^{\alpha} \frac{\partial}{\partial p_{1}^{\alpha}}
\big]
\widetilde{\cal T}(p_{a}, p_{b}, p_{1}, p_{2}) \,,\nonumber \\
&&I_{3 \mu} = \frac{\partial}{\partial p_{a \mu}}
\widetilde{\cal T}(p_{a}, p_{b}, p_{1}, p_{2}) \,,\nonumber \\
&&I_{4 \mu} = \frac{\partial}{\partial p_{1 \mu}}
\widetilde{\cal T}(p_{a}, p_{b}, p_{1}, p_{2}) \,.
\label{A7}
\end{eqnarray}
We shall now choose a simple trial function for $\widetilde{\cal T}$:
\begin{eqnarray}
\widetilde{\cal T}(p_{a}, p_{b}, p_{1}, p_{2}) =
h \big[ p_{a}^{2} + p_{1}^{2} - p_{b}^{2} - p_{2}^{2} \big]\,.
\label{A8}
\end{eqnarray}
On shell we have
\begin{eqnarray}
\widetilde{\cal T}(p_{a}, p_{b}, p_{1}, p_{2})_{\rm on\; shell} &=&
h \big[ 2 m_f^{2} - 2m_s^{2} \big] \nonumber \\
&=& {\rm const}\,.
\label{A9}
\end{eqnarray}
We shall assume that $h(2 m_f^{2} - 2 m_s^{2}) \neq 0$ and 
that also the derivative
\begin{eqnarray}
h' \big[ 2 m_f^{2} - 2 m_s^{2} \big] \neq 0\,,
\label{A10}
\end{eqnarray}
but otherwise arbitrary.
Evaluating $I_{1}, \ldots, I_{4}$ from (\ref{A7})
we get, using (\ref{3.20}),
\begin{eqnarray}
I_{1} &=& 
[ -(l_{1} \cdot p_{1}) + (l_{2} \cdot p_{2}) - (k \cdot p_{a}) ]\,
2h'( 2 m_f^{2} - 2m_s^{2} ) \nonumber \\
&=& - (k \cdot p_{a}) 2h' ( 2 m_f^{2} - 2m_s^{2} )\,, \nonumber\\
I_{2} &=& (k \cdot p_{1}) 2h' ( 2 m_f^{2} - 2m_s^{2} )\,,\nonumber \\
I_{3 \mu} &=& 2 p_{a \mu} \,h' ( 2 m_f^{2} - 2m_s^{2} )\,,\nonumber \\
I_{4 \mu} &=& 2 p_{1 \mu} \,h' ( 2 m_f^{2} - 2m_s^{2} )\,.
\label{A11}
\end{eqnarray}
We conclude, that the expression (20) of \cite{Gervais:2017yxv} which
is supposed to give Low's theorem up to order $\omega^{0}$
contains, in our simple example, the arbitrary quantity
$h' ( 2 m_f^{2} - 2m_s^{2} )$.
Thus, in our opinion, this equation has a problem.
On the other hand, inserting $\widetilde{\cal T}$ from (\ref{A8})
in our Eq.~(\ref{A1}) we get a sensible result.
Here, the on shell $\widetilde{\cal T}$ equals the constant (\ref{A9})
and on the r.h.s. of (\ref{A1}) 
only the terms proportional to 
${\cal M}^{(0)}$ survive, 
$\partial{\cal M}^{(0)}/\partial s_{L}$
and $\partial{\cal M}^{(0)}/\partial t$ being zero.

Finally we want to discuss the form of Low's theorem
presented in Eqs.~(2.8), (2.9) of \cite{Bern:2014vva}
and (25), (26) of \cite{Lysov:2014csa}.
For our reactions (\ref{2.1}) and (\ref{2.2}) these give for
${\cal M}_{\lambda}$ from our Eq.~(\ref{2.13})
\begin{eqnarray}
{\cal M}_{\lambda}
&=& 
\big\lbrace 
\frac{e}{(p_{a} \cdot k)} \big( p_{a \lambda} - i \eta_{a} k^{\nu} J_{a \lambda \nu} \big)
- \frac{e}{(p_{1} \cdot k)} \big( p_{1 \lambda} - i \eta_{1} k^{\nu} J_{1 \lambda \nu} \big)
\big\rbrace 
\widetilde{\cal T}(p_{a}, p_{b}, p_{1}, p_{2})
\nonumber \\
&&+ {\cal O}(\omega)\,.
\label{A12}
\end{eqnarray}
Here
\begin{eqnarray}
&&J_{a \lambda \nu} =
i \big( p_{a \lambda} \frac{\partial}{\partial p_{a}^{\nu}} -
        p_{a \nu}\frac{\partial}{\partial p_{a}^{\lambda}} \big)\,, \nonumber \\
&&J_{1 \lambda \nu} =
i \big( p_{1 \lambda} \frac{\partial}{\partial p_{1}^{\nu}} -
        p_{1 \nu}\frac{\partial}{\partial p_{1}^{\lambda}} \big)\,,
\label{A13}
\end{eqnarray}
and we have inserted factors $\eta_{a} = \pm 1$ and $\eta_{1} = \pm 1$ 
in (\ref{A12})
because we could not always find out the precise momentum orientations
used in \cite{Bern:2014vva} and \cite{Lysov:2014csa}.
But this will play no role in the following.

Now we shall use as trial function in (\ref{A12})
\begin{eqnarray}
\widetilde{\cal T}(p_{a}, p_{b}, p_{1}, p_{2}) =
\tilde{h} \big[ (p_{a} + p_{b})^{2} - (p_{1} + p_{2})^{2} \big]\,,
\label{A14}
\end{eqnarray}
where $\tilde{h}$ is a function satisfying
\begin{eqnarray}
\tilde{h}(0) \neq 0 \quad {\rm and} \quad \tilde{h}'(0) \neq 0
\label{A15}
\end{eqnarray}
but otherwise arbitrary. In the physical region we have,
on shell and off shell, $p_{a} + p_{b} = p_{1} + p_{2}$,
and therefore our ${\cal M}^{(0)}$ from (\ref{2.8}) is given by
\begin{eqnarray}
{\cal M}^{(0)} = \tilde{h}(0) = {\rm const}\,.
\label{A16}
\end{eqnarray}
Thus, from (\ref{A1}) we get our result as
\begin{eqnarray}
{\cal M}_{\lambda}
= e 
\big\lbrace  \frac{p_{a \lambda}}{(p_{a} \cdot k)} 
- \frac{p_{1 \lambda}}{(p_{1} \cdot k)}
- \frac{1}{(p_{1} \cdot k)^{2}}
\big[ p_{1 \lambda} (l_{1} \cdot k) - l_{1 \lambda} (p_{1} \cdot k)  \big] \big\rbrace \, \tilde{h}(0)
+ {\cal O}(\omega)\,. \qquad
\label{A17}
\end{eqnarray}
On the other hand, from (\ref{A12}) we find
\begin{eqnarray}
{\cal M}_{\lambda}
&=& e 
\big\lbrace  \big[ \frac{p_{a \lambda}}{(p_{a} \cdot k)} 
- \frac{p_{1 \lambda}}{(p_{1} \cdot k)} \big] \, \tilde{h}(0)
\nonumber \\
&&+ \big[ \eta_{a} \big( p_{a \lambda} \frac{(p_{b} \cdot k)}{(p_{a} \cdot k)} - p_{b \lambda} \big)
         +\eta_{1} \big( p_{1 \lambda} \frac{(p_{2} \cdot k)}{(p_{1} \cdot k)} - p_{2 \lambda} \big)
    \big] 2 \tilde{h}'(0) \big\rbrace
+ {\cal O}(\omega)\,.
\label{A18}
\end{eqnarray}
Clearly, the results (\ref{A17}) and (\ref{A18}) differ.
In (\ref{A18}) we also see the completely arbitrary quantity
$\tilde{h}'(0)$ occurring.
That is, at least for this example (\ref{A1}) and (\ref{A12})
are not equivalent.

The reader may wonder if our trial function (\ref{A14})
is reasonable since in $\pi^{-} \pi^{0} \to \pi^{-} \pi^{0}$
we have the same particles in the initial and the final state.
Should there be some symmetry requirement for $\tilde{h}(\cdot)$?
We can counter such an argumentation by considering instead of
$\pi^{-} \pi^{0} \to \pi^{-} \pi^{0}$
the reaction $\pi^{-} \pi^{0} \to K^{-} K^{0}$ where
there is no symmetry between the initial and the final state.
The results (\ref{A17}) and (\ref{A18}) stay the same.

With this we close our remarks on some papers from the literature.

\section{The cross section $d\sigma/d\omega$ for $\omega \to 0$}
\label{sec:appendixB}

In this appendix we shall discuss the cross section
$d\sigma/d\omega$ for $\omega \to 0$ for the reaction
$\pi^{-} \pi^{0} \to \pi^{-} \pi^{0} \gamma$
for real photon emission.
The results for charged-pion scattering are analogous.

We consider, thus, the reaction (\ref{2.2}) where the scattering
amplitude is defined in (\ref{2.13}).
The cross section is given as in (\ref{4.37}).
We work in the overall c.m. system, setting $k^{0} \equiv \omega$
and
\begin{eqnarray}
{\cal M}_{\lambda} \equiv
{\cal M}_{\lambda}^{(\pi^{-} \pi^{0} \to \pi^{-} \pi^{0} \gamma)}\,.
\label{B1}
\end{eqnarray}
\begin{eqnarray}
d\sigma({\pi^{-}\pi^{0} \to \pi^{-}\pi^{0} \gamma(k)}) &=&
\frac{1}{2\sqrt{s(s-4 m_{\pi}^{2})}}\,
\frac{d^{3}k}{(2 \pi)^{3} \,2 k^{0}}
\int \frac{d^{3}p_{1}'}{(2 \pi)^{3} \,2 p_{1}'^{0}}
\frac{d^{3}p_{2}'}{(2 \pi)^{3} \,2 p_{2}'^{0}} 
\nonumber \\
&&\times (2 \pi)^{4} \delta^{(4)}(p_{1}'+p_{2}'+k-p_{a}-p_{b})
(-1){\cal M}_{\lambda} ( {\cal M}^{\lambda} )^{*} \nonumber\\
&=&\frac{1}{\sqrt{s(s-4 m_{\pi}^{2})}}
\frac{1}{2^{4} (2 \pi)^{5}} \, \omega \,d\omega \,d\Omega_{\hat{k}}
\int d\Omega_{\hat{p}_{1}'}
\int_{0}^{\infty} d|\bpaap| \frac{|\bpaap|^{2}}{p_{1}'^{0}p_{2}'^{0}}\nonumber\\&&\times \delta(p_{1}'^{0}+p_{2}'^{0}+\omega-\sqrt{s})
(- {\cal M}_{\lambda}  {\cal M}^{\lambda*} )\,.
\label{B2}
\end{eqnarray}
Here we have
\begin{eqnarray}
p_{1}'^{0} &=& \sqrt{|\bpaap|^{2} + m_{\pi}^{2}}\,, \nonumber \\
\bpbbp &=& -\bpaap - \bk\,, \nonumber \\
p_{2}'^{0} &=& \sqrt{|\bpbbp|^{2} + m_{\pi}^{2}} \nonumber \\
           &=& \sqrt{|\bpaap|^{2} + 2 |\bpaap| \omega (\bhpaap \cdot \bhk) + \omega^{2}+m_{\pi}^{2}}\,,
\label{B3}
\end{eqnarray}
where $\bhpaap = \bpaap / |\bpaap|$ and
$\bhk = \bk / |\bk| = \bk / \omega$.
The $\delta$ function of the energies in (\ref{B2}) requires
\begin{eqnarray}
p_{1}'^{0}+p_{2}'^{0} = \sqrt{s}-\omega
\label{B4}
\end{eqnarray}
which allows us to calculate $|\bpaap|$ for given
$\sqrt{s}$, $\omega$, $\bhpaap$ and $\bhk$.
We get as solution
\begin{eqnarray}
|\bpaap| &=& \sqrt{ \frac{s(\sqrt{s}-2 \omega)^{2} - 4 (\sqrt{s}-\omega)^{2} m_{\pi}^{2}}{4 \left[ (\sqrt{s}-\omega)^{2} - \omega^{2} (\bhpaap \cdot \bhk)^{2} \right]} 
+ \frac{1}{4} \left[ \frac{\sqrt{s} (\sqrt{s} - 2 \omega) \omega (\bhpaap \cdot \bhk)}{(\sqrt{s} - \omega)^{2}-\omega^{2} (\bhpaap \cdot \bhk)^{2}} \right]^{2}} 
\nonumber\\
&&- \frac{1}{2} 
\frac{\sqrt{s} (\sqrt{s} - 2 \omega) \omega (\bhpaap \cdot \bhk)}
     {(\sqrt{s} - \omega)^{2}-\omega^{2} (\bhpaap \cdot \bhk)^{2}}\,.
\label{B5}
\end{eqnarray}
For $\omega = 0$ this gives, as it must be,
\begin{eqnarray}
\left.|\bpaap|\right|_{\omega = 0} = \frac{1}{2} \sqrt{s - 4 m_{\pi}^{2}} \,.
\label{B6}
\end{eqnarray}

Now we define the phase space function
\begin{eqnarray}
{\rm J}(s, \omega, \bhpaap, \bhk)&=&
\int_{0}^{\infty} d|\bpaap| \frac{|\bpaap|^{2}}{p_{1}'^{0}p_{2}'^{0}}
\,\delta(p_{1}'^{0}+p_{2}'^{0}+\omega-\sqrt{s})\nonumber\\
&=&
\frac{|\bpaap|^{2}}{|\bpaap| p_{2}'^{0} + p_{1}'^{0} 
\left[|\bpaap| + \omega (\bhpaap \cdot \bhk) \right]} \,.
\label{B7}
\end{eqnarray}
Here $p_{1}'^{0}$ and $p_{2}'^{0}$ have to be substituted according to
(\ref{B3}) and finally everywhere $|\bpaap|$ from (\ref{B5})
has to be inserted. For $\omega = 0$ we get
\begin{eqnarray}
{\rm J}(s, 0, \bhpaap, \bhk)=
\frac{1}{2} \sqrt{1 - \frac{4 m_{\pi}^{2}}{s}}\,.
\label{B8}
\end{eqnarray}

Collecting everything together we find
\begin{eqnarray}
\frac{d\sigma}{d\omega}(\pi^{-}\pi^{0} \to \pi^{-}\pi^{0} \gamma)=
\frac{1}{\sqrt{s(s-4 m_{\pi}^{2})}}
\frac{1}{2^{4} (2 \pi)^{5}} \,
\omega
\int d\Omega_{\hat{k}}\, d\Omega_{\hat{p}_{1}'}\,
{\rm J}(s, \omega, \bhpaap, \bhk)
(- {\cal M}_{\lambda}  {\cal M}^{\lambda*} ). \nonumber\\
\label{B9}
\end{eqnarray}

We are interested in the behaviour of $d\sigma / d\omega$
for $\omega \to 0$.
Therefore, we shall now consider the expansion of
${\cal M}_{\lambda}$ in powers of $\omega$ as given in (\ref{A1}).
We have used in (\ref{B9}) $\bhk$ and $\bhpaap$ as phase-space variables.
Therefore, we should choose the expansion of ${\cal M}_{\lambda}$
keeping $\bhpaap$ fixed, independent of $\omega$:
\begin{eqnarray}
\bhpaap = \bhpaa\,.
\label{B10}
\end{eqnarray}
This means that we choose in the expansion (\ref{A1})
\begin{eqnarray}
\blaperp = 0\,;
\label{B11}
\end{eqnarray}
see (\ref{3.19}). 
Then $k^{\mu}$, $l_{1}^{\mu}$ and $l_{2}^{\mu}$ are,
for fixed $\bhk$, strictly proportional to $\omega$.
Therefore, we can write from (\ref{A1}):
\begin{eqnarray}
{\cal M}_{\lambda} =
\frac{1}{\omega}\widehat{{\cal M}}_{\lambda}^{(0)}+
                \widehat{{\cal M}}_{\lambda}^{(1)}+
{\cal O}(\omega)\,,
\label{B12}
\end{eqnarray}
where
\begin{eqnarray}
\widehat{{\cal M}}_{\lambda}^{(0)} &=&
\widehat{{\cal M}}_{\lambda}^{(0)}(s, \bhpa, \bhpaa, \bhk) \nonumber\\
&=& e \omega \big[ \frac{p_{a \lambda}}{(p_{a} \cdot k)} 
-      \frac{p_{1 \lambda}}{(p_{1} \cdot k)} \big]
{\cal M}^{(0)}(s_{L},t,m_{\pi}^{2},m_{\pi}^{2},m_{\pi}^{2},m_{\pi}^{2})\,,
\label{B13}\\
\widehat{{\cal M}}_{\lambda}^{(1)} &=&
\widehat{{\cal M}}_{\lambda}^{(1)}(s, \bhpa, \bhpaa, \bhk) \nonumber\\
&=& e 
\big\lbrace 
-\frac{1}{(p_{1} \cdot k)^{2}}
\big[ p_{1 \lambda} (l_{1} \cdot k) - l_{1 \lambda} (p_{1} \cdot k)  \big]
+ 2 \big[ - \frac{(p_{b} \cdot k)}{(p_{a} \cdot k)} p_{a\lambda} + p_{b\lambda} \big]
\frac{\partial}{\partial s_{L}}  \nonumber \\
&&
- 2 \big[ (p_{a} - p_{1},k) - (p_{a} \cdot l_{1}) \big] 
\big[ \frac{p_{a\lambda}}{(p_{a} \cdot k)} - \frac{p_{1\lambda}}{(p_{1} \cdot k)} \big] 
\frac{\partial}{\partial t} 
\big\rbrace
{\cal M}^{(0)}(s_{L},t,m_{\pi}^{2},m_{\pi}^{2},m_{\pi}^{2},m_{\pi}^{2})\,. \nonumber\\
\label{B14}
\end{eqnarray}

Inserting (\ref{B12}) in (\ref{B9}) we find
\begin{eqnarray}
\frac{d\sigma}{d\omega}(\pi^{-}\pi^{0} \to \pi^{-}\pi^{0} \gamma)=
\frac{1}{\omega} \big[ {\cal A}^{(0)}(s,\omega)+
                       \omega {\cal A}^{(1)}(s,\omega)+
{\cal O}(\omega^{2}) \big]\,,
\label{B15}
\end{eqnarray}
where we have with (\ref{B10})
\begin{eqnarray}
&&{\cal A}^{(0)}(s,\omega)=
\frac{1}{\sqrt{s(s-4 m_{\pi}^{2})}}
\frac{1}{2^{4} (2 \pi)^{5}}
\int d\Omega_{\hat{k}}\, d\Omega_{\hat{p}_{1}}\,
{\rm J}(s, \omega, \bhpaa, \bhk)
(- \widehat{{\cal M}}^{(0)}_{\lambda} 
   \widehat{{\cal M}}^{(0)\lambda*} )\,,
\label{B16}\\
&&{\cal A}^{(1)}(s,\omega)=
\frac{1}{\sqrt{s(s-4 m_{\pi}^{2})}}
\frac{1}{2^{4} (2 \pi)^{5}}
\int d\Omega_{\hat{k}}\, d\Omega_{\hat{p}_{1}}\,
{\rm J}(s, \omega, \bhpaa, \bhk) \nonumber\\
&& \qquad \qquad \qquad \times
\big[- \widehat{{\cal M}}^{(1)}_{\lambda} (\widehat{{\cal M}}^{(0)\lambda})^{*}
     - \widehat{{\cal M}}^{(0)}_{\lambda} (\widehat{{\cal M}}^{(1)\lambda})^{*} \big]\,.
\label{B17}
\end{eqnarray}

We clearly have from (\ref{B16})
\begin{eqnarray}
{\cal A}^{(0)}(s,\omega) \neq 0\,.
\label{B18}
\end{eqnarray}
Furthermore, ${\cal A}^{(0)}(s,\omega)$ 
and ${\cal A}^{(1)}(s,\omega)$ are continuous functions
of $\omega$ in the region of interest for us.

Given the amplitude ${\cal M}^{(0)}$ for the basic process
$\pi^{-} \pi^{0} \to \pi^{-} \pi^{0}$ the result for
$d\sigma / d\omega$ in (\ref{B15}) is a strict consequence of QFT.
Thus, the experimental cross section should also be of the form
\begin{eqnarray}
\left.\frac{d\sigma}{d\omega}(\pi^{-}\pi^{0} \to \pi^{-}\pi^{0} \gamma)
\right|_{\rm exp} =
\frac{1}{\omega} \big[ {\cal A}^{(0)}(s,\omega)+
                       \omega {\cal A}^{(1)}(s,\omega)+
{\cal O}(\omega^{2}) \big]\,.
\label{B19}
\end{eqnarray}
Therefore, if we have a good ``standard'' representation
of the basic $\pi^{-} \pi^{0} \to \pi^{-} \pi^{0}$ process,
compatible with experiment,
and if the corresponding calculation of $d\sigma / d\omega$ respects
the rules of QFT, in particular the relation (\ref{A1}),
this standard result must also give the expansion (\ref{B15})
for $\omega \to 0$ and we shall then have
\begin{eqnarray}
R_{\rm exp}(\omega) = 
\frac{d\sigma_{\rm exp}/d\omega}{d\sigma_{\rm standard}/d\omega}
= 1 + {\cal O}(\omega^{2})\,.
\label{B20}
\end{eqnarray}
Equation (\ref{B20}) should, in particular, be true for our standard result
as discussed in Sec.~\ref{sec:4.1} 
for $\pi^{-} \pi^{0} \to \pi^{-} \pi^{0} \gamma$
and in Sec.~\ref{sec:4.2} for charged-pion scattering.
Of course, there we must stay in the relevant energy range,
that is for large $\sqrt{s}$ where pomeron exchange is dominant.

From Eq.~(\ref{B20}) we get immediately Eq.~(\ref{6.2}) for
which we have, thus, given a detailed derivation.
Finally we note that (\ref{B20}) and (\ref{6.2}) will also hold
if cuts in phase space are introduced, e.g., of the following forms:
\begin{eqnarray}
|\bhpa \cdot \bhk| \leqslant c < 1\,,
\label{B21}
\end{eqnarray}
or
\begin{eqnarray}
|\bhpaa \cdot \bhk| \leqslant c < 1\,,
\label{B22}
\end{eqnarray}
or
\begin{eqnarray}
(|\bhpa \cdot \bhk| \leqslant c < 1) \wedge
(|\bhpaa \cdot \bhk| \leqslant c < 1)\,.
\label{B23}
\end{eqnarray}
Applying, for instance, the cut (\ref{B21}) we have to replace
in (\ref{B16}) and (\ref{B17}) the kinematic function
${\rm J}(s, \omega, \bhpaa, \bhk)$ from (\ref{B7}) by
\begin{eqnarray}
{\rm J}(s, \omega, \bhpaa, \bhk) \,\theta(c-|\bhpa \cdot \bhk|)\,.
\label{B24}
\end{eqnarray}
If this cut is applied both to the standard calculation
and to the experimental determination of $d\sigma / d\omega$
the result is again (\ref{B20}).
The analogous statements hold for the cuts of the type
(\ref{B22}) and (\ref{B23}).

Finally we note that relations for the radiative cross sections
for $\omega \to 0$ were discussed already a long time ago by
Burnett and Kroll \cite{Burnett:1967km}.
They considered, in particular, the case where the charged particle
carries spin 1/2. 
The critique which we have
to formulate here is that their results contain derivatives
of the non-radiative amplitudes with respect to
one momentum keeping the other ones fixed, i.e.
derivatives where one has to extrapolate into the unphysical region; see (\ref{A4}).
We have shown in Appendix~\ref{sec:appendixA} that
in this way one can get essentially any arbitrary result.
We have demonstrated in our present paper that such extrapolations
into the unphysical region of the basic amplitude are
never necessary when our methods are used.

\section*{Remarks}
Here we add some clarifying remarks concerning our paper.

\begin{enumerate}

\item 
It should be noted that the soft-photon theorems 
in the versions of both Low and Weinberg 
hold also for spin-flip amplitudes in hadronic high-energy scattering.
An example is the photon emission in pion-proton scattering
in the soft-photon limit discussed 
from a general quantum field theory (QFT) 
point of view in \cite{Lebiedowicz:2023ell}.
There is no model dependence
for the terms of orders $\omega^{-1}$ and $\omega^{0}$ 
in the corresponding radiative amplitude.

\item 
In the first paragraph of Sec.~\ref{sec:4}, 
the reader may find some historical remarks on Regge theory and the pomeron.
There we also write that the tensor-pomeron model which we use
in this article is for \underline{soft} hadronic high-energy reactions.
For extensive discussions of this model 
see \cite{Ewerz:2013kda,Ewerz:2016onn}.
This model is convenient for us to use,
but there is no claim that this model is in any sense unique.
In our present paper there is also no intention
to apply \underline{this} tensor-pomeron model 
to \underline{hard} diffractive reactions. 
For these latter reactions other concepts and models
should be and are used.
In particular, methods of perturbative QCD can be applied there;
see for instance \mbox{\cite{Crittenden:1997yz,Ivanov:1998gk}}.
But let us note that in \cite{Britzger:2019lvc} 
an extension of the above tensor-pomeron model
for soft reactions \cite{Ewerz:2013kda},
including in addition a hard tensor pomeron, 
gave a very satisfactory description of the low-$x$ 
deep inelastic structure functions
for $0 \leq Q^{2} \leq 50$~GeV$^{2}$.
This latter model was also successfully applied 
for a description of the HERA data 
on Deeply Virtual Compton Scattering (DVCS) 
in \cite{Lebiedowicz:2022xgi}.

\end{enumerate}

\bibliography{refs}

\end{document}